\documentclass[prc,superscriptaddress,twocolumn,showpac]{revtex4-2}

\usepackage[body={18cm,25cm},top=1.5cm]{geometry}
\usepackage{amsmath,amssymb,bm,bbm}
\usepackage{graphicx}
\usepackage{multirow}
\usepackage{rotating}
\usepackage{xcolor}
\usepackage{tikz}
\usepackage{hyperref}
\usepackage{CJKutf8}
\usepackage{longtable}
\usepackage{ulem}

\definecolor{darkgreen}{rgb}{0,0.5,0}
\definecolor{green3}{rgb}{0.20,0.60,0.20}
\definecolor{Black}{rgb}{0.000000,0.000000,0.000000}
\definecolor{Blue}{rgb}{0.000000,0.000000,1.000000}
\definecolor{Cyan}{rgb}{0.000000,1.000000,1.000000}
\definecolor{Red}{rgb}{1.000000,0.000000,0.000000}
\definecolor{Green}{rgb}{0.000000,1.000000,0.000000}
\definecolor{Magenta}{rgb}{1.000000,0.000000,1.000000}
\definecolor{White}{rgb}{1.000000,1.000000,1.000000}
\definecolor{Yellow}{rgb}{1.000000,1.000000,0.000000}
\definecolor{Violet}{rgb}{0.552941,0.219608,0.788235}

\newcommand{\eq}[1]{\begin{equation} #1 \end{equation}}
\newcommand{\ket}[1]{ | #1 \rangle }

\newcommand{\elmx}[3]{\langle #1 | #2 | #3 \rangle}
\newcommand{\vect}[1]{\mathbf{#1}}

\newcommand{\EGM}{\Delta E_{\rm GM}}

\makeatletter
\def\p@subsection{}
\makeatother

\begin{document}

\begin{CJK*}{UTF8}{gbsn}

% Title

\title{Gallagher--Moszkowski splitting in deformed odd-odd nuclei
  within a microscopic approach}

\author{L. Bonneau}
\email[Corresponding author: ]{bonneau@lp2ib.in2p3.fr}
\affiliation{LP2i Bordeaux, UMR 5797, Universit\'e de Bordeaux, CNRS, F-33170,
  Gradignan, France} 

\author{N. Kontowicz}
\affiliation{LP2i Bordeaux, UMR 5797, Universit\'e de Bordeaux, CNRS, F-33170,
  Gradignan, France} 

\author{J. Bartel}
\affiliation{IPHC, UMR 7178, Universit\'e de Strasbourg, CNRS,
  F-67000, Strasbourg, France} 

\author{H. Molique}
\affiliation{IPHC, UMR 7178, Universit\'e de Strasbourg, CNRS,
  F-67000, Strasbourg, France} 

\author{Meng-Hock Koh (辜 明 福)}
\affiliation{Department of Physics, Faculty of Science, Universiti
  Teknologi Malaysia, 81310 Johor Bahru, Johor, Malaysia} 
\affiliation{UTM Centre for Industrial and Applied Mathematics, 81310
  Johor Bahru, Johor, Malaysia} 

\author{N. Minkov}
\affiliation{Institute of Nuclear Research and Nuclear Energy,
  Bulgarian Academy of Sciences, Tzarigrad Road 72, BG-1784, Sofia,
  Bulgaria} 

\date{\today}

% Abstract

\begin{abstract}
  \begin{description}
  \item[Background] Low-lying bandhead states in axially prolate deformed
    odd-odd nuclei have long been described essentially within the
    rotor+two-quasiparticle picture. This approach allows one to
    explain the appearance of so-called Gallagher--Moszkowski doublets
    of bandheads with $K=\Omega_n\pm\Omega_p$, sum and difference of
    neutron and proton angular momentum projections on the symmetry
    axis. According to an empirical rule stated by Gallagher and
    Moszkowski the spin-aligned configuration lies lower in energy than
    the spin-anti-aligned configuration. A recent study by Robledo,
    Bernard and Bertsch in [Phys. Rev. C{\bf 89}, 021303(R) (2014)]
    within the Gogny energy-density functional with selfconsistent
    blocking of the unpaired nucleons showed that calculations fail to
    reproduce this rule in about half of the cases and points to the
    density-dependent term of the functional as responsible of this 
    failure. 
  \item[Purpose] In this paper we aim at pushing further this analysis to
    exhibit the mechanism underlying the energy splitting in a
    Gallagher--Moszkowski doublet.
  \item[Method] We work in the framework of the Skyrme energy-density
    functional approach, including BCS pairing correlations with
    selfconsistent blocking. We use the SIII parametrization with
    time-odd terms and seniority pairing matrix elements extending a
    previous study of $K$-isomeric states in even-even nuclei [Phys. Rev. C
      \textbf{105}, 044329 (2022)].
  \item[Results] We find that the energy splitting results from
a competition between the spin-spin, density-dependent and
current-current terms of the Skyrme energy-density functional.
  \item[Conclusions] In doublets where the larger $K$ 
value is lower in energy the Gallagher--Moszkowski rule is always
satisfied by the SIII Skyrme energy-density functional. In
doublets, on the contrary, where the smaller $K$ value lies lower, the
energy splittings are calculated to be rather small and often a 
disagreement with the Gallagher--Moszkowski rule occurs.
\end{description}
\end{abstract}

\maketitle

\end{CJK*}

%-------------------------------------------------------------------------
%
%
%                           Introduction
%
%
%-------------------------------------------------------------------------

\section{Introduction}

\begin{figure}[b]
  \includegraphics[width=0.3\textwidth]{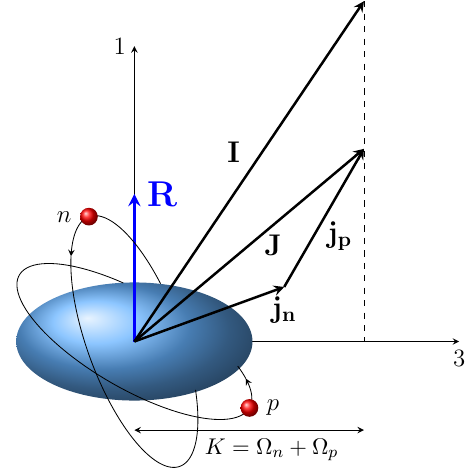}
  \caption{Schematic representation of collective $\vect R$, intrinsic
    $\vect J = \vect j_n + \vect j_p$ and total $\vect I = \vect R +
    \vect J$ angular momenta in an axially deformed odd-odd nucleus
    within the deformation-alignment scheme.\label{fig_rotor+2qp}}
\end{figure}

Deformed doubly-odd nuclei have long been described within the
rotor+two-quasiparticle picture as a rotating core coupled through the
so-called Coriolis term to one neutron and one proton moving in a
spheroidal mean field~\cite{Bohr-Mottelson53}. Axial deformation 
in such nuclei provides $K$, the projection on the symmetry axis of
the intrinsic angular momentum, as a good quantum number. In the
deformation-alignment scheme, the core rotates around an axis
orthogonal to the symmetry axis of the nucleus and the projection on
the symmetry axis of the total angular momentum of the nucleus is thus
equal to $K$. Moreover $K$ is the sum of the neutron $\Omega_n$ and
the proton $\Omega_p$ angular momentum projections on the symmetry
axis (see Fig.~\ref{fig_rotor+2qp}).
Bohr and Mottelson~\cite{Bohr-Mottelson53} and Peker~\cite{Peker57}
showed that the ground-state angular momentum of many deformed odd-odd
nuclei can be accounted for by the coupling of the unpaired proton and
neutron angular momenta. Gallagher and Moszkowski then proposed 
in Ref.~\cite{Gallagher-Moszkowski58}
a spin-spin coupling rule to
determine whether the lower lying configuration corresponds to
$\Omega_n+\Omega_p$ or $|\Omega_n-\Omega_p|$. According to this
rule, the lower-lying configuration is 
the one in which spins of the neutron and the proton are parallel.
This rule is remarkably successfull as, even to date, it suffers from
only one exception, in the $^{166}$Ho nucleus. As reviewed by Boisson
at al.~\cite{Boisson76}, this success was accompanied by several early
theoretical investigations~\cite{Pyatov63,DePinho65_PLB15,Nunberg65} to
understand its origin in terms of the residual interaction between the
unpaired neutron and proton in the framework of the
two-quasiparticle+rotor model. About two decades later, Jain and  
collaborators provided in Ref.~\cite{Jain98_RMP} a comprehensive
review of the structure of deformed odd-odd rare-earth nuclei,
including a systematic analysis of the neutron-proton
residual interaction in the two-quasiparticle+rotor model. Note that,
however, at the time of this  review, the work on the tensor
contribution to the neutron-proton residual interaction, in the same
model, just published by Covello and collaborators~\cite{Covello97}
was not mentionned in Ref.~\cite{Jain98_RMP}.
More recently two systematic studies of odd-odd nuclei were
performed. On the one hand Robledo, Bernard and
Bertsch~\cite{Robledo14} investigated the Gallagher--Moszkowski 
splitting with the Gogny energy-density functional (EDF) with
selfconsistent blocking. They identified the density-dependent term of
the functional as responsible for the disagreement with the
Gallagher--Moszkowski rule where it occured. 
On the other hand Ward and collaborators~\cite{Ward19} performed
systematic calculations of ground-state spins and parities of odd-odd
nuclei across the nuclear chart within the macroscopic-microscopic
finite-range droplet model combined with a two-quasiparticle+rotor
model with various residual interactions.

Overall very few EDF based calculations in deformed odd-odd
nuclei have been performed so far. In Ref.~\cite{Bennour87} Bennour
and collaborators studied the spectroscopic properties of axially
deformed doubly-odd nuclei in the rare-earth and actinide regions in
the two-quasiparticle+rotor picture with Skyrme-EDF intrinsic
solutions. The SIII parametrization of the effective Skyrme
interaction was used to generate the neutron and proton quasiparticle
energies and their interaction matrix elements (playing the role
of the neutron-proton residual interaction). More
precisely they described the bandhead states of the considered odd-odd
nuclei by the creation of two quasiparticles on the (fully paired)
Hartree--Fock--BCS ground state of a neighboring even-even nucleus. In
contrast Robledo and collaborators~\cite{Robledo14} later used the
Gogny EDF with selfconsistent blocking (SCB), hence breaking the
time-reversal symmetry at the one-body level. In the present paper we
extend the recent study of two-quasiparticle states in even-even
nuclei~\cite{Minkov22} within the Skyrme-EDF framework with BCS
pairing and selfconsistent blocking to deformed odd-odd nuclei. 
Even though several codes based on Skyrme EDF with superior 
capabilities already exist~\cite{Maruhn14,Schunck17,Ryssens21}, we are 
not aware of any published applications to deformed doubly-odd
nuclei, neither within the Skyrme nor within the Gogny EDF, after those
of Ref.~\cite{Robledo14}.

In the present work we focus on Gallagher--Moszkowski doublets 
and aim at understanding the mechanism of their energy splitting
within the Skyrme-EDF framework with BCS pairing and selfconsistent
blocking, pushing further the analysis performed by Robledo and
collaborators~\cite{Robledo14}. Because we focus on deciphering the
microscopic mechanism at work behind the Gallager-Moszkowski empirical
rule, we do not perform our study on a systematic basis,
but in selected strongly deformed nuclei in two mass regions and
for which sufficient data on Gallagher--Moszkowski doublets
are experimentally known: 
\begin{itemize}
  \item rare-earth mass region and beyond: odd-odd nuclei surrounding the
    even-even $^{156,158}$Gd, $^{174}$Yb and $^{178}$Hf nuclei (see
    Fig.~\ref{fig_nucleid_chart_rare-earths}). 
\begin{figure}[h]
  \includegraphics[width=0.43\textwidth]{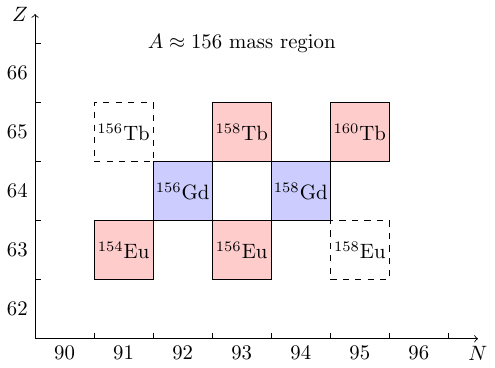}
  \includegraphics[width=0.43\textwidth]{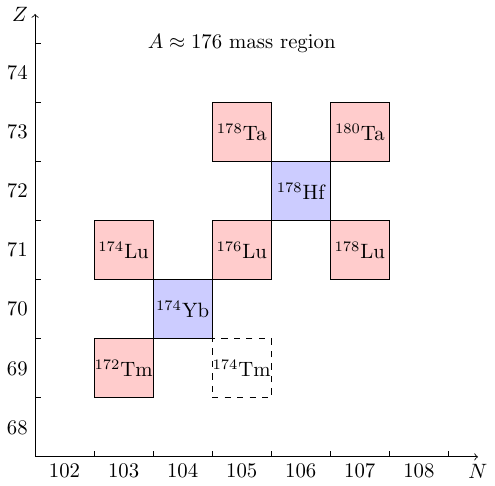}
  \caption{\label{fig_nucleid_chart_rare-earths} Portions of the
    nuclid chart in the vicinity of studied rare-earth nuclei. Dashed
    boxes represent odd-odd nuclei for which too few data are
    available and are thus not studied here.}
\end{figure}
  \item actinide mass region between $A\approx230$ and $A\approx250$:
    odd-odd nuclei surrounding the even-even nuclei $^{230,232}$Th,
    $^{240,242}$Pu and $^{250,252}$Cf (see
    Fig.~\ref{fig_nucleid_chart_actinides}).
\begin{figure*}[t]
  \includegraphics[width=0.9\textwidth]{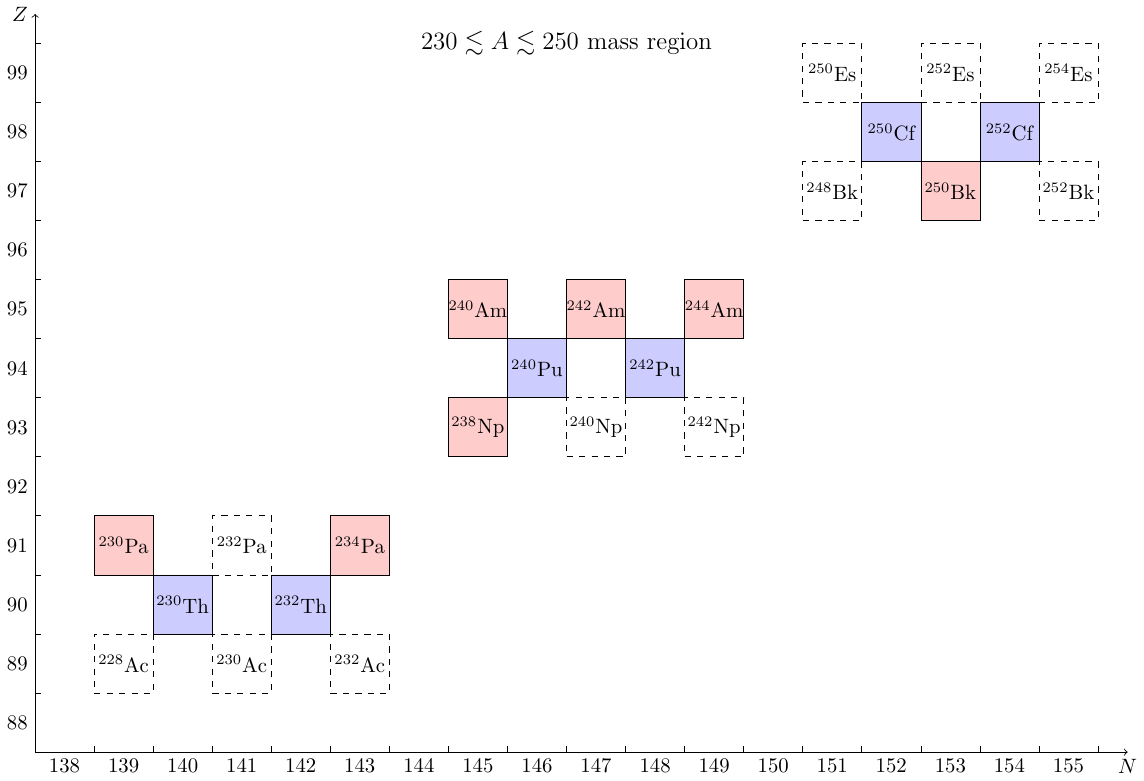}
  \caption{\label{fig_nucleid_chart_actinides} Same as
    Fig.~\ref{fig_nucleid_chart_rare-earths} for the actinide nuclei
    of interest.}
\end{figure*}
\end{itemize}

After a brief presentation of the theoretical framework and
calculational details in the next section, we present in section
III the resulting bandhead spectra obtained within selfconsistent
blocking for the Gallagher--Moszkowski doublets known
experimentally. Then we show the relevance of perturbative blocking
for the investigation of the splitting mechanism in section
IV. Finally we draw conclusions and give perspectives in section V. 

%-------------------------------------------------------------------------
%
%
%            Theoretical framework and calculational details
%
%
%-------------------------------------------------------------------------

\section{Theoretical framework and calculational details}

All two-quasiparticle states of the considered nuclei are
assumed to have axially symmetric and left-right symmetric shapes in
the intrinsic (body-fixed) frame. They are described within the
Skyrme+BCS energy-density functional with selfconsistent blocking (of
one neutron and one proton single-particle states) as explained in 
Ref.~\cite{Bonneau15}. As in a previous study of $K$-isomeric
two-quasiparticle states~\cite{Minkov22} we use the SIII 
parametrization~\cite{Beiner75} of the Skyrme energy-density
functional and constant pairing matrix elements (often called
``seniority'' or BCS pairing interaction).

Because of the selfconsistent blocking, time-reversal symmetry is
broken at the one-body level and the Kramers degeneracy is removed in
the single-particle energy spectrum. Moreover, as explained in
Refs.~\cite{Bonneau15,Koh17}, we work in the ``minimal'' scheme of the SIII
Skyrme parametrization, in which the only time-odd fields retained in
the Hartree--Fock Hamiltonian are the spin and current vector
fields. Including the other terms would introduce a bias as they are
accompanied by terms involving time-even densities that were not taken
into account in the SIII fitting protocol~\cite{Beiner75}. In the
notation of the appendix of Ref.~\cite{Koh17} this corresponds to
keeping the $B_3$ (current-current term), $B_9$ (spin-orbit term),
$B_{10,11}$ (spin-spin terms), and $B_{12,13}$ (density-dependent
spin-spin terms).

The pairing contribution to the energy-density functional is
calculated from the expectation value of the pairing interaction in a
BCS state including blocking. The parametrization of the nucleon-number
dependence of the pairing matrix elements is the same as in
Ref.~\cite{Bonneau15} and the fitting protocol of their strength is
presented in Refs.~\cite{Nor19,Minkov22}. In this work we use the same
values as in Ref.~\cite{Nor19} for the nuclei around $A=178$
($^{174,176,178}$Lu, $^{178,180}$Ta) and the same values as in
Ref.~\cite{Minkov22} for the actinide nuclei ($^{230,234}$Pa,
$^{238}$Np, $^{242}$Am). In the medium-heavy rare-earth nuclei around
$A=156$, we adjust the neutron pairing strength parameter $G_n$ 
to the first $2^+$ excitation energy in $^{156}$Gd 
keeping that of protons $G_p$ equal to $0.9 \times G_n$ as done in
the other mass regions (see Ref.~\cite{Nor19} for a precise definition of
$G_n$ and $G_p$). This observable is shown in Ref.~\cite{Nor19} to be a
relevant measure of pairing correlations equivalent to the odd-even
binding-energy differences. We find $G_n=17.1$~MeV. The BCS equations
are then solved for all single-particle states with a smearing factor
$f(e_i) = \big[1+\exp\big((e_i-X-\lambda_q)/\mu\big)\big]^{-1}$ where
$e_i$ is the energy of the single-particle state $\ket i$, $X$ plays
the role of a limiting factor with $X \!=\! 6\;$MeV, $\lambda_q$ is
the chemical potential for the charge state $q=n,p$ and $\mu
\!=\!0.2$~MeV is a diffuseness parameter. 
\begin{figure}[h]
  \hspace*{-0.5cm}
  \includegraphics[width=0.515\textwidth]{
    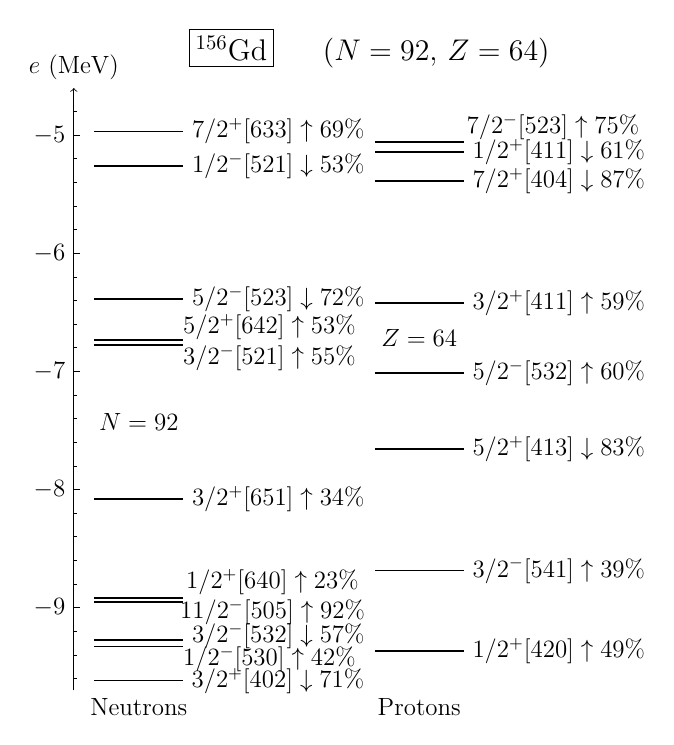} 
  \caption{Neutron and proton single-particle energies $e$ (in MeV)
    calculated in the ground state of $^{156}$Gd. %(left panel) and $^{158}$Gd.
    The dominant Nilsson quantum numbers and the weight of the
    corresponding contribution are indicated. Up and down arrows
    correspond respectively to spin projections $+1/2$ and $-1/2$ (in $\hbar$
    unit) on the symmetry axis.
    \label{fig_Gd156_Gd158_sp_spectra}}   
\end{figure}

Finally the single-particle states are expanded in a truncated
cylindrical harmonic-oscillator basis with parity symmetry. In
nuclei around $A=178$, we use the same basis parameters as
in Ref.~\cite{Nor19} and in actinides we take the parameters of
Ref.~\cite{Minkov22}. Around $A=156$, the optimal basis parameters
are found to be $b=0.495$ and $q=1.18$ in the notation of
Ref.~\cite{Flocard73}, with $N_0+1=15$ spherical oscillator
major shells. As in Ref.~\cite{Minkov22} all integrations are
performed by Gauss--Hermite quadratures with 30 mesh points in the $z$
direction of the symmetry axis and Gauss--Laguerre quadratures with 15
mesh points in a direction orthogonal to the symmetry axis. 

%-------------------------------------------------------------------------
%
%
%                     Results for bandhead spectra
%
%
%-------------------------------------------------------------------------

\section{Results for bandhead spectra}

In this section we report on the bandhead spectra calculated within
the SIII Skyrme energy-density functional with selfconsistent blocking
as explained above and present the results by mass region.
We report only on experimentally observed Gallagher--Moszkowski
doublets and for each studied odd-odd nucleus, we take the
two-quasiparticle $K^{\pi}$ configuration of the experimental ground
state as a reference state to build the energy spectrum of calculated
doublets. For completeness we also give the lowest-energy calculated
bandhead when it is not among observed Gallagher--Moszkowski doublets.

\subsection{Rare-earth nuclei around $A = 156$}

The odd-odd nuclei surrounding the $^{156,158}$Gd isotopes are
described by two-quasiparticle configurations with respect to the
ground state of these even-even nuclei. Sufficient experimental data
on Gallagher--Moszkowski doublets are available in $^{154,156}$Eu and
$^{158,160}$Tb. It is worth noting that the $^{160}$Tb nucleus was 
studied in Ref.~\cite{Bennour87} where calculations in the
rotor+two-quasiparticle picture with Skyrme-EDF intrinsic solutions
were compared with experimental data for several Gallagher--Moszkowski
doublets.
\begin{figure*}[t]
  (a)\includegraphics[width=0.9\textwidth]{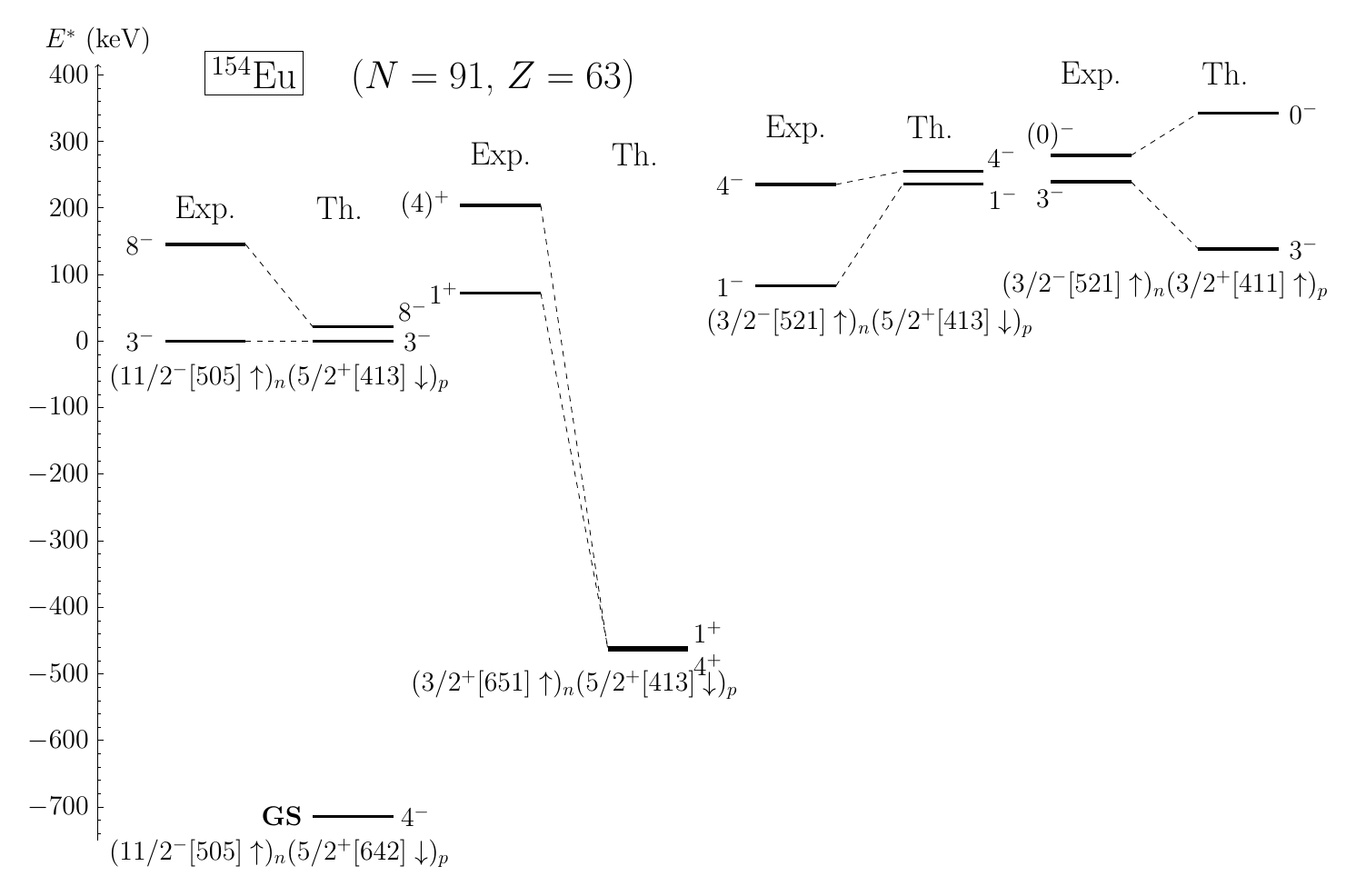}
  \\[0.25cm]
  (b)\includegraphics[width=0.9\textwidth]{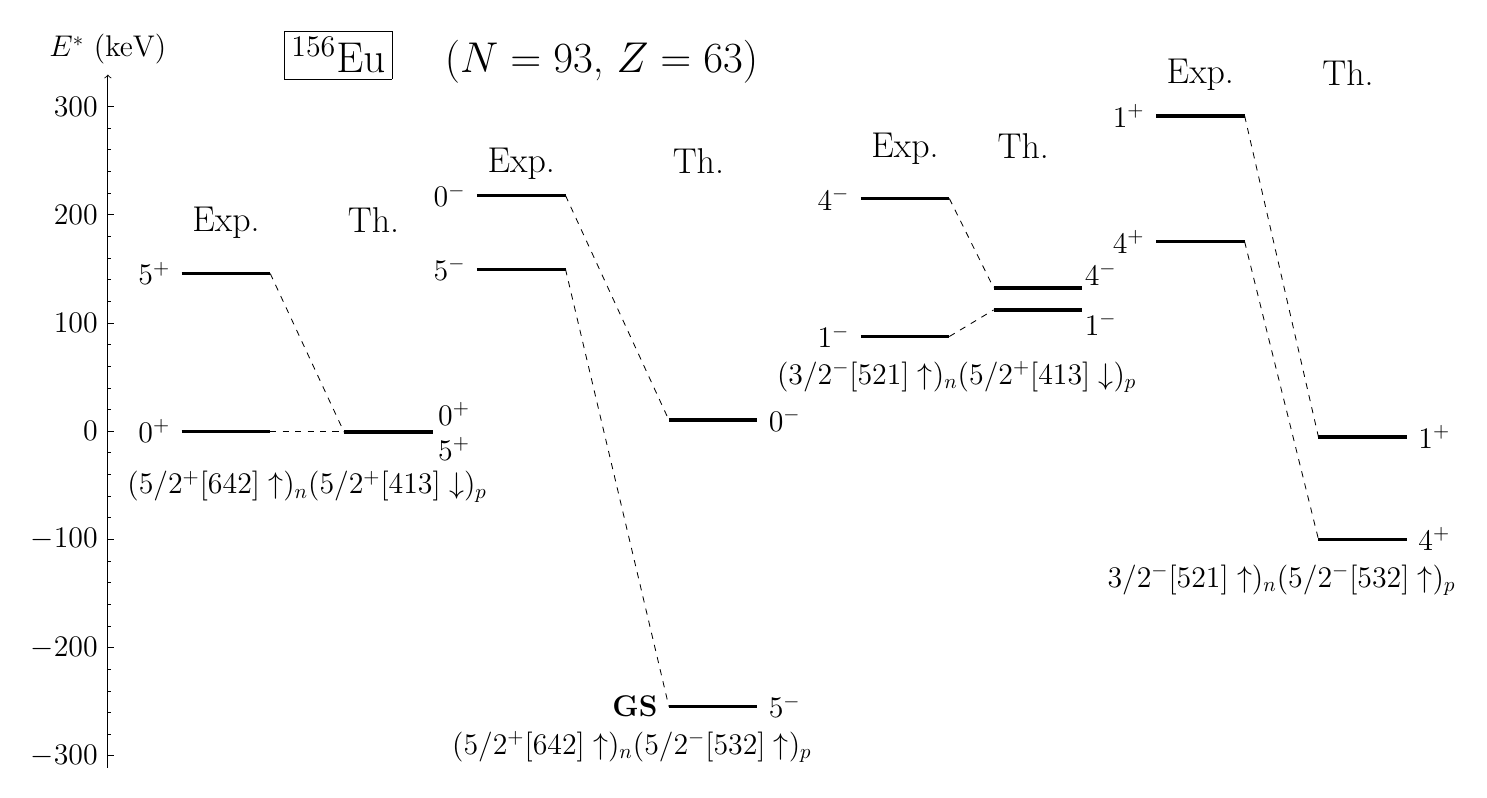}
  \caption{Bandhead spectra of $^{154}$Eu (a) and $^{156}$Eu
    (b) restricted to Gallagher--Moszkowski doublets
    experimentally observed. For each doublet the left level bars
    correspond to experiment (labelled ``Exp.''), while the right ones
    correspond to the SCB calculations (labelled ``Th. (SCB)'').
  \label{fig_Eu154_Eu156}}
\end{figure*}

First we show in Fig.~\ref{fig_Gd156_Gd158_sp_spectra} the neutron and
proton single-particle spectra calculated in the ground state of the 
$^{156}$Gd nucleus as an example. In the neutron single-particle spectrum, the
second largest cylindrical harmonic-oscillator contribution to the
$1/2^+$ state around $e=-9$~MeV has $[400]\uparrow$ Nilsson quantum
numbers, in addition to the $[640]\uparrow$ contribution displayed in
Fig.~\ref{fig_Gd156_Gd158_sp_spectra}. From a theoretical point of
view we thus expect low-lying bandhead states (up to about 1~MeV
excitation energy) in neighboring odd-odd nuclei with
configurations involving on the one hand the
$11/2^-[505]\uparrow$, $1/2^+$ ($[640]\uparrow$ or $[400]\uparrow$),
$3/2^+[651]\uparrow$, $3/2^-[521]\uparrow$, $5/2^+[642]\uparrow$,
$5/2^-[523]\downarrow$ neutron single-particle states, and, on the
other hand, the $5/2^+[413]\downarrow$, $5/2^-[532]\uparrow$, 
$3/2^+[411]\uparrow$, $7/2^+[404]\downarrow$ proton single-particle
states. \\
\begin{figure*}[t]
  (a) \includegraphics[width=0.725\textwidth]{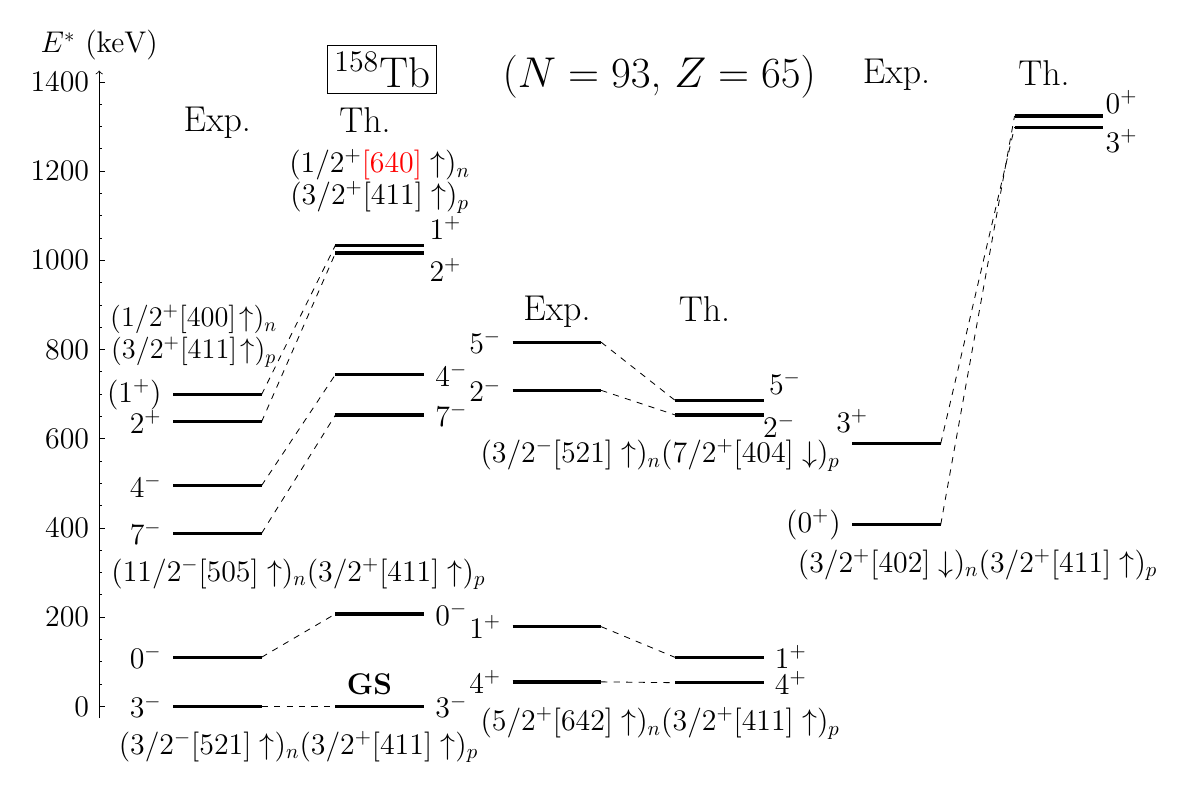}
  \\
  (b) \includegraphics[width=0.725\textwidth]{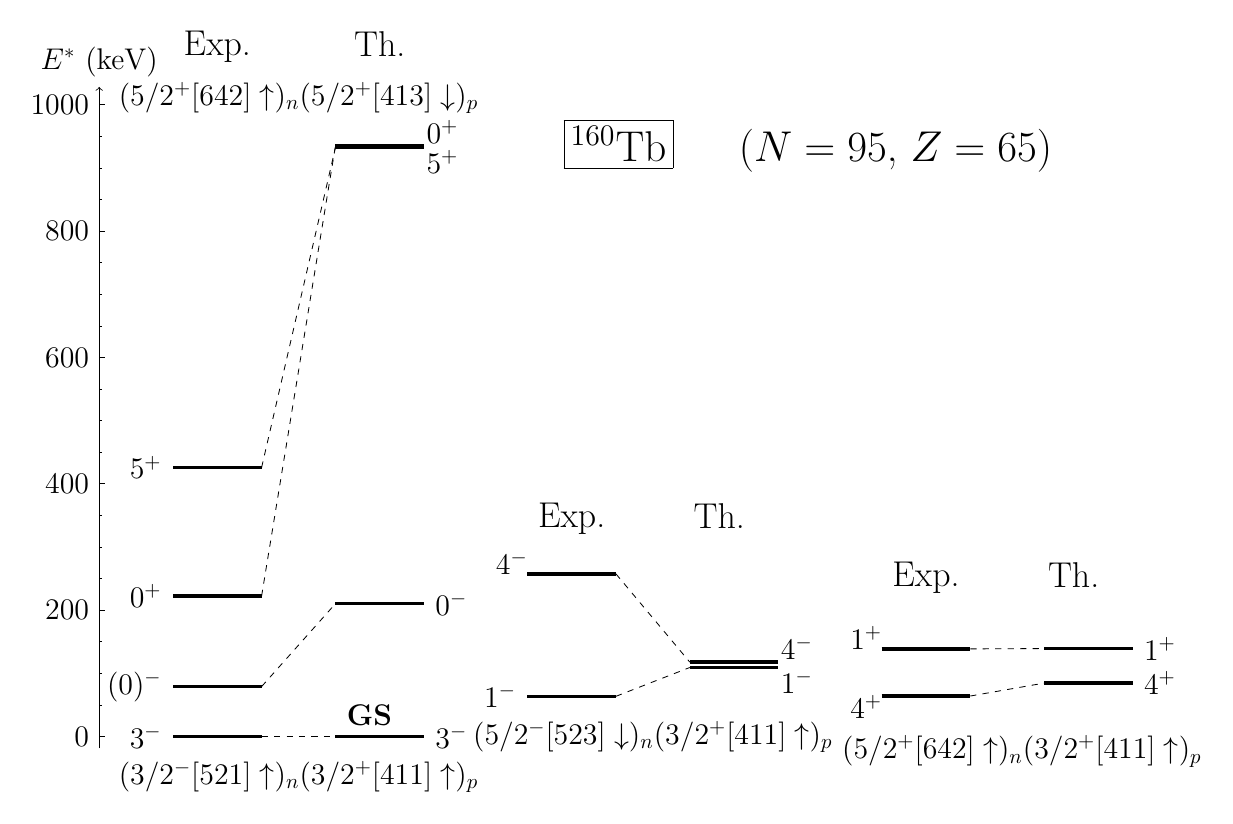}
  \caption{Same as Fig.~\ref{fig_Eu154_Eu156} for the $^{158}$Tb (a)
    and $^{160}$Tb (b) nuclei. 
    \label{fig_Tb158_Tb160_exp_th}}
\end{figure*}

{\itshape $^{154,156}$Eu nuclei ($Z\!=\!63$ and $N\!=\!91,\;93$).}
In panel~(a) of Fig.~\ref{fig_Eu154_Eu156} we display the bandhead
spectrum of $^{154}$Eu restricted to the four Gallagher--Moszkowski
doublets experimentally observed. Theoretical excitation energies are
all calculated with respect to the $K^{\pi} = 3^-$ state with the
configuration $(11/2^-[505]\uparrow)_n (5/2^+[413]\downarrow)_p$,
which corresponds to the experimental ground state. Overall the energy
ordering of calculated doublets follows the expectation from the
Koopmans approximation, except for the $(4^+,1^+)$ doublet built on the
$(3/2^+[651]\uparrow)_n$ $(5/2^+[413]\downarrow)_p$ configuration,
calculated to lie below the $K^{\pi} = 3^-$ reference state. This
suggests that, in the neutron single-particle spectrum, the
$11/2^-[505]$ level should be located above the $3/2^+[651]$ level.
While experimental excitation energies are all below 300 keV, the
calculated bandhead states of the considered doublets, except the
discrepant one, agree with experimental data within less than 200~keV,
which can be deemed as a rather good result. \\

With the addition of two neutrons to the
previous nucleus, the lowest energy configurations of $^{156}$Eu
are now expected to involve the $3/2^-$ or $5/2^+$ neutron state
together with the $5/2^-$ or $5/2^+$ proton state. The resulting four
doublets turn out to be precisely 
the ones experimentally observed. They are displayed in panel (b) of 
Fig.~\ref{fig_Eu154_Eu156} where the
excitation energies are defined with respect to the experimental
ground state $K^{\pi}=0^+$ with configuration
$({5}/2^+[642]\uparrow)_n$ $(5/2^+[413]\downarrow)_p$. 
Again the calculated excitation energies agree rather well with the
experimental ones, with a discrepancy below 200 keV except for the
$K^{\pi}=5^-$ state which is found to be the theoretical ground state
almost 300 keV below the $K^{\pi}=0^+$ state. This results from the
position of the $5/2^-$ level above the $5/2^+$ level in the proton
single-particle spectrum consistently with the above discussion for
the $^{154}$Eu nucleus. 

As discussed in Ref.~\cite{Jain98_RMP}, the octupole degree of freedom
is expected to play a role in $^{154}$Eu, maybe less so in
$^{156}$Eu. However, according to Afanasjev and
Ragnarsson~\cite{Afanasjev95}, the existence of static octupole
deformation in this nucleus (and neighboring nuclei) is not supported
by Woods--Saxon calculations of polarization energies of
octupole-driving orbitals. Later, systematic axial
reflection-asymmetric calculations within the covariant density 
functional theory~\cite{Agbemava16} showed that even-even 
neighbors of the odd-odd $^{154,156}$Eu isotopes do not have octupole
ground-state shape. Finally, a recent global survey of
pear-shaped even-even nuclei within nonrelativistic and relativistic
EDF approaches~\cite{Cao20} lead to the same conclusion. 
Even if we did not break intrinsic parity to incorporate the octupole
degree of freedom in the studied odd-odd nuclei, the basic ingredients
for a proper description of bandhead states of two-quasiparticle
character are present in our approach, especially the ``parity'' doublets
of single-neutron and single-proton states with $\Omega_{n,p}=3/2$ and
$\Omega_{n,p}=5/2$ giving rise to the $(4,1)^{\pm}$ doublets in
$^{154,156}$Eu and $(5,0)^{\pm}$ in $^{156}$Eu. \\

{\itshape $^{158,160}$Tb nuclei (Z=65 and N=103, 105).} The
bandhead spectrum for the six experimentally observed
Gallagher--Moszkowski doublets in the $^{158}$Tb nucleus is displayed
in Fig.~\ref{fig_Tb158_Tb160_exp_th}. Note that, in $^{158}$Tb,
the $1/2^+$ neutron single-particle state involved in the
calculated $(2^+,1^+)$ doublet closest to the experimental one is the
$1/2^+[640]$ instead of the $1/2^+[400]$ proposed in
Ref.~\cite{Nica17_A=158}.

The theoretical ground state $K^{\pi} = 3^-$ and first excited
bandhead state $K^{\pi} = 4^+$, both involving the $3/2^+$ proton
single-particle level just above the proton Fermi level of $^{156}$Gd,
agree very well with the experimental ones. The very small excitation
energy of the $4^+$ state confirms the quasi-degeneracy of the $3/2^-$
and $5/2^+$ levels seen in the neutron single-particle spectrum of
$^{156}$Gd. Moreover the rather good position of the theoretical
$(2^-,5^-)$ doublet seems to indicate a correct shell gap between the
$3/2^+$ and $7/2^+$ levels in the proton single-particle spectrum in
$^{156}$Gd. 
\begin{figure}[t]
  \hspace*{-0.5cm}
  \includegraphics[width=0.5\textwidth]{
    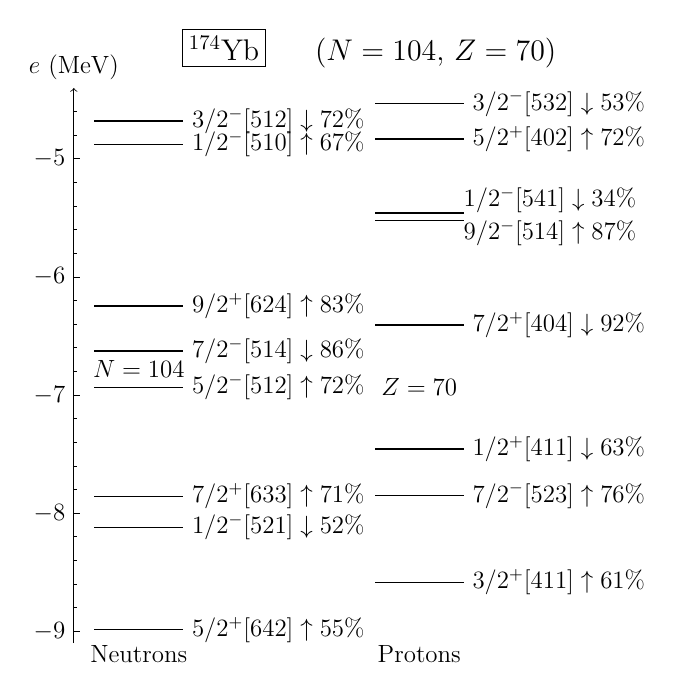} \\
  \hspace*{-0.5cm}
  \includegraphics[width=0.5\textwidth]{
    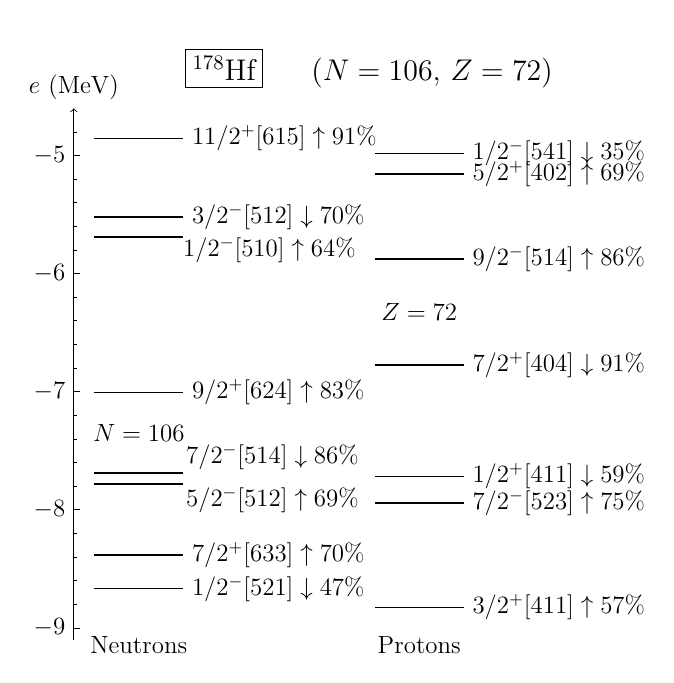} 
  \caption{Same as Fig.~\ref{fig_Gd156_Gd158_sp_spectra} for
    $^{174}$Yb and $^{178}$Hf.
    \label{fig_Yb174_Hf178_sp_spectra}} 
\end{figure}

In contrast the doublets involving the $11/2^-$ or $1/2^+$ neutron
states are calculated to be too high, which is consistent with too low
$11/2^-$ or $1/2^+$ levels around $-9$~MeV in the neutron single-particle
spectrum of $^{156}$Gd.

The bandhead spectrum for the
four experimentally observed Gallagher--Moszkowski doublets in the
$^{160}$Tb nucleus is displayed in the bottom panel of
Fig.~\ref{fig_Tb158_Tb160_exp_th}. Similarly to what has been obtained
in the $^{158}$Tb nucleus, the ground state is found to be the $3^-$
state in agreement with experiment and the excitation energies of the
$4^+$, $1^+$ and $1^-$ are in excellent agreement with the
experimental ones. Apart from the $(5^+,0^+)$ doublet,
for which the discrepancy is of the order of 600~keV, 
the remaining excitation energies are reproduced within less than the
typical 200~keV. \\

The overall picture emerging from the energy spectra of these odd-odd
nuclei around $^{156,158}$Gd is that the single-particle
states with the quantum numbers relevant for the low-lying
two-quasiparticle bandhead states and Gallagher--Moszkowski doublets
are present around the neutron and proton Fermi levels in the region
$88 \leqslant N \leqslant 98$ and $62 \leqslant Z \leqslant
68$. However some single-particle hole states appear too high and
are expected to generate bandhead states experimentally unobserved in
the low-lying spectrum of the considered odd-odd nucleus.
\begin{figure}[t]
  \includegraphics[width=0.475\textwidth]{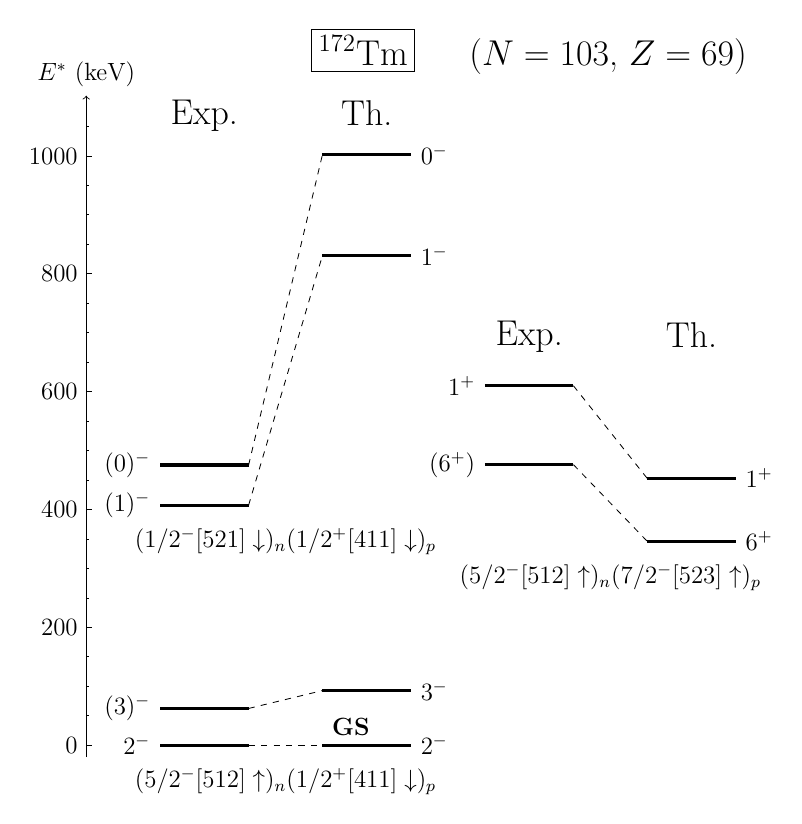} \\
  \includegraphics[width=0.475\textwidth]{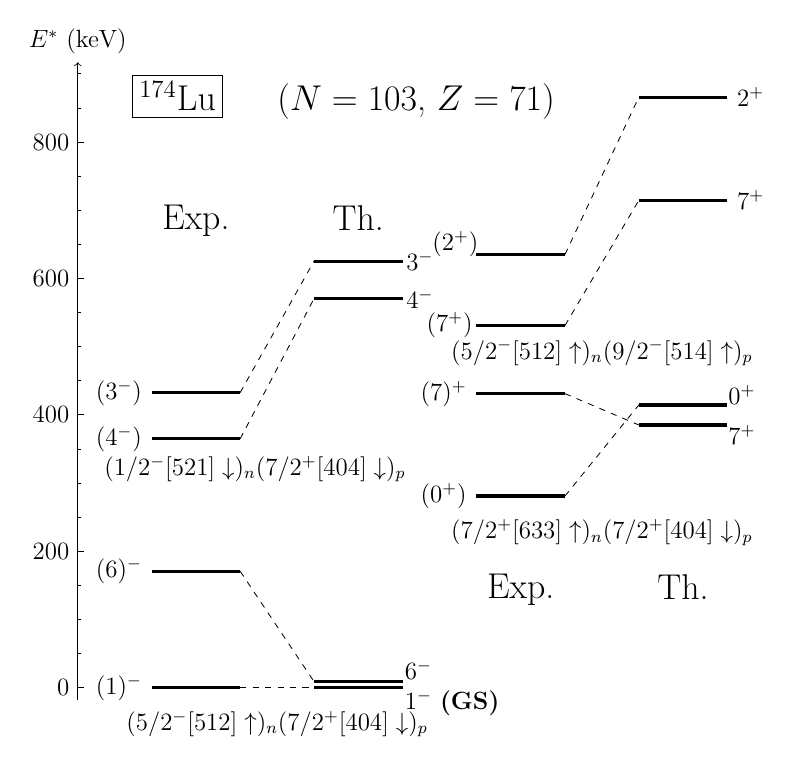}
  \caption{Same as Fig.~\ref{fig_Eu154_Eu156} for the $^{172}$Tm 
    and $^{174}$Lu nuclei. \label{fig_Tm172_Lu174_exp_th}}
\end{figure}

\subsection{Rare-earth nuclei around $A = 176$}

The odd-odd nuclei surrounding $^{174}$Yb and $^{178}$Hf are described by
two-quasiparticle configurations with respect to the ground state of
these even-even nuclei. A lot of experimental data on 
Gallagher--Moszkowski doublets are available in $^{172}$Tm,
$^{174,176,178}$Lu and $^{178,180}$Ta nuclei.
\begin{figure*}[t]
  \includegraphics[width=0.9\textwidth]{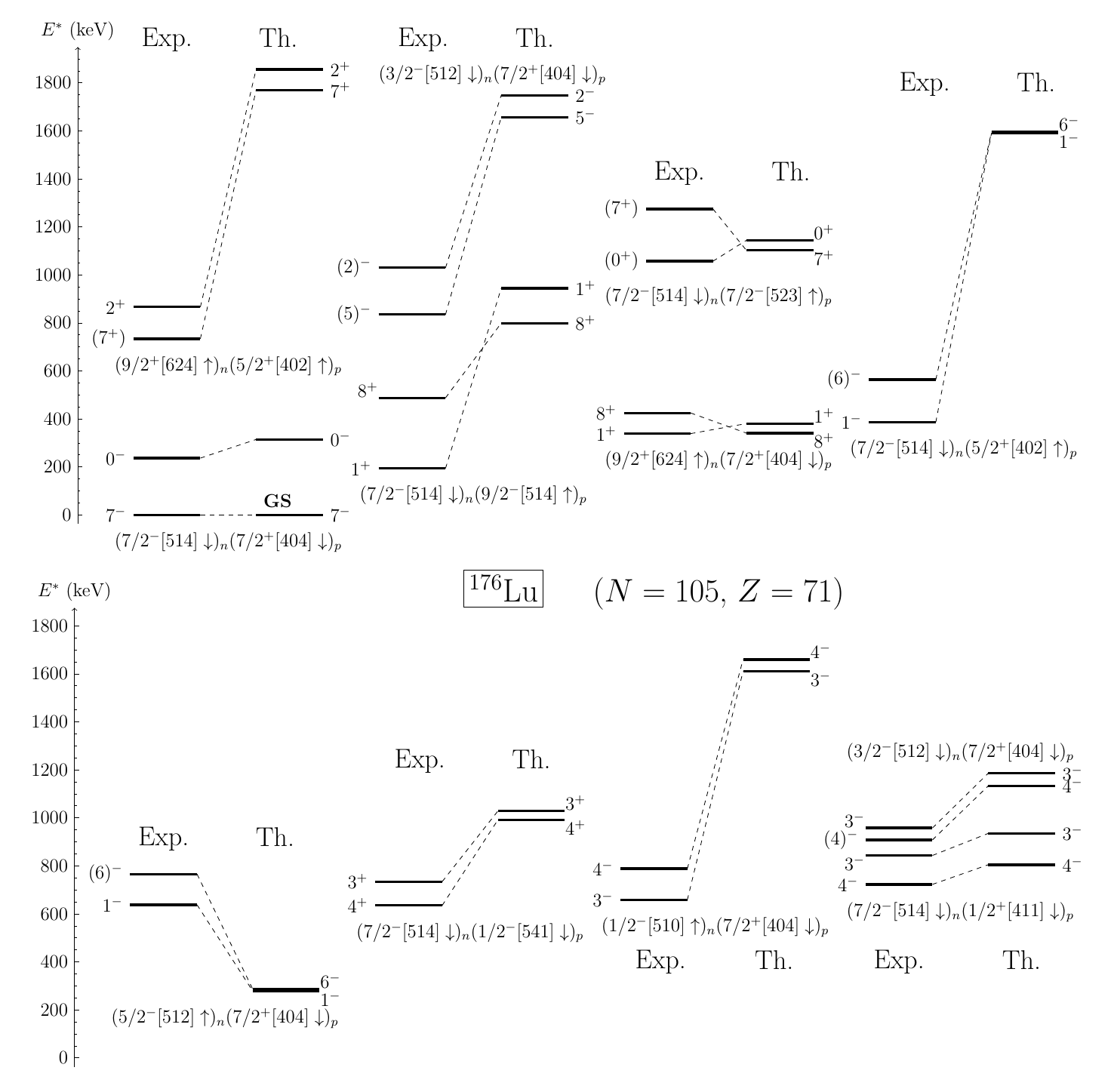}
  \caption{Same as Fig.~\ref{fig_Eu154_Eu156} for the $^{176}$Lu
    nucleus. \label{fig_Lu176_exp_th}}
\end{figure*}

First we show in Fig.~\ref{fig_Yb174_Hf178_sp_spectra} the neutron and
proton single-particle spectra calculated in the ground state of the 
$^{174}$Yb and $^{178}$Hf. We expect low-lying bandhead states
(up to about 1~MeV excitation energy) in neighboring odd-odd nuclei
with configurations involving on the one hand the
$7/2^+[633]\uparrow$, $5/2^-[512]\uparrow$, 
$7/2^-[514]\downarrow$, $9/2^+[624]\uparrow$
neutron single-particle states, and, on the other hand, the
$7/2^-[523]\uparrow$, $1/2^+[411]\downarrow$, $7/2^+[404]\downarrow$,
$9/2^-[514]\uparrow$, $5/2^+[402]\uparrow$, $1/2^-[541]\downarrow$
proton single-particle states. \\

{\itshape $^{172}$Tm and $^{174}$Lu nuclei (N=103 and Z=69, 71).}
The bandhead spectra for the experimentally observed
Gallagher--Moszkowski doublets in the $^{172}$Tm and $^{174}$Lu nuclei
are displayed in Fig.~\ref{fig_Tm172_Lu174_exp_th}. Theoretical
excitation energies of $^{172}$Tm are all calculated with respect
to the $K^{\pi} = 2^-$ state with the configuration
$(5/2^-[512]\uparrow)_n (1/2^+[411]\downarrow)_p$, which 
corresponds to the experimental ground state, whereas in $^{174}$Lu
the calculated ground state is the $K^{\pi} = 1^-$ state in agreement
with experiment with the configuration $(5/2^-[512]\uparrow)_n
(7/2^+[404]\downarrow)_p$. Overall the calculated excitation energies
in these two nuclei agree with experimental data
within less than about 200~keV with the
notable exception of the $(0^-,1^-)$ doublet in $^{172}$Tm with an
overestimation of about 450~keV. The corresponding configuration
differs from that of the ground state by the $1/2^-[521]$
neutron state, which suggests that this calculated $1/2^-$ neutron
state lies too far below the $5/2^-[512]$ state. \\

\begin{figure}[h]
  \includegraphics[width=0.4\textwidth]{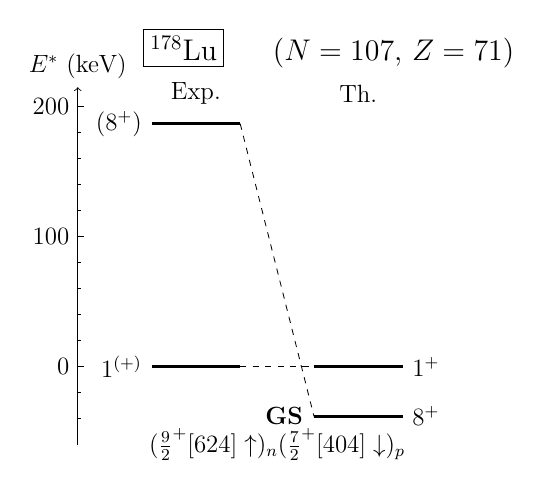}
  \caption{Same as Fig.~\ref{fig_Eu154_Eu156} for the $^{178}$Lu
    nucleus. \label{fig_Lu178_exp_th}} 
\end{figure}

\begin{figure*}[t]
  \includegraphics[width=0.7\textwidth]{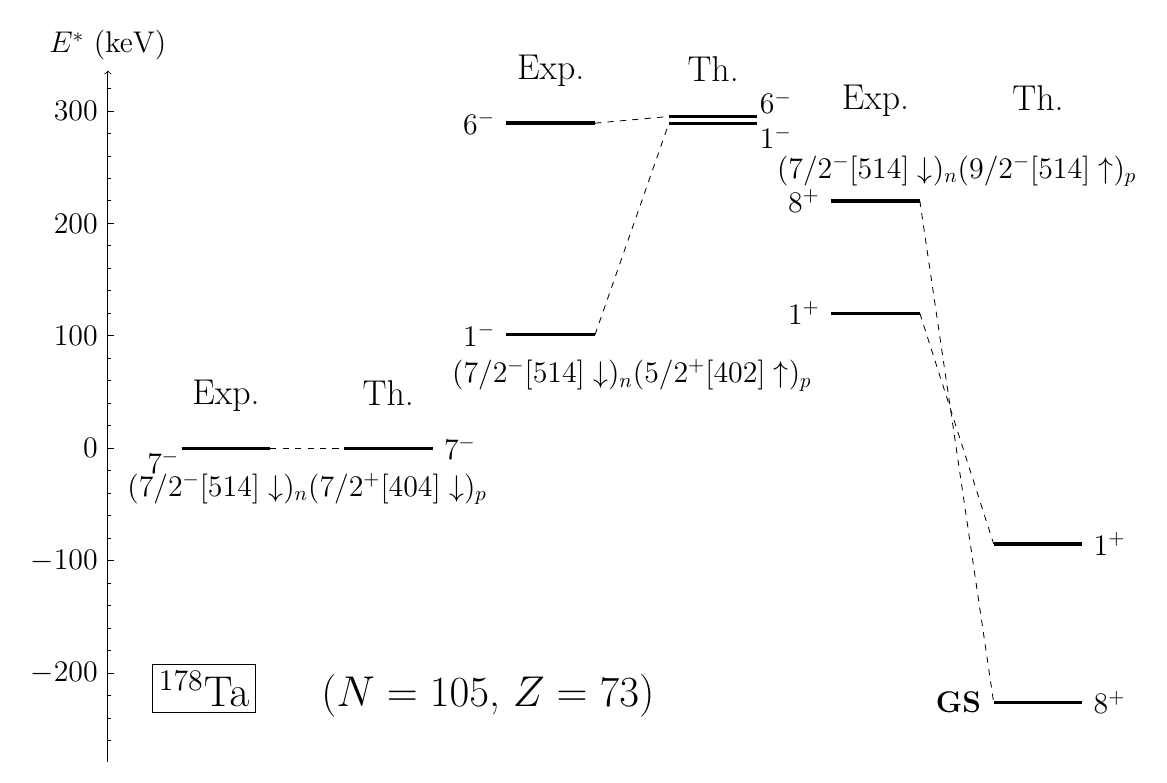}
  \\[0.25cm]
  \includegraphics[width=0.7\textwidth]{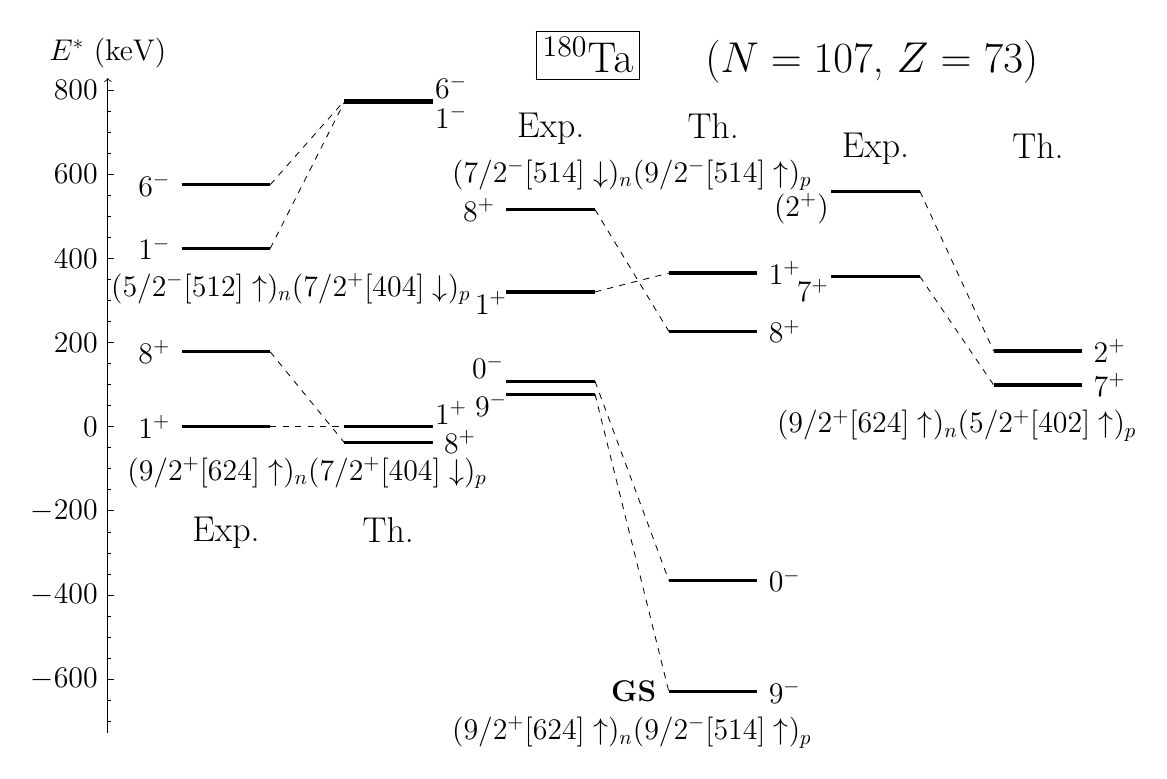}
  \caption{Same as Fig.~\ref{fig_Eu154_Eu156} for $^{178}$Ta and $^{180}$Ta 
    nuclei. In $^{178}$Ta the ground state of taken to be the $7^-$
    state as in Refs.~\cite{A=178_NDS110_2009,Nubase2020} and we use the
    excitation energy of the $1^+$ isomer estimated from systematics
    $100$~keV given by Ref.~\cite{Nubase2020}.
    \label{fig_Ta178_Ta180_exp_th}}
\end{figure*}

{\itshape $^{176,178}$Lu (N=105, 107 and Z=71) and $^{178,180}$Ta
  nuclei (N=105, 107 and Z=73).} The bandhead 
spectra for the experimentally observed Gallagher--Moszkowski
doublets in the $^{176,178}$Lu and $^{178,180}$Ta nuclei surrounding
the even-even $^{178}$Hf nucleus are displayed in
Figs.~\ref{fig_Lu176_exp_th} to \ref{fig_Ta178_Ta180_exp_th}.

The $^{176}$Lu nucleus is the experimentally most studied one with
12 observed doublets and with configurations assigned in
Refs.~\cite{Balodis72,Klay91}. Its ground state with $K^{\pi}=7^-$ is
reproduced by our calculations. Excellent agreement of calculated
excitation energies with experimental data is obtained for the
following Gallagher--Moszkowski doublets and configurations: 
\begin{itemize}
\item ($1^+,8^+$), $(9/2^+[624]\uparrow)_n
  (7/2^+[404]\downarrow)_p$; 
\item ($0^+,7^+$), $(7/2^-[514]\downarrow)_n (7/2^-[523]\uparrow)_p$; 
\item ($4^-,3^-$), $(7/2^-[514]\downarrow)_n (1/2^+[411]\downarrow)_p$.
\end{itemize}

\noindent From the agreements for the last two doublets,
we can deduce that $7/2^-[523]$ and $1/2^+[411]$ proton states are
correctly located below the $7/2^+[404]$ state. On the neutron side,
the agreement for the first doublet of the above list suggests the
good relative position of the $9/2^+[624]$ state with respect to the
$7/2^-[514]$ state. 

In contrast the large overestimation of the excitation energies for
the doublet $(7/2^-[514]\downarrow)_n(9/2^-[514]\uparrow)_p$ 
indicates a too large $Z=72$ gap. Moreover the doublets involving
the $5/2^+[402]$ proton state combined with the $7/2^-[514]$ or
$9/2^+[624]$ neutron states are calculated to be much too high above
the corresponding experimental ones. This suggests that the
$5/2^+[402]$ proton state is significantly too high above the
$9/2^-[514]$ state.

On the neutron side, the large underestimation of the excitation
energies for the $(6^-,1^-)$ doublet involving the $5/2^-[512]$
neutron state indicate that this state is too close below the
$7/2^-[514]$ state. When combined with the too high $9/2^-[514]$
proton state, this yields some compensation, hence a moderate
discrepancy for the $(7^+)$ excitation energy with the
$(5/2^-[512])_n (9/2^-[514])_p$ configuration in $^{174}$Lu
In the same spirit the $3/2^-[512]$ neutron state appears to be too
high above the $9/2^+[624]$ state according to the large
overestimation of the excitation energy for the $(5^-,2^-)$ doublet
in $^{176}$Lu involving this $3/2^-$ neutron state.

Moreover the rather fair agreement for the $(4^-,3^-)$ doublet 
in $^{174}$Lu or $^{176}$Lu built
on the $(1/2^-[521])_n$ $(7/2^+[404])_p$ configuration suggests a
reasonable relative position of the involved state in the neutron
spectrum. Therefore the large overestimation of the excitation
energies for the $(5^-,2^-)$ doublet with the $(3/2^-[512])_n$
$(7/2^+[404])_p$ configuration and the $(3^-,4^-)$ doublet with the
$(1/2^-[510])_n$ $(7/2^+[404])_p$ configuration can be
understood as a too large $N=108$ energy gap (over 1.2~MeV in
$^{174}$Yb and $^{178}$Hf as can be seen in
Fig.~\ref{fig_Yb174_Hf178_sp_spectra}). \\

The bandhead spectra of $^{178}$Lu and $^{178,180}$Ta are shown in
Figs.~\ref{fig_Lu178_exp_th} and \ref{fig_Ta178_Ta180_exp_th},
respectively. The Gallagher--Moszkowski doublets considered in these
three nuclei involve the same single-particle states than in
$^{176}$Lu and lead to the same conclusions regarding the relative
positions of these states. It should be noted that in
$^{178}$Ta the ground state taken to be the $7^-$ state as in
Refs.~\cite{A=178_NDS110_2009,Nubase2020}. 
The experimental excitation energies of the $1^-$, $6^-$ members
of the doublet built from the $(7/2^-[514])_n(5/2^+[402])_p$
configuration and of the $8^+$ member of the doublet built from the
$(7/2^-[514])_n(9/2^-[514])_p$ configuration are measured with
respect to the $7^-$ state, whereas the $1^+$ member of the doublet
built from the $(7/2^-[514])_n(9/2^-[514])_p$ configuration is
estimated, from systematics, to lie $100$~keV above the $7^-$ state
by Ref.~\cite{Nubase2020}.

Similarly to the conclusion drawn in the end of the previous
subsection, the overall picture emerging from the energy spectra of
the odd-odd nuclei surrounding $^{174}$Yb and $^{178}$Hf is that the
single-particle states with the quantum numbers relevant to the
low-lying two-quasiparticle bandhead states and Gallagher--Moszkowski
doublets are present around the neutron and proton Fermi levels in the
region $100 \leqslant N \leqslant 108$ and
$68 \leqslant Z \leqslant 78$. However the level density in the proton
spectrum is a bit too low and the $5/2^-$ neutron hole is a bit too
high.

\subsection{Actinides between $A=230$ and $A=250$}

\begin{figure}[t]
  \hspace*{-0.5cm}
  \includegraphics[width=0.5\textwidth]{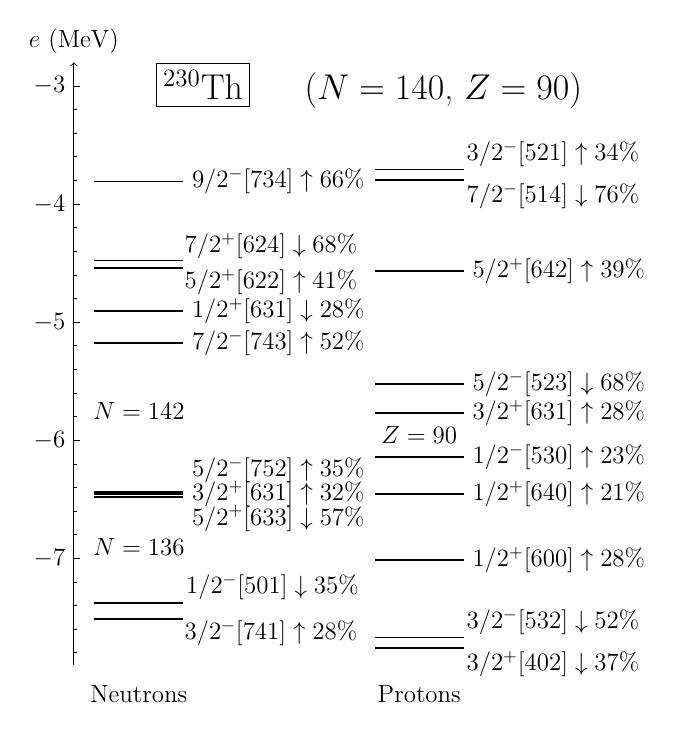} 
  \caption{Same as Fig.~\ref{fig_Gd156_Gd158_sp_spectra} for
    the $^{230}$Th nucleus. \label{fig_Th230_Th232_sp_spectra}}
\end{figure}

\begin{figure}[t]
  \includegraphics[width=0.5\textwidth]{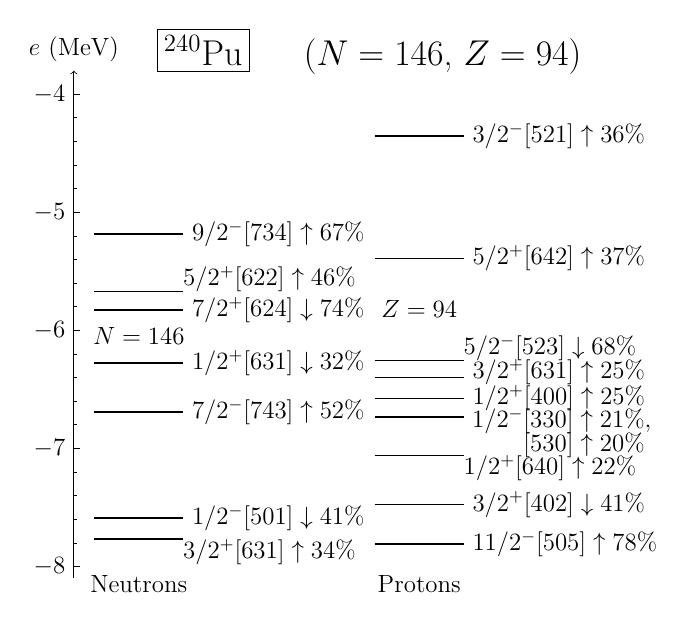}
  \includegraphics[width=0.5\textwidth]{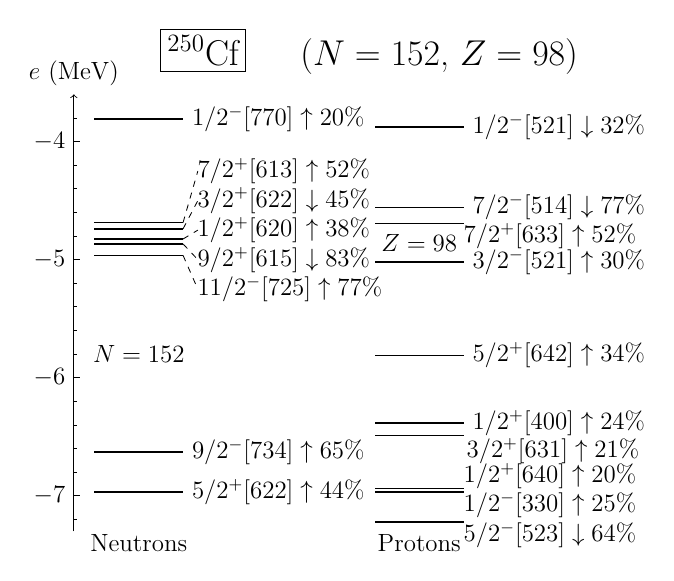}
  \caption{Same as Fig.~\ref{fig_Gd156_Gd158_sp_spectra} for
    the $^{240}$Pu and $^{250}$Cf nuclei. In $^{240}$Pu, the second
    largest component of the $1/2^-$ proton state is also given, below
    the dominant one.
    \label{fig_Pu240_Cf250_sp_spectra}}
\end{figure}

The considered odd-odd actinides are neighbors of the $^{230,232}$Th,
$^{240,244}$Pu and $^{250}$Cf even-even nuclei. The single-particle
spectra of $^{230}$Th, $^{240}$Pu and $^{250}$Cf are displayed in
Figs.~\ref{fig_Th230_Th232_sp_spectra} and
\ref{fig_Pu240_Cf250_sp_spectra}. The neutron spectrum in this Th
isotope has the remarkable feature of a quasi-degeneracy of three
states around the Fermi level,
namely $5/2^+[633]$, $3/2+[631]$ and $5/2^-[752]$. It
is worth noting that the first two of these quasi-degenerate levels
are involved in the $^{229m}$Th clock isomer.
Moreover a large shell gap of the order of 1~MeV is obtained for
$N=142$ and an even larger gap of about 1.5~MeV appears for $N=152$
in $^{250}$Cf.

Some neutron single-particle states with
high $\Omega$ quantum numbers (projection on the symmetry axis of the
total angular momentum of a nucleon) are found near the Fermi level of
the above even-even nuclei, with $\Omega=7/2$ near $N=146$,
$\Omega=9/2$ and $11/2$ near $N=152$. On the proton side, the largest
$\Omega$ value found near the Fermi level of the above Th and Pu
isotopes is $5/2$, whereas, for the Cf isotopes, states with
$\Omega=7/2$ are found just above $Z=98$. This is why $^{250}$Bk has
a couple of Gallagher--Moszkowski doublets involving a state with a
rather large $K$ quantum number up to 7 (see below), in contrast to
the other studied odd-odd nuclei.

Contrary to the mass region of $^{154,156}$Eu for which no octupole
deformation is found in the ground state of neighboring even-even
nuclei, even-even $^{224-228}$Th isotopes are known to exhibit
signatures of reflection-asymmetric ground-state shape (see, for
example Ref.~\cite{Cocks99_NPA645}) and the Skyrme EDF calculations of 
Ref.~\cite{Cao20} predict octupole deformation in even-even
$^{220-228}$Th isotopes. Therefore the odd-odd nuclei studied in the
present work are likely to also exhibit octupole deformation 
in the observed Gallagher--Moszkowski doublets. However the breaking
of intrinsic-parity symmetry to incorporate the octupole degree of
freedom is beyond the scope of our calculations. \\

\begin{figure*}[t]
  \includegraphics[width=0.8\textwidth]{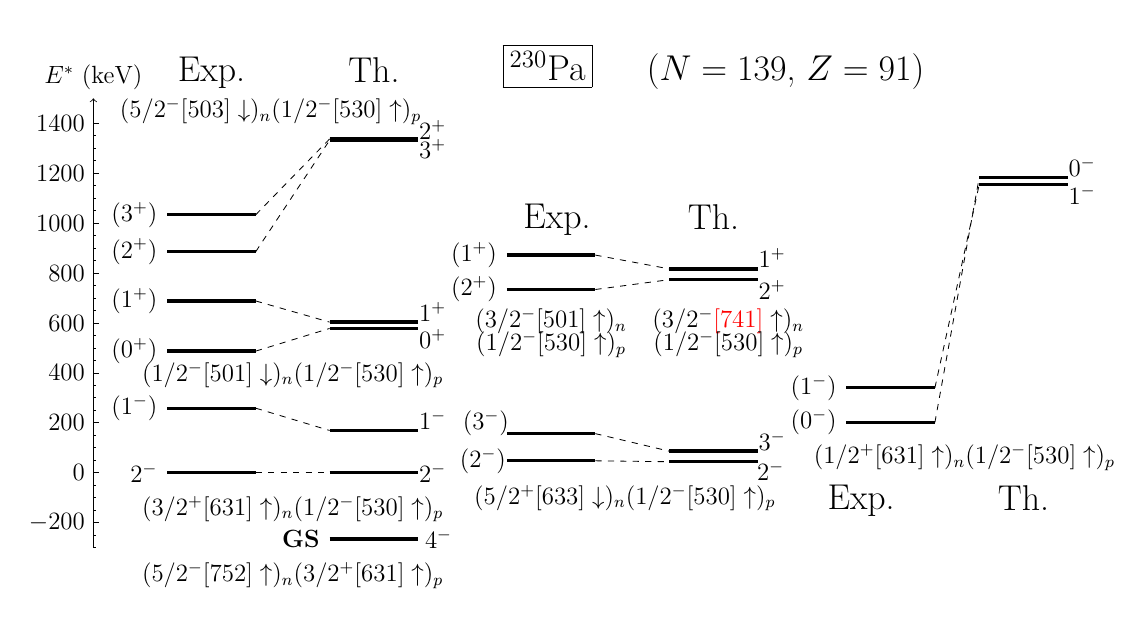}
  \includegraphics[width=0.55\textwidth]{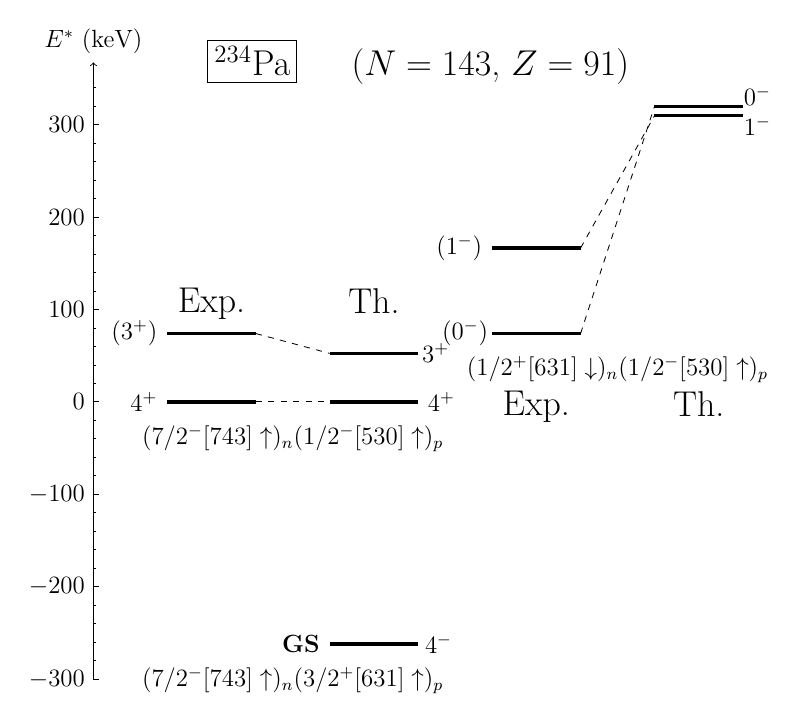}
  \caption{Same as Fig.~\ref{fig_Eu154_Eu156} for the $^{230,234}$Pa
    nuclei. \label{fig_Pa230_Pa234_exp_th}}
\end{figure*}

{\itshape $^{230,234}$Pa nuclei.} Figure~\ref{fig_Pa230_Pa234_exp_th}
shows the spectra of these two Protactinium isotopes. Note that the
calculated doublet $(2^+,1^+)$ in $^{230}$Pa is built on the
$(3/2^-[741]\uparrow)_n$ $(1/2^-[530]\uparrow)_p$ configuration and
agrees well with experimental data which had been tentatively
interpreted as a $(3/2^-[501]\uparrow)_n(1/2^-[530]\uparrow)_p$
configuration in Ref.~\cite{Kotthaus13_PRC87}. 
In the neutron single-particle spectrum of $^{230}$Th, a
$3/2^-[501]\uparrow$ neutron hole state lies about 0.5~MeV below the
$3/2^-[741]$ state and does not appear in
Fig.~\ref{fig_Th230_Th232_sp_spectra}. Therefore the configuration  
$(3/2^-[501]\uparrow)_n(1/2^-[530]\uparrow)_p$
is expected to yield much larger excitation energies for the
members of the corresponding $(2^+,1^+)$ doublet.

Among the selected Gallagher--Moszkowski doublets in these two
isotopes of Pa, the $K^{\pi}$ quantum numbers of the calculated
lowest-lying state agree with experiment. Moreover, apart from two
cases, namely the $(0^-,1^-)$ and $(2^+,3^+)$ doublets, the calculated
excitation energies agree very well with experiment, within about
100~keV at most. The largest dicrepancy between theory and experiment
is obtained for the $(0^-,1^-)$ doublet with an overstimation of the
excitation energies of about 1~MeV in $^{230}$Pa and about 250~keV in
$^{234}$Pa and the $(2^+,3^+)$ doublet in $^{230}$ with an
overstimation of about 300 to 400 keV. Since the excitation energies
of the doublets involving the $1/2^-[530]$ proton single-particle
states other than $(2^+,3^+)$ (e.g.\ the $(2^-,3^-)$
Gallagher--Moszkowski doublets in $^{230}$Pa) are well reproduced, the
above overestimation is essentially caused by the $1/2^+[631]$ neutron
state clearly too far above the neutron Fermi level of $^{230}$Pa.
As this is less so in $^{234}$Pa, we can ascribe this to an
overstimated $N=142$ shell gap. \\

\begin{figure*}[t]
  \hspace*{-0.25cm}
  \begin{tabular}{cc}
  \includegraphics[width=0.5\textwidth]{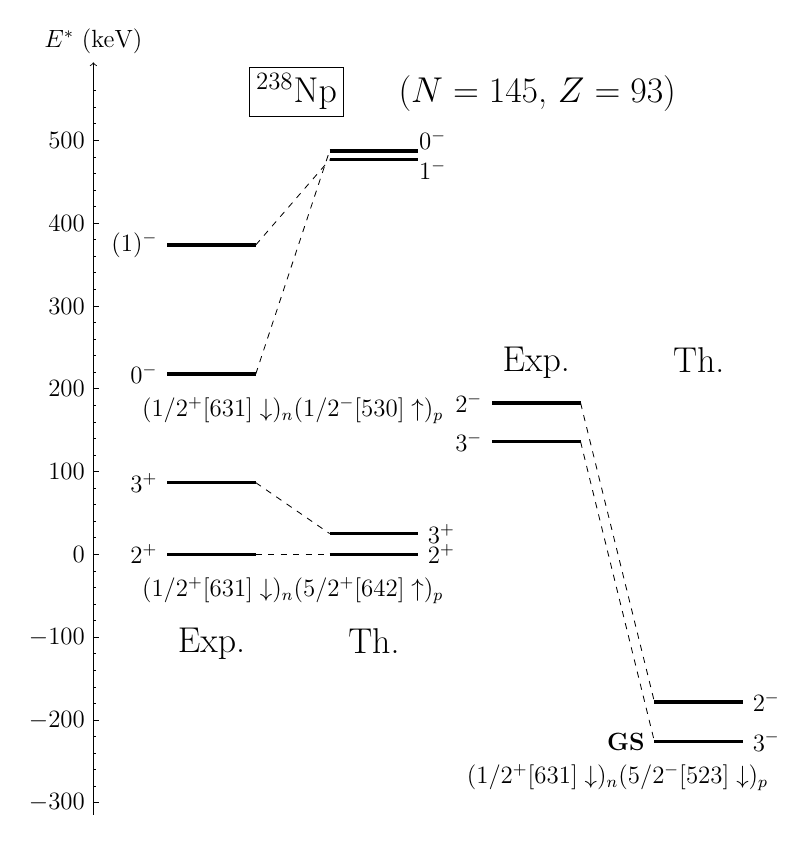} &
  \includegraphics[width=0.5\textwidth]{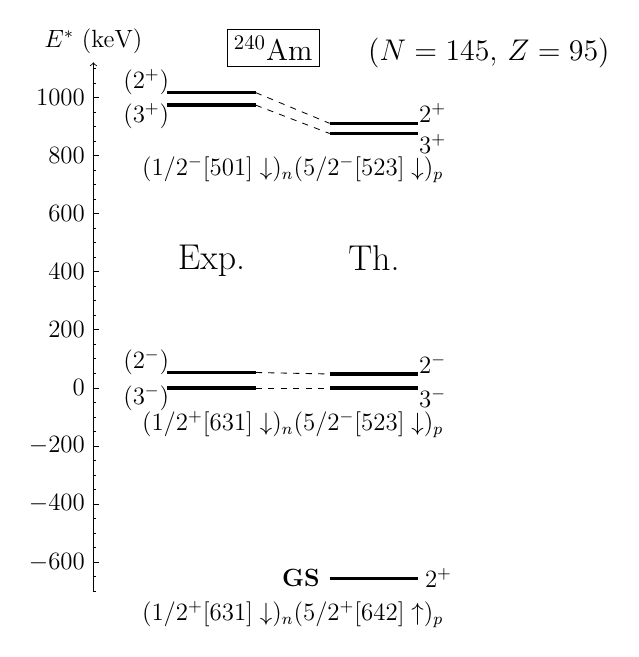} \\
  \includegraphics[width=0.55\textwidth]{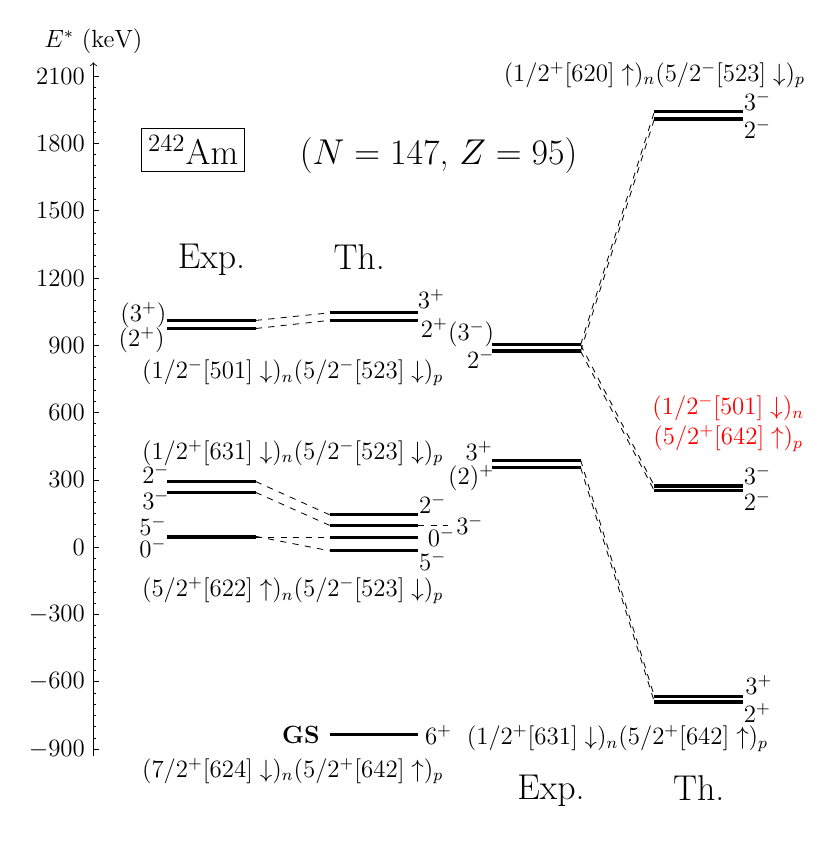} &
  \includegraphics[width=0.35\textwidth]{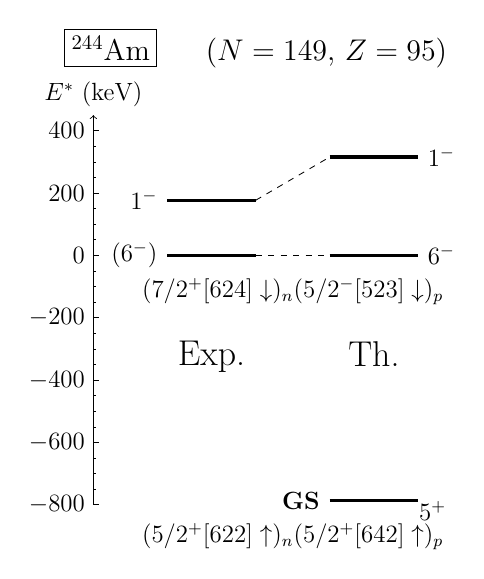}
  \end{tabular}
  \caption{Same as Fig.~\ref{fig_Eu154_Eu156} for the $^{238}$Np and
    $^{240,242,244}$Am nuclei. \label{fig_Np238_Am_exp_th}}
\end{figure*}

{\itshape $^{238}$Np and $^{240,242,244}$Am nuclei.} Figure
\ref{fig_Np238_Am_exp_th} shows the spectra of
these doubly odd nuclei around $A=240$. Note that in $^{242}$Am, the
ground state (not reported in Fig.~\ref{fig_Np238_Am_exp_th})
is the $1^-$ member of the $K^{\pi}=0^-$ band, so we have
set the excitation energy of the $0^-$ state to the experimental value 
(44.092~keV) and placed all the other calculated states in the
$^{242}$Am spectrum with respect to the calculated $K^{\pi}=0^-$
state.

In $^{238}$Np we find the experimentally observed $(2^-,3^-)$ doublet
about 200 keV below the $(3^+,2^+)$ doublet involving the $2^+$
experimental ground state. The configurations of these two doublets
differ by the proton states ${5}/2^-[523]\downarrow$ and
${5}/2^+[642]\uparrow$ which have the same $\Omega$ quantum number
but opposite parities, the $1/2^+[631]$ neutron blocked state
being common to these two doublets. As shown by
Fig.~\ref{fig_Pu240_Cf250_sp_spectra}, the ${5}/2^-[523]$
quasiparticle state is calculated to be of hole character in
$^{240}$Pu whereas the ${5}/2^+[642]$ quasiparticle state is of
particle character. This explains why the $(2^-,3^-)$ doublet lies
below the $(3^+,2^+)$ doublet in $^{238}$Np. Therefore these two
$\Omega=5/2$ proton states are inverted in the calculated
single-particle spectrum. This interpretation is supported by the
reverse ordering of the $(2^-,3^-)$ and $(3^+,2^+)$ doublets, based on
the above configurations, observed in $^{242}$Am. Despite this
inversion of the ${5}/2^+[642]$ and ${5}/2^-[523]$ states in
the proton single-particle spectrum, the calculated energy splitting
in the above $(2^-,3^-)$ and $(3^+,2^+)$ doublets agrees very well
with experiment in the nuclei where they are observed, namely 
$^{238}$Np and $^{240,242}$Am.

Another large discrepancy is observed in the $^{242}$Am for the
$(2^-,3^-)$ doublet based on the $({1}/2^+[620]\uparrow)_n$
$(5/2^-[523]\downarrow)_p$ configuration, with an overestimation of
the experimental excitation energies of both members of the doublet
by about 1~MeV. Because these excitation energies are compared with
those of the $(0^-,5^-)$ doublet which involves the same  
$5/2^-[523]$ proton state and because the $1/2^+[620]$ neutron state
in the $(2^-,3^-)$ doublet is calculated to lie above the $N=152$
shell gap (see the neutron single-particle spectrum of $^{250}$Cf in
Fig.~\ref{fig_Pu240_Cf250_sp_spectra}), we deduce that the resulting
disagreement is attributable to an overestimation of the $N=152$
gap. The inversion of the ${5}/2^+[642]$ and ${5}/2^-[523]$ states
across the proton Fermi level in a neighboring Pu nucleus makes thus
more favorable to form a $(2^-,3^-)$ doublet with the proton state
${5}/2^+[642]$ of particle character and a $1/2^-$ neutron near the
Fermi level of $^{240}$Pu. As displayed in
Fig.~\ref{fig_Pu240_Cf250_sp_spectra}, the nearest $1/2^-$ neutron
state is $1/2^-[501]$, just below the $N=142$ shell gap. 
Blocking this neutron state and the ${5}/2^+[642]$
proton eventually gives a $(2^-,3^-)$ doublet with an excitation
energy of a little bit less than 300~keV, hence much lower than the
one based on the proposed $({1}/2^+[620]\uparrow)_n$
$(5/2^-[523]\downarrow)_p$ configuration. Not surprisingly it is also
calculated to lie well below the $(2^+,3^+)$ doublet based on the
$(1/2^-[501]\downarrow)_n$ $(5/2^-[523]\downarrow)_p$, as a
consequence of the inversion of the two 5/2 proton states. However,
it is remarkable that the energy splittings in the $(2^-,3^-)$
doublets built on
$(1/2^-[501]\downarrow)_n(5/2^+[642]\uparrow)_p$ and
$(1/2^+[620]\uparrow)_n(5/2^-[523]\downarrow)_p$ are virtually
the same. Part of the reason could be that both configurations lead to
spin anti-alignment in the $3^-$ state. \\
$\phantom{h}$ \\

\begin{figure*}[t]
  \includegraphics[width=\textwidth]{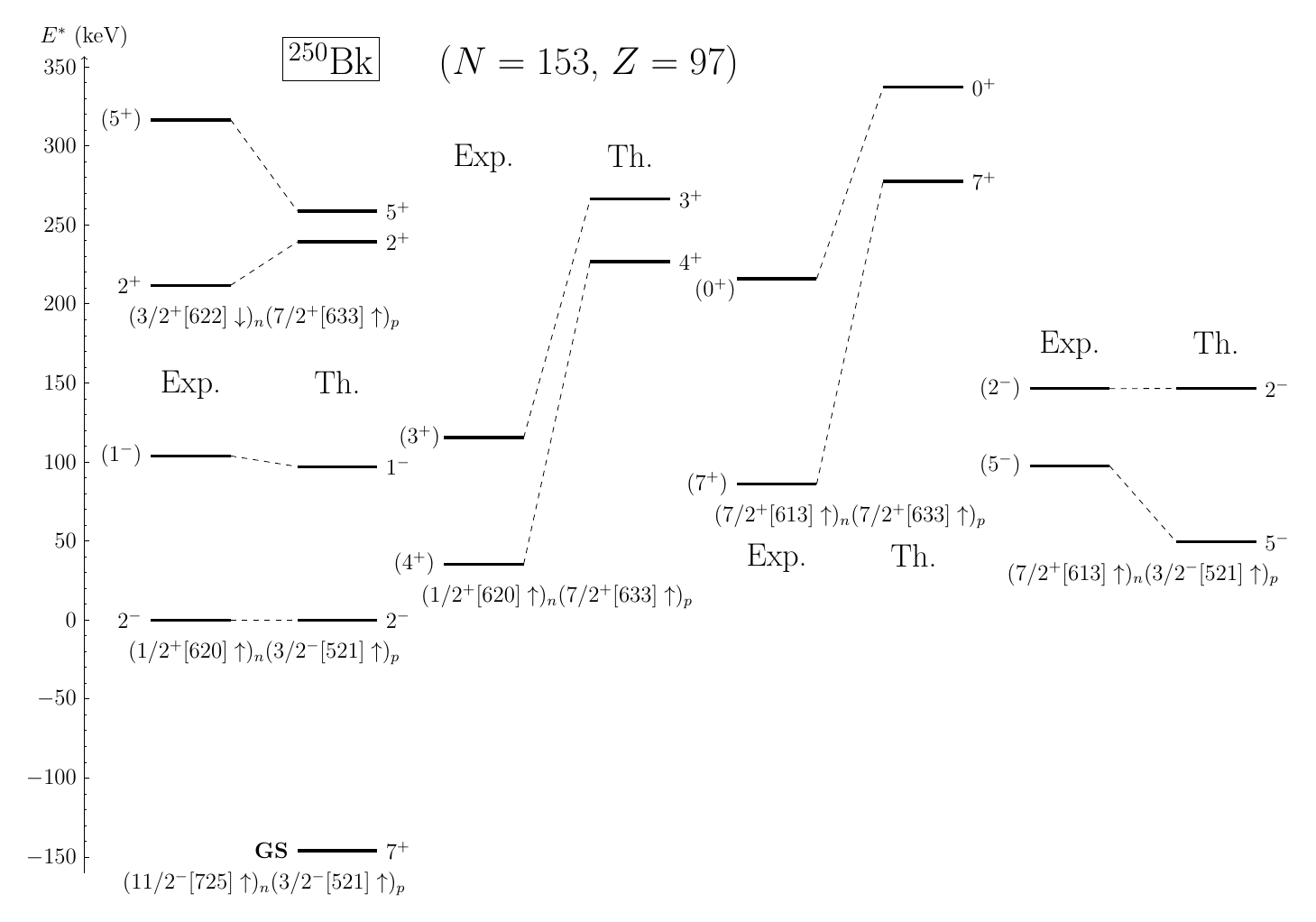}
  \caption{Same as Fig.~\ref{fig_Eu154_Eu156} for the $^{250}$Bk
    nucleus. \label{fig_Bk250_exp_th}} 
\end{figure*}

{\itshape $^{250}$Bk nucleus.} The spectrum of this nucleus 
is shown in Figure~\ref{fig_Bk250_exp_th}, with an energy scale
much less compressed than in previous spectra. This nucleus
is particularly interesting because it has one more neutron than the
``deformed magic number'' $N=152$ so that 
the unpaired (blocked) neutron state falls into a region of
very high level density as one can infer from the 
neutron single-particle level spectrum of $^{250}$Cf in
Fig.~\ref{fig_Pu240_Cf250_sp_spectra}. 
In contrast the proton level density
just below the $Z=98$ gap in the single-particle spectrum is
low. Therefore many configurations can be expected to yield
Gallagher--Moszkowski doublets at low excitation energies, so that
comparison with experiment will prove to be a demanding test of the
model.

The experimentally observed doublets all involve a neutron state just
above the $N=152$ shell gap, namely $1/2^+[620]$, $3/2^+[622]$ and
$7/2^+[613]$ (they are among the five levels grouped around $e=-5$~MeV
in the single-particle spectrum of $^{250}$Cf). On the proton side,
the observed doublets involve either the $3/2^-[521]$ state (of hole
character in $^{250}$Cf) or the $7/2^+[633]$ state (of particle
character in $^{250}$Cf). Among these doublets, the lowest lying
includes the ground state $2^-$ of $^{250}$Bk and is interpreted as a
$(1/2^+[620]\uparrow)_n$ $(3/2^-[521]\uparrow)_p$ two-quasiparticle
configuration. 
Although Fig.~\ref{fig_Bk250_exp_th} seems to exhibit large
discrepancies between theory and experiment for some states, our
calculations reproduce extremely well the excited $2^-$ state among
the nine considered excited states and produce the eight
other excited states with less than 350~keV excitation energy.

Moreover, the largest discrepancy
between the experimental excitation energy and the calculated one is
about 200~keV. It occurs in the $(4^+,3^+)$ and $(7^+,0^+)$ doublets,
whereas in the other doublets the discrepancy is below 80 keV or so,
which would look very good at the scale of the previous
spectra. The particularly low excitation energies of these doublets
make less justified the conclusions that can be drawn about the energy
width of the $Z=98$ gap because beyond mean-field correlations could
have an effect of the same order of magnitude as the above
discrepancies. However, the energy difference between the $(4^+)$ and
$(7^+)$ states being well reproduced, we can deduce that the (small)
energy spacing between the $1/2^+[620]$ and $7/2^+[613]$ neutron
single-particle states is correct.

%-------------------------------------------------------------------------
%
%
%                 Gallagher--Moszkowski energy splitting
%
%
%-------------------------------------------------------------------------

\section{Gallagher--Moszkowski energy splitting}

We now focus on the energy splitting $\EGM$ in
Gallagher--Moszkowski doublets studied in the previous section. We
recall that it is defined as the difference between the excitation
energies of the spin-antialigned and spin-aligned configurations,
$\EGM = E_{\uparrow\downarrow} - E_{\uparrow\uparrow}$, and that it
should be positive according to the Gallagher--Moszkowski
rule~\cite{Gallagher-Moszkowski58}. This quantity
strongly depends on the two-quasiparticle configuration on which the
doublet is built and is a relative excitation energy, so we expect it
to be weakly dependent of the potential discrepancies between the
calculated and measured energy spectra of odd-odd nuclei.

\subsection{Selfconsistent blocking}

The $\EGM$ values resulting from the selfconsistent blocking
calculations reported in the previous section are displayed in the
column ``SCB'' of Table~\ref{tab_GM_splitting}.
\begin{widetext}
  
\begin{longtable}{*{3}c*{7}rc}
  \caption{Energy difference $\EGM$ between spin-anti-aligned and
    spin-aligned states of Gallagher--Moszkowski
    doublets.
    The column labelled ``Total'' corresponds to
      $\Delta E_{\rm GM}^{(\rm pert)}$ whereas the contributions
      $\Delta E_{\rm GM}^{(t_0)}$, $\Delta E_{\rm GM}^{(t_3)}$,
      $\Delta E_{\rm GM}^{(t_1+t_2)}$ and $\Delta E_{\rm GM}^{(W)}$
      defined in Eqs.~(\ref{eq_EGM_t0}) to (\ref{eq_EGM_W})
      are reported in the four preceding columns.
      Experimental data are taken from references quoted in
    the last column. In $^{178}$Ta the ground state of taken to be the $7^-$
    state as in Refs.~\cite{A=178_NDS110_2009,Nubase2020} and we use the
    excitation energy of the $1^+$ isomer estimated from systematics
    $100$~keV given by Ref.~\cite{Nubase2020}.
    \label{tab_GM_splitting}} \\
\hline \hline
\multirow{3}{*}{Nucleus} & \multirow{3}{*}{Configuration}
& \multirow{3}{*}{$(K_{\uparrow\uparrow},K_{\uparrow\downarrow})^{\pi}$}
& \multicolumn{7}{c}{$\Delta E_{\rm GM}$ (keV)} \\
\cline{4-10}
&&& \multicolumn{5}{c}{Perturb.} & \multirow{2}{*}{SCB} &
\multirow{2}{*}{Exp.} & \multirow{2}{*}{Refs.} \\
\cline{4-8}
&&& $t_0$ & $t_3$ & $t_1+t_2$ & $W$ & Total &  &  \\
\hline
\endfirsthead
\multicolumn{10}{c}{TABLE \thetable: Continued.} \\
\hline
\multirow{3}{*}{Nucleus} & \multirow{3}{*}{Configuration}
& \multirow{3}{*}{$(K_{\uparrow\uparrow},K_{\uparrow\downarrow})^{\pi}$}
& \multicolumn{7}{c}{$\EGM$ (keV)} \\
\cline{4-10}
&&& \multicolumn{5}{c}{Perturb.} & \multirow{2}{*}{SCB} &
\multirow{2}{*}{Exp.} & \multirow{2}{*}{Refs.} \\
\cline{4-8}
&&& $t_0$ & $t_3$ & $t_1+t_2$ & $W$ & Total &  &  \\
\hline
\endhead
%
%\hline
\endfoot
\hline\hline
\endlastfoot
$^{154}$Eu & $ ({11}/2^-[505]\uparrow
93\%)_n({5}/2^+[413]\downarrow 84\%)_p $ & $ (3,8)^- $ & $ 281.1 $
& $ -140.1 $ & $ -135.4 $ & $ 47.3 $ & $ 53.0 $ & $ 21.7 $ & 145.3 &
\multirow{4}{*}{\cite{Eu154_NDS110_2009,Rotter84_NPA417_Eu154,Balodis87_NPA472_Eu154}} \\
$^{154}$Eu & $ ({3}/2^+[651]\uparrow 35\%)_n({5}/2^+[413]\downarrow 83\%)_p $ & $ (1,4)^+ $ & $ 117.9 $ & $ -64.3 $ & $ -42.2 $ & $ 0.7 $ & $ 12.1 $ & $ -2.8 $ & 131.905 \\
$^{154}$Eu & $ ({3}/2^-[521]\uparrow 55\%)_n({5}/2^+[413]\downarrow 83\%)_p $ & $ (1,4)^- $ & $ 119.4 $ & $ -41.9 $ & $ -42.5 $ & $ 0.0 $ & $ 35.0 $ & $ 18.9 $ & 152.459 \\
$^{154}$Eu & $ ({3}/2^-[521]\uparrow 54\%)_n({3}/2^+[411]\uparrow 58\%)_p $ & $ (3,0)^- $ & $ 290.4 $ & $ -155.0 $ & $ 74.5 $ & $ 9.3 $ & $ 219.3 $ & $ 203.8 $ & 39.7488 \\
\hline
$^{156}$Eu & $ ({5}/2^+[642]\uparrow
53\%)_n({5}/2^+[413]\downarrow 84\%)_p $ & $ (0,5)^+ $ & $ 183.6 $
& $ -96.1 $ & $ -75.5 $ & $ 12.6 $ & $ 24.6 $ & $ -0.8 $ & 145.682 &
\multirow{4}{*}{\cite{Eu156_NDS113_2012}} \\
$^{156}$Eu & $ ({3}/2^-[521]\uparrow 56\%)_n({5}/2^+[413]\downarrow 83\%)_p $ & $ (1,4)^- $ & $ 119.4 $ & $ -42.5 $ & $ -42.5 $ & $ 0.0 $ & $ 34.5 $ & $ 20.4 $ & 127.441 \\
$^{156}$Eu & $ ({5}/2^+[642]\uparrow 53\%)_n({5}/2^-[532]\uparrow 60\%)_p $ & $ (5,0)^- $ & $ 336.5 $ & $ -154.5 $ & $ 139.4 $ & $ 4.5 $ & $ 325.9 $ & $ 264.6 $ & 68.1036 \\
$^{156}$Eu & $ ({3}/2^-[521]\uparrow
56\%)_n({5}/2^-[532]\uparrow 60\%)_p $ & $ (4,1)^+ $ & $ 123.8 $ &
$ -46.9 $ & $ 45.3 $ & $ -3.4 $ & $ 118.8 $ & $ 94.6 $ & 116.154 \\
\hline
$^{158}$Tb & $ ({3}/2^-[521]\uparrow
55\%)_n({3}/2^+[411]\uparrow 59\%)_p $ & $ (3,0)^- $ & $ 299.2 $ &
$ -160.5 $ & $ 74.7 $ & $ 8.5 $ & $ 221.9 $ & $ 207.5 $ & 110.3 &
\multirow{6}{*}{\cite{Nica17_A=158}} \\
$^{158}$Tb & $ ({5}/2^+[642]\uparrow 53\%)_n({3}/2^+[411]\uparrow 59\%)_p $ & $ (4,1)^+ $ & $ 110.0 $ & $ -46.1 $ & $ 19.2 $ & $ -4.1 $ & $ 79.1 $ & $ 56.8 $ & 123.86 \\
$^{158}$Tb & $ ({11}/2^-[505]\uparrow 92\%)_n({3}/2^+[411]\uparrow 60\%)_p $ & $ (7,4)^- $ & $ 162.4 $ & $ -64.8 $ & $ 25.3 $ & $ -12.5 $ & $ 110.4 $ & $ 90.8 $ & 107.01 \\
$^{158}$Tb & $ ({3}/2^+[402]\downarrow 74\%)_n({3}/2^+[411]\uparrow 60\%)_p $ & $ (0,3)^+ $ & $ 286.1 $ & $ -152.6 $ & $ -75.7 $ & $ -32.7 $ & $ 25.1 $ & $ -25.4 $ & 181.9 \\
$^{158}$Tb & $ ({1}/2^+[640]\uparrow 25\%)_n({3}/2^+[411]\uparrow 58\%)_p $ & $ (2,1)^+ $ & $ 48.2 $ & $ -23.4 $ & $ 4.6 $ & $ 4.1 $ & $ 33.5 $ & $ 17.7 $ & 60.9 \\
$^{158}$Tb & $ ({3}/2^-[521]\uparrow 54\%)_n({7}/2^+[404]\downarrow 87\%)_p $ & $ (2,5)^- $ & $ 101.2 $ & $ -39.5 $ & $ -27.1 $ & $ 10.5 $ & $ 45.1 $ & $ 32.6 $ & 108 \\
\hline
$^{160}$Tb & $ ({3}/2^-[521]\uparrow 55\%)_n({3}/2^+[411]\uparrow 59\%)_p $ & $ (3,0)^- $ & $ 299.2 $ & $ -161.7 $ & $ 74.7 $ & $ 8.5 $ & $ 220.7 $ & $ 210.3 $ & 79.0925 &
\multirow{4}{*}{\cite{Nica17_A=160}} \\
$^{160}$Tb & $ ({5}/2^-[523]\downarrow 73\%)_n({3}/2^+[411]\uparrow 59\%)_p $ & $ (1,4)^- $ & $ 141.5 $ & $ -56.2 $ & $ -43.3 $ & $ -16.3 $ & $ 25.7 $ & $ 8.6 $ & 193.855 \\
$^{160}$Tb & $ ({5}/2^+[642]\uparrow 52\%)_n({3}/2^+[411]\uparrow 59\%)_p $ & $ (4,1)^+ $ & $ 110.0 $ & $ -46.7 $ & $ 19.2 $ & $ -4.1 $ & $ 78.5 $ & $ 54.7 $ & 74.6254 \\
$^{160}$Tb & $ ({5}/2^+[642]\uparrow 52\%)_n({5}/2^+[413]\downarrow 83\%)_p $ & $ (0,5)^+ $ & $ 185.8 $ & $ -99.0 $ & $ -75.3 $ & $ 11.6 $ & $ 23.1 $ & $ -2.6 $ & 203.371 \\
\hline
$^{172}$Tm & $ ({5}/2^-[512]\uparrow 73\%)_n({1}/2^+[411]\downarrow 65\%)_p $ & $ (2,3)^- $ & $ 400.9 $ & $ -208.0 $ & $ -80.4 $ & $ 58.6 $ & $ 171.1 $ & $ 92.9 $ & 62.529 &
\multirow{2}{*}{\cite{Tm172_NDS75_1995}} \\
$^{172}$Tm & $ ({1}/2^-[521]\downarrow 54\%)_n({1}/2^+[411]\downarrow 64\%)_p $ & $ (1,0)^- $ & $ 242.0 $ & $ -122.7 $ & $ 45.6 $ & $ 17.0 $ & $ 181.9 $ & $ 171.1 $ & 68.108 \\
$^{172}$Tm & $ ({5}/2^-[512]\uparrow 73\%)_n({7}/2^-[523]\uparrow 77\%)_p $ & $ (6,1)^+ $ & $ 143.2 $ & $ -57.2 $ & $ 51.6 $ & $ -6.2 $ & $ 131.4 $ & $ 106.7 $ & 133.862 & \cite{Hughes08_PRC77} \\
\hline
$^{174}$Lu & $ ({5}/2^-[512]\uparrow 72\%)_n({7}/2^+[404]\downarrow 92\%)_p $ & $ (1,6)^- $ & $ 139.0 $ & $ -56.7 $ & $ -55.7 $ & $ -2.5 $ & $ 24.1 $ & $ 9.1 $ & 170.83 &
\multirow{4}{*}{\cite{Lu174_NDS87_1999}} \\
$^{174}$Lu & $ ({7}/2^+[633]\uparrow 71\%)_n({7}/2^+[404]\downarrow 92\%)_p $ & $ (0,7)^+ $ & $ 197.3 $ & $ -106.6 $ & $ -99.5 $ & $ 6.2 $ & $ -2.6 $ & $ -29.3 $ & 150.242 \\
$^{174}$Lu & $ ({1}/2^-[521]\downarrow 51\%)_n({7}/2^+[404]\downarrow 92\%)_p $ & $ (4,3)^- $ & $ 86.6 $ & $ -37.4 $ & $ 12.8 $ & $ 8.0 $ & $ 70.0 $ & $ 55.1 $ & 67.697 \\
$^{174}$Lu & $ ({5}/2^-[512]\uparrow 72\%)_n({9}/2^-[514]\uparrow 87\%)_p $ & $ (7,2)^+ $ & $ 257.5 $ & $ -99.5 $ & $ 67.4 $ & $ -4.6 $ & $ 220.8 $ & $ 151.3 $ & 104.1 \\
\hline
$^{176}$Lu & $ ({7}/2^-[514]\downarrow 86\%)_n({7}/2^+[404]\downarrow 91\%)_p $ & $ (7,0)^- $ & $ 305.6 $ & $ -163.1 $ & $ 180.5 $ & $ 10.7 $ & $ 333.7 $ & $ 314.6 $ & 236.908 &
\multirow{12}{*}{\cite{Lu176_NDS107_2006}} \\
$^{176}$Lu & $ ({7}/2^-[514]\downarrow 85\%)_n({9}/2^-[514]\uparrow 86\%)_p $ & $ (1,8)^+ $ & $ 487.9 $ & $ -251.9 $ & $ -246.8 $ & $ 8.9 $ & $ -1.9 $ & $ -145.4 $ & 293.482 \\
$^{176}$Lu & $ ({9}/2^+[624]\uparrow 83\%)_n({7}/2^+[404]\downarrow 91\%)_p $ & $ (1,8)^+ $ & $ 269.0 $ & $ -141.7 $ & $ -137.8 $ & $ 19.9 $ & $ 9.3 $ & $ -39.7 $ & 86.0468 \\
$^{176}$Lu & $ ({7}/2^-[514]\downarrow 86\%)_n({5}/2^+[402]\uparrow 69\%)_p $ & $ (1,6)^- $ & $ 159.1 $ & $ -69.0 $ & $ -61.8 $ & $ -11.6 $ & $ 16.6 $ & $ 4.6 $ & 177.357 \\
$^{176}$Lu & $ ({5}/2^-[512]\uparrow 70\%)_n({7}/2^+[404]\downarrow 91\%)_p $ & $ (1,6)^- $ & $ 139.0 $ & $ -57.7 $ & $ -55.7 $ & $ -2.5 $ & $ 23.1 $ & $ 7.7 $ & 127.911 \\
$^{176}$Lu & $ ({7}/2^-[514]\downarrow 86\%)_n({1}/2^-[541]\downarrow 33\%)_p $ & $ (4,3)^+ $ & $ 42.7 $ & $ -23.3 $ & $ 18.1 $ & $ 6.3 $ & $ 43.8 $ & $ 37.0 $ & 99.162 \\
$^{176}$Lu & $ ({1}/2^-[510]\uparrow 65\%)_n({7}/2^+[404]\downarrow 91\%)_p $ & $ (3,4)^- $ & $ 117.6 $ & $ -49.1 $ & $ -2.5 $ & $ 2.8 $ & $ 68.9 $ & $ 49.3 $ & 129.779 \\
$^{176}$Lu & $ ({7}/2^-[514]\downarrow 87\%)_n({1}/2^+[411]\downarrow 61\%)_p $ & $ (4,3)^- $ & $ 267.5 $ & $ -123.0 $ & $ 34.3 $ & $ 8.1 $ & $ 186.9 $ & $ 130.5 $ & 120.506 \\
$^{176}$Lu & $ ({9}/2^+[624]\uparrow 83\%)_n({5}/2^+[402]\uparrow 70\%)_p $ & $ (7,2)^+ $ & $ 141.8 $ & $ -59.7 $ & $ 38.1 $ & $ -3.0 $ & $ 117.1 $ & $ 86.9 $ & 132.323 \\
$^{176}$Lu & $ ({3}/2^-[512]\downarrow 71\%)_n({7}/2^+[404]\downarrow 91\%)_p $ & $ (5,2)^- $ & $ 139.2 $ & $ -58.5 $ & $ 42.6 $ & $ 2.3 $ & $ 125.5 $ & $ 90.9 $ & 194.861 \\
$^{176}$Lu & $ ({1}/2^-[521]\downarrow 49\%)_n({7}/2^+[404]\downarrow 91\%)_p $ & $ (4,3)^- $ & $ 86.6 $ & $ -37.9 $ & $ 12.8 $ & $ 8.0 $ & $ 69.5 $ & $ 52.5 $ & 49.642 \\
$^{176}$Lu & $ ({7}/2^-[514]\downarrow 86\%)_n({7}/2^-[523]\uparrow 76\%)_p $ & $ (0,7)^+ $ & $ 292.8 $ & $ -152.7 $ & $ -140.3 $ & $ 15.0 $ & $ 14.8 $ & $ -41.4 $ & 217.5 \\
\hline
$^{178}$Lu & $ ({9}/2^+[624]\uparrow 82\%)_n({7}/2^+[404]\downarrow 91\%)_p $ & $ (1,8)^+ $ & $ 264.8 $ & $ -139.0 $ & $ -134.6 $ & $ 20.8 $ & $ 12.0 $ & $ -38.3 $ & 187 & \cite{A=178_NDS110_2009,Burke93_PRC47} \\
\hline
%% $^{178}$Ta & $ ({7}/2^-[514]\downarrow
%% 86\%)_n({7}/2^+[404]\downarrow 91\%)_p $ & $ (7,0)^- $ & $ 305.9 $
%% & $ -164.4 $ & $ 178.6 $ & $ 10.9 $ & $ 330.9 $ & $ 299.1 $ & \\
$^{178}$Ta & $ ({7}/2^-[514]\downarrow
86\%)_n({5}/2^+[402]\uparrow 67\%)_p $ & $ (1,6)^- $ & $ 159.1 $ &
$ -69.6 $ & $ -60.8 $ & $ -10.9 $ & $ 17.8 $ & $ 6.2 $ & 188.1 &
\cite{A=178_NDS110_2009,Kondev98_NPA632} \\
$^{178}$Ta & $ ({7}/2^-[514]\downarrow
86\%)_n({9}/2^-[514]\uparrow 86\%)_p $ & $ (1,8)^+ $ & $ 478.3 $ &
$ -247.9 $ & $ -237.9 $ & $ 10.5 $ & $ 3.0 $ & $ -141.0 $ & 100 & \cite{Nubase2020} \\
\hline
$^{180}$Ta & $ ({9}/2^+[624]\uparrow 81\%)_n({7}/2^+[404]\downarrow 90\%)_p $ & $ (1,8)^+ $ & $ 264.8 $ & $ -140.4 $ & $ -134.6 $ & $ 20.8 $ & $ 10.6 $ & $ -36.9 $ & 177.87 &
\multirow{5}{*}{\cite{Ta180_NDS126_2015}} \\
$^{180}$Ta & $ ({9}/2^+[624]\uparrow 82\%)_n({9}/2^-[514]\uparrow 85\%)_p $ & $ (9,0)^- $ & $ 310.9 $ & $ -151.9 $ & $ 173.5 $ & $ -10.8 $ & $ 321.7 $ & $ 264.1 $ & 30.58 \\
$^{180}$Ta & $ ({7}/2^-[514]\downarrow 86\%)_n({9}/2^-[514]\uparrow 85\%)_p $ & $ (1,8)^+ $ & $ 478.3 $ & $ -251.0 $ & $ -237.9 $ & $ 10.5 $ & $ -0.2 $ & $ -138.4 $ & 195.57 \\
$^{180}$Ta & $ ({9}/2^+[624]\uparrow 83\%)_n({5}/2^+[402]\uparrow 66\%)_p $ & $ (7,2)^+ $ & $ 138.6 $ & $ -59.4 $ & $ 37.3 $ & $ -3.2 $ & $ 113.3 $ & $ 80.9 $ & 202.49 \\
$^{180}$Ta & $ ({5}/2^-[512]\uparrow 68\%)_n({7}/2^+[404]\downarrow 90\%)_p $ & $ (1,6)^- $ & $ 140.5 $ & $ -60.6 $ & $ -55.3 $ & $ -4.3 $ & $ 20.3 $ & $ 3.7 $ & 151.43 \\
\hline
$^{230}$Pa & $ ({5}/2^-[503]\downarrow 52\%)_n({1}/2^-[530]\uparrow 24\%)_p $ & $ (2,3)^+ $ & $ 54.3 $ & $ -23.8 $ & $ -4.1 $ & $ -16.7 $ & $ 9.7 $ & $ 5.4 $ & 148 &
\multirow{7}{*}{\cite{A=230_NDS113_2012,Kotthaus13_PRC87}} \\
$^{230}$Pa & $ ({3}/2^+[631]\uparrow 33\%)_n({1}/2^-[530]\uparrow 24\%)_p $ & $ (2,1)^- $ & $ 358.2 $ & $ -204.0 $ & $ 5.1 $ & $ 71.1 $ & $ 230.3 $ & $ 169.7 $ & 259 \\
$^{230}$Pa & $ ({5}/2^+[633]\downarrow 57\%)_n({1}/2^-[530]\uparrow 24\%)_p $ & $ (2,3)^- $ & $ 125.0 $ & $ -57.6 $ & $ -4.0 $ & $ 1.0 $ & $ 64.4 $ & $ 42.4 $ & 109 \\
$^{230}$Pa & $ ({1}/2^+[631]\downarrow 29\%)_n({1}/2^-[530]\uparrow 24\%)_p $ & $ (0,1)^- $ & $ 214.5 $ & $ -111.7 $ & $ -3.3 $ & $ -59.9 $ & $ 39.5 $ & $ -27.9 $ & 139 \\
$^{230}$Pa & $ ({1}/2^-[501]\downarrow 37\%)_n({1}/2^-[530]\uparrow 24\%)_p $ & $ (0,1)^+ $ & $ 105.8 $ & $ -59.1 $ & $ -2.0 $ & $ -7.1 $ & $ 37.6 $ & $ 25.5 $ & 201 \\
$^{230}$Pa & $ ({3}/2^-[741]\uparrow 29\%)_n({1}/2^-[530]\uparrow 21\%)_p $ & $ (2,1)^+ $ & $ 122.6 $ & $ -47.4 $ & $ 3.1 $ & $ 3.6 $ & $ 81.8 $ & $ 43.4 $ & 138 \\
\hline
$^{234}$Pa & $ ({7}/2^-[743]\uparrow 52\%)_n({1}/2^-[530]\uparrow 22\%)_p $ & $ (4,3)^+ $ & $ 149.9 $ & $ -61.1 $ & $ 2.6 $ & $ -2.3 $ & $ 89.1 $ & $ 52.4 $ & 73.92 &
\cite{A=234_NDS108_2007,Godart73_NPA217} \\
$^{234}$Pa & $ ({1}/2^+[631]\downarrow 30\%)_n({1}/2^-[530]\uparrow 22\%)_p $ & $ (0,1)^- $ & $ 225.3 $ & $ -120.9 $ & $ -3.2 $ & $ -60.1 $ & $ 41.1 $ & $ -10.0 $ & 92.38 & \cite{A=234_NDS108_2007,Godart73_NPA217,DePinho65_NP65} \\
\hline
$^{238}$Np & $ ({1}/2^+[631]\downarrow 33\%)_n({5}/2^+[642]\uparrow 38\%)_p $ & $ (2,3)^+ $ & $ 110.4 $ & $ -46.0 $ & $ -17.9 $ & $ 2.5 $ & $ 49.0 $ & $ 25.2 $ & 86.6738 &
\multirow{3}{*}{\cite{A=238_NDS127_2015}} \\
$^{238}$Np & $ ({1}/2^+[631]\downarrow 33\%)_n({5}/2^-[523]\downarrow 69\%)_p $ & $ (3,2)^- $ & $ 89.5 $ & $ -37.1 $ & $ 16.6 $ & $ 1.2 $ & $ 70.3 $ & $ 47.8 $ & 46.8325 \\
$^{238}$Np & $ ({1}/2^+[631]\downarrow 32\%)_n({1}/2^-[530]\uparrow 21\%)_p $ & $ (0,1)^- $ & $ 252.6 $ & $ -135.5 $ & $ -3.1 $ & $ -59.1 $ & $ 55.0 $ & $ -10.7 $ & 155.735 \\
\hline
$^{240}$Am & $ ({1}/2^+[631]\downarrow 34\%)_n({5}/2^-[523]\downarrow 69\%)_p $ & $ (3,2)^- $ & $ 89.5 $ & $ -37.6 $ & $ 16.7 $ & $ 1.2 $ & $ 69.8 $ & $ 47.9 $ & 53 &
\multirow{2}{*}{\cite{A=240_NDS109_2008}} \\
$^{240}$Am & $ ({1}/2^-[501]\downarrow 43\%)_n({5}/2^-[523]\downarrow 69\%)_p $ & $ (3,2)^+ $ & $ 66.1 $ & $ -31.7 $ & $ 7.7 $ & $ -1.5 $ & $ 40.6 $ & $ 34.9 $ & 43 \\
\hline
$^{242}$Am & $ ({5}/2^+[622]\uparrow 46\%)_n({5}/2^-[523]\downarrow 68\%)_p $ & $ (0,5)^- $ & $ 158.7 $ & $ -75.8 $ & $ -60.3 $ & $ -32.8 $ & $ -10.2 $ & $ -59.7 $ & 4.508 &
\multirow{6}{*}{\cite{A=242_NDS186_2022}} \\
$^{242}$Am & $ ({1}/2^+[631]\downarrow 33\%)_n({5}/2^-[523]\downarrow 68\%)_p $ & $ (3,2)^- $ & $ 90.0 $ & $ -37.9 $ & $ 16.3 $ & $ 1.4 $ & $ 69.8 $ & $ 47.2 $ & 48.48 \\
$^{242}$Am & $ ({1}/2^+[631]\downarrow 32\%)_n({5}/2^+[642]\uparrow 36\%)_p $ & $ (2,3)^+ $ & $ 111.2 $ & $ -46.6 $ & $ -17.5 $ & $ 2.0 $ & $ 49.1 $ & $ 24.7 $ & 32.49 \\
$^{242}$Am & $ ({1}/2^+[620]\uparrow 37\%)_n({5}/2^-[523]\downarrow 69\%)_p $ & $ (2,3)^- $ & $ 142.9 $ & $ -70.5 $ & $ -2.8 $ & $ -1.9 $ & $ 67.7 $ & $ 32.9 $ & 28.42 \\
$^{242}$Am & $ ({1}/2^-[501]\downarrow 43\%)_n({5}/2^+[642]\uparrow 36\%)_p $ & $ (2,3)^- $ & $ 52.0 $ & $ -27.1 $ & $ -5.2 $ & $ 1.4 $ & $ 21.0 $ & $ 21.9 $ & 28.42 \\
$^{242}$Am & $ ({1}/2^-[501]\downarrow 43\%)_n({5}/2^-[523]\downarrow 68\%)_p $ & $ (3,2)^+ $ & $ 65.5 $ & $ -31.8 $ & $ 7.8 $ & $ -1.6 $ & $ 39.9 $ & $ 33.6 $ & -36 \\
\hline
$^{244}$Am & $ ({7}/2^+[624]\downarrow
75\%)_n({5}/2^-[523]\downarrow 68\%)_p $ & $ (6,1)^- $ & $ 389.2 $
& $ -219.3 $ & $ 145.5 $ & $ 33.5 $ & $ 348.9 $ & $ 315.5 $ & 177.2 & \cite{A=244_NDS146_2017,vonEgidy84_PRC29} \\
\hline
$^{250}$Bk & $ ({1}/2^+[620]\uparrow 38\%)_n({3}/2^-[521]\uparrow 30\%)_p $ & $ (2,1)^- $ & $ 204.2 $ & $ -105.6 $ & $ 4.4 $ & $ 33.8 $ & $ 136.7 $ & $ 97.1 $ & 103.83 &
\multirow{5}{*}{\cite{A=250_NDS94_2001}} \\
$^{250}$Bk & $ ({1}/2^+[620]\uparrow 38\%)_n({7}/2^+[633]\uparrow 52\%)_p $ & $ (4,3)^+ $ & $ 91.2 $ & $ -38.2 $ & $ 2.6 $ & $ -0.3 $ & $ 55.3 $ & $ 39.6 $ & 80.06 \\
$^{250}$Bk & $ ({7}/2^+[613]\uparrow
51\%)_n({7}/2^+[633]\uparrow 53\%)_p $ & $ (7,0)^+ $ & $ 82.5 $ &
$ -36.7 $ & $ 34.0 $ & $ -4.3 $ & $ 75.5 $ & $ 59.6 $ & 129.94 \\
$^{250}$Bk & $ ({7}/2^+[613]\uparrow 51\%)_n({3}/2^-[521]\uparrow 30\%)_p $ & $ (5,2)^- $ & $ 109.1 $ & $ -55.7 $ & $ 36.0 $ & $ 14.0 $ & $ 103.4 $ & $ 97.1 $ & 48.94 \\
$^{250}$Bk & $ ({3}/2^+[622]\downarrow 45\%)_n({7}/2^+[633]\uparrow 52\%)_p $ & $ (2,5)^+ $ & $ 111.4 $ & $ -47.2 $ & $ -30.1 $ & $ 1.6 $ & $ 35.7 $ & $ 19.5 $ & 104.64 \\
\hline\hline
\end{longtable}

\end{widetext}

The first global observation that can be made relates to the sign of
the calculated $\EGM$ values. For the lighter
rare-earth nuclei considered in this work, from $^{154}$Eu to
$^{172}$Tm, the calculated Gallagher--Moszkowski energy splittings are
all positive except in four cases. Out of these four exceptions three
correspond to a very small value of $\big|\EGM\big| < 3$~keV and one 
(in $^{158}$Tb) is such that $\big|\EGM\big| \approx 25$~keV. One can
thus conclude that our calculations are in an overall qualitative
agreement with the empirical rule in all five investigated
nuclei. More quantitatively this corresponds to an about 80\% of
agreement with experiment. In contrast, in the other studied
rare-earth-and-beyond nuclei, from $^{174}$Lu to $^{180}$Ta, we find 
8 cases with $\EGM \leqslant 0$ out of 24 experimentally
observed doublets in these Lu and Ta nuclei, hence a percentage of
agreement of about  65\% in this mass region. Finally, in the studied 
actinides, from $^{230}$Pa to $^{250}$Bk, the sign of the energy
splitting $\EGM$ is much better reproduced than in the Lu and
Ta doublets, with only 4 discrepant cases out of a total of 25
experimentally observed doublets (84\% of agreement),
slightly better than in the lighter studied rare-earth nuclei.

Another global observation about the calculated energy splittings
relates to their amplitude. In the majority of cases we obtain
$\EGM$ between 0 and 100~keV, but it can reach 300~keV in
doublets with a large difference in $K$ values. The cases breaking the
Gallagher--Moszkowski rule correspond to $\big|\EGM\big| 
\lesssim 50$~keV except for one doublet $(1^+,8^+)$ in three nuclei
($^{176}$Lu, $^{178,180}$Ta) for which $\EGM \approx -140$~keV.

To better understand these observations and try to unravel the
mechanism of energy splitting in Gallagher--Moskkowski doublets within
the Skyrme energy-density functional framework, we consider a
perturbative approach to blocking.

\subsection{Perturbative blocking}

Perturbative blocking corresponds to performing one Hartree--Fock--BCS
iteration with blocking on top of the converged ground-state solution
of a neighboring even-even nucleus (called a ``core''). Therefore
there is some arbitrariness in the choice of this even-even core among
four possible neighboring even-even nuclei. Because we selected the
studied odd-odd nuclei as being neighbors of chosen even-even ones, as
shown in Figs.~\ref{fig_nucleid_chart_rare-earths} and
\ref{fig_nucleid_chart_actinides}, we decide to choose the cores among
these doubly-even nuclei. Table~\ref{tab_list_cores} lists the
retained even-even cores.

The observable we are studying is essentially
dependent on the neutron and proton single-particle wave functions
involved in the blocked configuration. Indeed we find that, assuming
perturbative blocking 
\eq{
  \EGM = \elmx{n\overline p}{\widehat V_{\rm Sk}}
         {\widetilde{n\overline p}} \,,
}
where $\widehat V_{\rm Sk}$ is the Skyrme effective potential
(including the density-dependent term with the nucleon density of the
core), $\ket{n}$ is the blocked neutron single-particle state,
$\ket{\overline p}$ is the time reversal of the proton blocked state
and $\ket{\widetilde{n\overline p}} =
\ket{n\overline p} - \ket{\overline p \, n}$. Therefore we expect that
the dependence of the perturbatively calculated $\EGM$ on the
core is weak. 
\begin{table}[h]
  \caption{List of the retained even-even core for each studied
    odd-odd nucleus.
    \label{tab_list_cores}}
  \begin{tabular}{cc}
    \hline \hline
    Odd-odd nuclei & Even-even core \\
    \hline
    $^{154,156}$Eu & $^{156}$Gd \\
    $^{158,160}$Tb & $^{158}$Gd \\
    $^{172}$Tm, $^{174,176}$Lu & $^{174}$Yb \\
    $^{178}$Lu, $^{178,180}$Ta & $^{178}$Hf \\
    $^{230}$Pa & $^{230}$Th \\
    $^{234}$Pa & $^{232}$Th \\
    $^{240}$Am & $^{240}$Pu \\
    $^{242,244}$Am & $^{242}$Pu \\
    $^{250}$Bk & $^{250}$Cf \\
    \hline \hline
  \end{tabular}
\end{table}

\begin{figure}[h]
  \includegraphics[width=0.475\textwidth]{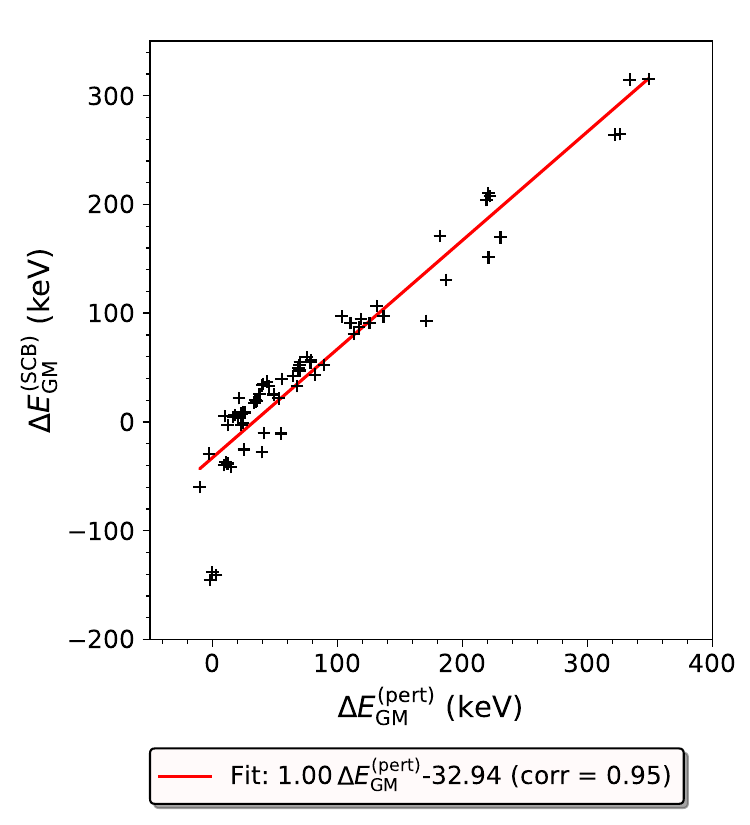}
  \caption{Correlation plot of $\EGM^{\rm (SCB)}$ ands a function of $\EGM^{\rm
    (pert)}$ for the 70 Gallagher--Moszkowski doublets considered in
    the present work. The correlation coefficient of the linear
    regression is denoted by ``corr''.
    \label{fig_correlation_SCB_perturb}}
\end{figure}

First we check whether calculations with perturbative blocking 
on top of the mean field of a neighboring even-even nucleus lead to
an energy splitting, denoted by $\EGM^{(\rm pert)}$ comparable to 
the one obtained with selfconsistent blocking, denoted by
$\EGM^{(\rm SCB)}$. Table~\ref{tab_GM_splitting} displays the
$\EGM$ values obtained with perturbative (in the column labelled
``Total'') and selfconsistent blocking (in the column labelled
``SCB'') for all the 70 Gallagher--Moszkowski doublets considered in
the present work, and Figure~\ref{fig_correlation_SCB_perturb} shows the
correlation plot of $\EGM^{(\rm pert)}$ and $\EGM^{(\rm SCB)}$ for all
these doublets as well as the linear regression equation and
correlation coefficient.
The latter is found to be almost equal to 1, whereas three 
outliers are clearly visible in the plot. These outliers
correspond precislely to the $(1^+,8^+)$ doublet in the three 
aforementioned nuclei $^{176}$Lu and $^{178,180}$Ta. Moreover the
intercept in the equation of linear regression is calculated to be
about $-33$~keV, which means that the $\EGM$ value calculated with
selfconsistent blocking tends to be a bit smaller than the one
obtained with perturbative blocking. Thus in most cases $\EGM^{(\rm
  SCB)}$ and $\EGM^{(\rm pert)}$ have the same sign and order of
magnitude, and the discrepancies occur in doublets for which the
splitting is small (below a few tens of keV). This statistic analysis
establishes the relevance of perturbative calculations to investigate
the mechanism of Gallagher--Moszkowski splitting. This will be done in
the next subsection through an analysis of the contributions to $\EGM$
from the various terms of the Skyrme energy-density functional.

\subsection{Mechanism of the Gallagher--Moszkowski splitting}

We now address the mechanism by which the Gallagher--Moszkowski energy
splitting $\EGM$ occurs in the perturbative-blocking
framework. In the appendix we derive the following expressions of the
contributions of the Skyrme EDF to $\EGM$ 
\eq{
  \EGM^{(\rm pert)} =
  \EGM^{(t_0)} + \EGM^{(t_3)} + \EGM^{(t_1+t_2)} + \EGM^{(W)}
}
where
\begin{subequations}
\begin{align}
  \label{eq_EGM_t0}
  \EGM^{(t_0)} & =  -t_0\, x_0
  \int d^3 \vect r \; \vect s_n \cdot \vect s_p \\
  \EGM^{(t_3)} & =  -\frac{t_3}6 \int d^3 \vect r \; \rho^{\alpha}
  \, \vect s_n \cdot \vect s_p \\
  \EGM^{(t_1+t_2)} & = (t_1+t_2)
    \int d^3 \vect r \; \vect j_n \cdot \vect j_p \\
  \EGM^{(W)} & =  W \int d^3 \vect r \, \big(\vect s_n \cdot
      \bm{\nabla}\times\vect j_p + \vect s_p \cdot
      \bm{\nabla}\times\vect j_n\big) \,.
      \label{eq_EGM_W}
  \end{align}
\end{subequations}
By definition $\vect s_n$ and $\vect s_p$ are the neutron and proton
local spin densities of the neutron-spin-up--proton-spin-up
configuration (which is thus spin-aligned). Similarly $\vect j_n$ and
$\vect j_p$ are the neutron and proton local current densities in this
configuration. Therefore the central $\EGM^{(t_0)}$
and density-dependent $\EGM^{(t_3)}$ zero-range contributions have 
the signs of $-t_0x_0$ and $-t_3$, respectively. According to the
values of the SIII parameters, one always has $\EGM^{(t_0)} > 0$ and
$\EGM^{(t_3)} < 0$. The signs of $\EGM^{(t_1+t_2)}$ and $\EGM^{(W)}$,
in which $t_1+t_2 > 0$ and $W > 0$, depend on the contrary on the
spin and orbital contents of the blocked states.

\begin{figure}[h]
  \includegraphics[width=0.475\textwidth]{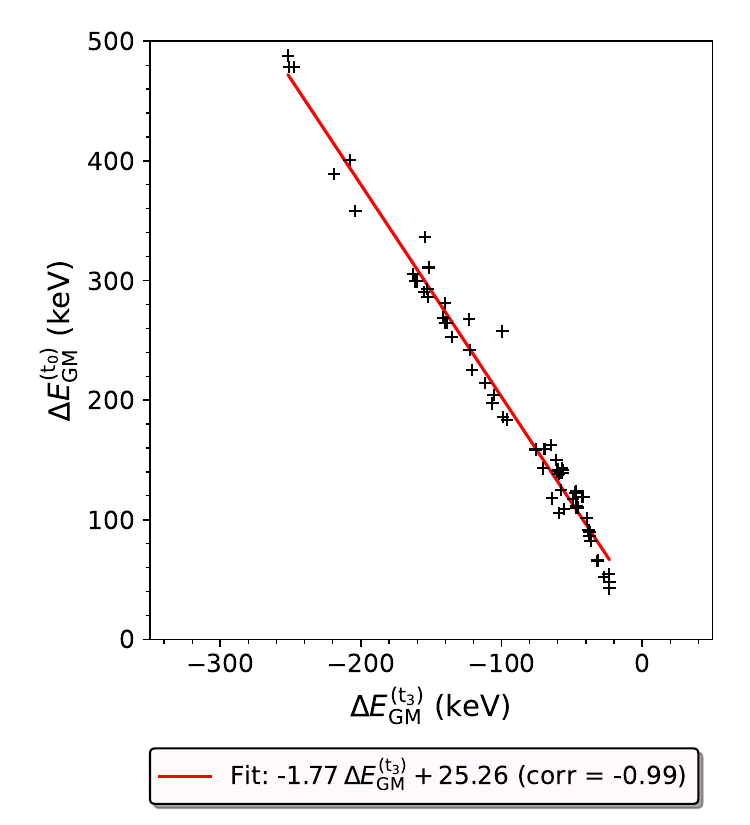}
  \includegraphics[width=0.475\textwidth]{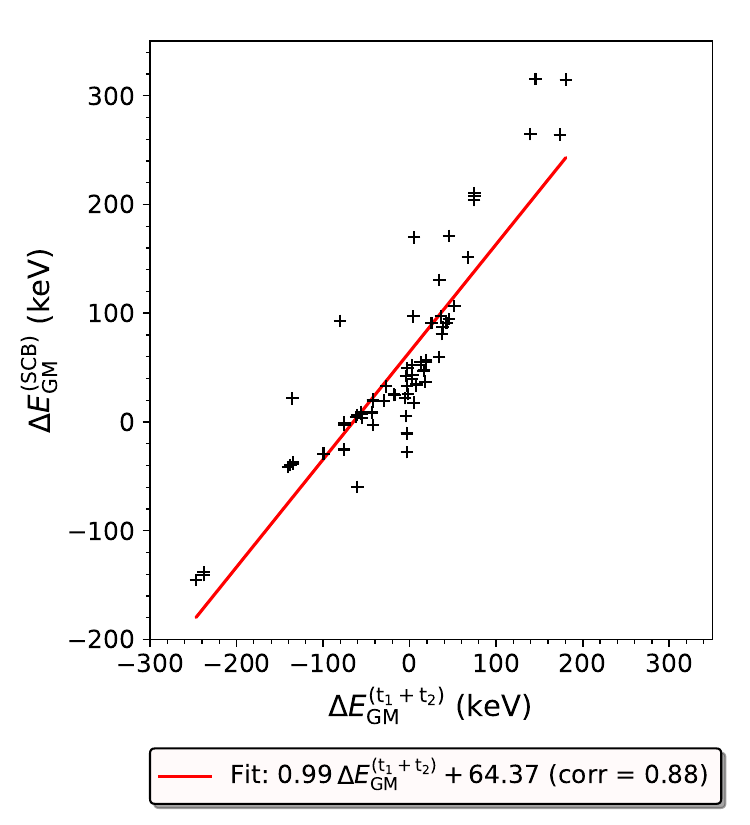}
  \caption{Same as Fig.~\ref{fig_correlation_SCB_perturb} for the
    correlation between $\EGM^{(t_0)}$ and $\EGM^{(t_3)}$ (upper
    panel), and between $\EGM^{(t_1+t_2)}$ and $\EGM^{(\rm SCB)}$
    (lower panel). \label{fig_corr_plots}}
\end{figure}

The numerical results for the various contributions to $\Delta E_{\rm
  GM}$ in the perturbative-blocking calculations are displayed in
Table~\ref{tab_GM_splitting}, and are compared
with the Gallagher--Moszkowski energy splittings obtained with
selfconsistent blocking. Several systematic trends in perturbative
results emerge from the large number of considered doublets: 
\begin{itemize}
\item $\EGM^{(t_0)}$ and $\EGM^{(t_3)}$ are the largest two
  contributions in absolute value and are related by
  $\EGM^{(t_0)} \approx -2 \, \EGM^{(t_3)}$, so that the sum
  of these two contributions is always positive, of the order of
  $\EGM^{(t_0)} + \EGM^{(t_3)} \approx 50$ to 150~keV.
\item The $\EGM^{(t_1+t_2)}$ contribution is positive whenever the
  spin-antialigned configuration corresponds to the smaller of the two
  $K$ values, in other words, when the lower-lying configuration has
  the larger $K$ value.
\item The $\EGM^{(W)}$ contribution is small (less than 25 keV in
  absolute value) in most configurations, with 8 exceptions only for
  which $|\EGM^{(W)}|$ ranges from about 35 to 70~keV.
\end{itemize}
The first two points can be substantiated by the correlation plots of
figure~\ref{fig_corr_plots}.

The main conclusion from these observations is that the
Gallagher--Moszkowski rule is always satisfied in doublets for which
the spin-aligned configuration corresponds to the larger $K$
value. This is so because the $t_1+t_2$ (central gradient)
positive contribution adds up to the zero-range contributions,
and is not counter-balanced by the too small spin-orbit
contribution. \\

It is possible to go further in the analysis of the contributions
to $\Delta E_{\rm GM}$ of each term of the Skyrme effective potential
if we approximate the neutron spin-up state $\ket n$ and proton
spin-up state $\ket p$ of the corresponding blocked levels (see their
definition in the appendix) by the dominant axially-deformed
harmonic-oscillator basis state characterized by the Nilsson quantum
numbers $N_q$, $n_{z,q}$, $\Lambda_q$ and $\Sigma_q$.
Under this approximation we establish in the appendix that
\begin{itemize}
  \item $\EGM^{(t_0)}$ and $\EGM^{(t_3)}$ are respectively
    proportional to and of the sign of $-t_0x_0
    \Sigma_n\Sigma_p$ and $-\frac{t_3}6 \Sigma_n\Sigma_p$;
  \item $\EGM^{(t_1+t_2)}$ is proportional to and of the sign of
    $(t_1+t_2) \Lambda_n\Lambda_p$;
  \item $\EGM^{(W)}$ is proportional to and of the sign of
    $W (\Lambda_n - \Lambda_p)I_{pn}$ where $I_{pn}$ is an integral
    over $\mathbb R_+$ involving products of generalized Laguerre
    polynomials and their first derivatives defined by Eq.~(\ref{Ipn}).
\end{itemize}
Because $x_0>0$ and the $t_0$ term of the Skyrme effective
potential is the dominant and attractive term, it always tends to
satisfy the Gallagher--Moszkowski rule. 
The density-dependent term being repulsive, it always tends to 
counterbalance the $\EGM^{(t_0)}$ term as already noticed
by Robledo, Bernard and Bertsch within the
Gogny--EDF~\cite{Robledo14}. 

Moreover $K_{\uparrow\uparrow} = \Lambda_n + \Lambda_p$ while
$K_{\uparrow\downarrow} = \Lambda_n - \Lambda_p$. If $\Lambda_n$ and
$\Lambda_p$ have the same sign, then $|K_{\uparrow\downarrow}| <
|K_{\uparrow\uparrow}|$ and $\EGM^{(t_1+t_2)} > 0$. In other words,
the sign of $\EGM^{(t_1+t_2)}$ is the sign of $\Sigma_{k_n}
\Sigma_{k_p}$, where $k_n$ and $k_p$ are such that $\Omega_{k_n} > 0$
and $\Omega_{k_p} > 0$. This explains the above conclusion that our
calculations comply with the Gallagher--Moszkowski rule in doublets
for which the larger $|K|$ value occurs in the spin-aligned
configuration. In addition, this explains why $\EGM^{(t_1+t_2)}$ is
very small in absolute value (a few keV) for doublets involving
blocked states with $\Omega =  1/2$ and large for doublets in which
$\Omega_n$ and $\Omega_p$ are large. The magnitude of
$\EGM^{(t_1+t_2)}$ also depends on the overlap of neutron and proton
wavefunctions. Because of the large $\Lambda_q$ values involved and
because neutrons and protons spatial wavefunctions are the same in the
$({7}/2^-[514]\downarrow)_n ({9}/2^-[514]\uparrow)_p$ configuration,
the energy splitting in the $(1,8)^+$ doublet is thus negative and
especially large in absolute value. It amounts to $\EGM^{(t_1+t_2)}
\approx -240$~keV in $^{176}$Lu and $^{178,180}$Ta nuclei. 

Finally, as shown in the appendix, the spin-orbit contribution
$\EGM^{(W)}$ is difficult to analyze because it strongly depends on
the nodal and azimuthal structure of the blocked neutron and proton
wavefunctions. Its sign cannot be simply related to the sign of
$\Lambda_n - \Lambda_p$ (see Eq.~(\ref{EGM_W_HO}), so we
do not comment on it further. \\

%-------------------------------------------------------------------------
%
%
%                       Conclusions and perspectives
%
%
%-------------------------------------------------------------------------

\section{Conclusions and perspectives}

Within the Skyrme energy-density functional approach, including BCS
pairing correlations with selfconsistent blocking, we calculate the
total energy of 70 Gallagher–Moszkowski doublets of bandheads in
rare-earth nuclei around mass numbers $A\approx 156$ and 176 as well as
in actinide nuclei around mass numbers $A\approx 230$, 240 and 250. Each
bandhead state is described as a one-neutron, one-proton blocked
configuration, and axial and intrinsic parity symmetries are
assumed. We use the SIII Skyrme parameterization in the particle-hole
channel and the seniority pairing interaction. The strengths of the
latter interaction for neutrons and for protons are adjusted using the
procedure described in Ref.~\cite{Nor19} and successfully applied in
two-quasiparticle $K$ isomers of actinides in Ref.\cite{Minkov22}.

First we analyze the bandhead spectra of 10 rare-earth and 7 actinide 
odd-odd nuclei by comparison with experimental data. To determine the
level scheme of these nuclei from the total energies calculated for
each bandhead state, we calculate the excitation energy of a given
bandhead state with respect to the experimental lowest-lying bandhead
state. All calculated doublets but two are obtained with a
two-quasiparticle configuration identical with the one proposed in the
litterature to interpret the experimental data. The two exceptions are
the $(2^+,1^+)$ doublet of $^{158}$Tb with a calculated configuration
$(1/2^+[400])_n$ $(3/2^+[411])_p$ instead of $(1/2^+[620])_n$
$(3/2^+[411])_p$, and the $(2^+,1^+)$ doublet of $^{230}$Pa with a
calculated configuration $(3/2^-[741])_n$ $(1/2^-[530])_p$ instead of
$(3/2^-[501])_n$ $(1/2^-[530])_p$. Moreover the typical discrepancy
between the calculated and experimental excitation energies is of the
order of 100 to 200~keV, which shows the overall relevance of the
calculated single-particle spectra. However there are a few exceptions
with large discrepancies which are interpreted by misplaced
single-particle states. These occur in particular, but not only,
around an exceeding shell gap, such as the $N=142$ and $N=152$ gaps in
actinides. Also an inversion of $5/2^+$ and $5/2^-$ proton states is
observed across the $Z=94$ gap.

Then we focus on the energy difference $\EGM$ between the two members
of a Gallagher--Moszkowski doublet and address the mechanism in the
Skyrme-EDF framework by which the Gallagher--Moszkowski rule
operates. To do so, we first establish the relevance of a study based
on calculations with perturbative blocking as compared to
selfconsistent blocking and then analyze the contributions to $\EGM$
from the time-odd part of the Skyrme SIII energy-density functional,
namely the central zero-range $\Delta E_{t_0}$ term, the central
gradient $\EGM^{(t_1+t_2)}$ term, the central zero-range
density-dependent $\EGM^{(t_3)}$ term and the zero-range spin-orbit
$\EGM^{(W)}$ term. Given the signs of the involved SIII parameters, we
show that the Gallagher--Moszkowski rule is always satisfied in
doublets for which the spin-aligned configuration corresponds to the
larger $K$ value. This is so because the $\EGM^{(t_1+t_2)}$
contribution is positive in that case and adds up to the zero-range
contributions, and is not counter-balanced by the too small
spin-orbit contribution. More precisely, we find that $\EGM^{(t_0)}$
and $\EGM^{(t_3)}$ are the largest two contributions in absolute value
and are approximately related by $\EGM^{(t_0)} \approx 2\, \EGM^{(t_3)}$,
so that the sum of these two contributions is always positive, of the
order of $\EGM^{(t_0)} + \EGM^{(t_3)} \approx 50$ to 150~keV. They can be
counterbalanced, and sometimes exceed in absolute value, by the
$\EGM^{(t_1+t_2)}$ contribution only when the spin-aligned member of a 
doublet corresponds to the lower of the two $K$ values. This mechanism
can be anticipated from the dominant Nilsson quantum numbers of the
two-quasiparticle configuration of the doublet, when the involved
single-particle wave functions are not too much fragmented in the
cylindrical harmonic-oscillator basis. If
$\Omega_q[N_qn_{z,q}\Lambda_q]$ denote the Nilsson quantum numbers of
the dominant contribution to the charge state $q$, with $\Omega_q >0$
by definition, then cases likely to break the Gallagher--Moszkowski
rule are such that (i) $\Sigma_n\Sigma_p <0$, with
$\Sigma_q=\Omega_q-\Lambda_q$ is the spin projection on the symmetry
axis, (ii) $\Lambda_n\Lambda_p$ is large (this product is always
positive or vanishing by definition of $\Omega_q$).

While the above conclusions are derived from the Skyrme SIII
parametrization, a similar study using other parametrizations could
be attempted to highlight the interplay between the various terms
entering the Skyrme EDF. 
Indeed, in the above perturbative approach, the contributions to
$\EGM$ arise from the time-odd terms of the Skyrme EDF and are
proportional to combinations of the Skyrme parameters for fixed local
densities. Moreover we recall that the $\vect s \cdot \vect T$ terms,
involving the spin-kinetic $\vect T$ local density in the notation of
Ref.~\cite{Hellemans12}, the $\sum\limits_{\mu,\nu}J_{\mu\nu}^2$
terms, involving the spin-current $J_{\mu\nu}$ local density, and the
$\vect s \cdot \Delta \vect s$ terms are neglected in our so-called
``minimal'' scheme (as explained in Refs.~\cite{Bonneau15,Koh17}).
Therefore different Skyrme parametrizations might yield 
different competition mechanisms between the time-odd terms. A
comparative study of various Skyrme energy-density functionals would
thus allow to potentially reveal general features of the
Gallagher--Moszkowski energy splitting with the Skyrme-EDF
approach. It is worth mentioning that a similar detailed work, pushing
further the analysis made by Robledo, Bernard and Bertsch~\cite{Robledo14}, is
in principle possible for Gogny-type energy-density functionals if the Gogny
effective potential is broken down into partial-wave terms. Indeed in
such an expansion, the $S$-wave terms would include the central
zero-range $t_0$ and $t_3$ terms of the Skyrme effective potential,
whereas the $P$-wave terms would include the central gradient
$t_1+t_2$ SIII term and the zero-range spin-orbit term. It would then
be interesting to compare with the corresponding Skyrme terms and
to quantify the contribution to $\EGM$ of $D$-waveterms and beyond.

In the framework of mean-field type of approaches, the present study
has two main limitations: the nuclear shapes are restricted by
intrinsic axial and parity symmetries on the one hand, particle-number
nonconserving treatment of pairing correlations on the other
hand. Indeed some nuclei fall in mass regions of observed octupole
deformation, essentially the $A\approx 230$ nuclei considered here. The
case of $^{154}$Eu is unclear as this nucleus does exhibit parity
doublet bands but static octupole deformation is not supported
according to Ref.~\cite{Afanasjev95}. However octupole vibrations
should play a role in the low-energy structure of this odd-odd
nucleus~\cite{Jain98_RMP}. In contrast, no triaxial static deformation
is expected in the considered sample of nuclei. Therefore a
natural extension of the present study is to repeat the above analysis
with axial and reflection-asymmetric Skyrme-EDF solutions. Regarding
the treatment of pairing correlations, it is known that
particle-number non-conserving Bogoliubov approach, including its
presently used BCS approximation, fail in weak pairing regimes.
This happens in particular in BCS calculations with the blocking
procedure because the blocked energy levels are excluded from pairing
correlations. This calls for a particle-number conserving description
of multi-quasiparticle states, such as the particle-number
projection~\cite{Dietrich64,Egido82,Heenen93,Bender08,Bender09,Simenel10,Rodriguez11},
an equation-of-motion method for the pairing-density matrix operators
in Heisenberg representation~\cite{Jia13}, or a
multi-particle-multi-hole configuration mixing~\cite{Zeng83,Pillet02,Pillet08}.

More generally the coupling of single-particle and collective motions
is not taken into account in our approach and contributes in the
structure of low-lying states of odd-odd nuclei. That is especially
true for $K=0$ states. They deserve to be addressed to assess their
potential role in the Gallagher--Moszkowski splitting mechanism. This
could be done in a fully microscopic way within the generator
coordinate method based on angular-momentum and parity projected
Hartree--Fock--Bogoliubov intrinsic states using the gaussian overlap
approximation, but this is challenging from the computational point of
view. A~simpler approach would be a semi-microscopic framework relying
on the Bohr--Mottelson unified model in which the intrinsic wave 
function would be, for example, of the Hartree--Fock--Bogolyubov type
(at the price of some redundancy in the collective variables). This
framework would allow to incorporate at once the core polarization in
the intrinsic wave functions and the particle-rotation,
particle-vibration, rotation-vibration couplings, hence restoring
intrinsic symmetry breaking in a less demanding way than in the fully
microscopic approach.

Finally, despite the above mentioned limitations, the present work,
through the evidenced mechanism of section~IV, may serve to put
constraints on the parameters entering the time-odd terms of the
Skyrme energy-density functional. Moreover the study of excitation
energies of nuclear states of multi-quasiparticle nature, as done in
this paper, may provide information on the ordering of the
single-particle levels which is important to identify the location of
deformed magic numbers. This is especially relevant in the rare-earth
region to locate waiting points in the nucleosynthesis r-process.

\begin{acknowledgments}

We thank Prof. Philippe Quentin for his reading of the manuscript.
This work is partly supported by CNRS and the Bulgarian National
Science Fund (BNSF) under Contract No. KP-06-N48/1. M.H.K would like
to acknowledge the French Embassy in Malaysia for the financial
assistance under the Mobility Programme to Support French-Malaysian
Cooperation in Research and Higher Education 2023 to facilitate a
research visit at both LP2iB and IHPC.

\end{acknowledgments}

\appendix

\setcounter{equation}{0}

\renewcommand{\theequation}{A.\arabic{equation}}
\renewcommand{\thesubsection}{A\arabic{subsection}} 
%-------------------------------------------------------------------------
%
%
%      Appendix: Derivation of the Gallagher--Moszkowski energy
%                  splitting in Skyrme-EDF
%
%
%-------------------------------------------------------------------------

\section*{Appendix: Derivation of the Gallagher--Moszkowski energy splitting in Skyrme-EDF}

Let us denote by $\{\ket{\psi_k^{(q)}},\ket{\psi_{\overline
    k}^{(q)}},k=1,2...\}$ the single-particle (canonical) basis for
the charge state $q =n$ (neutrons) or $p$ (protons) in the
Skyrme--Hartree--Fock--BCS ground state solution of 
the underlying even-even nucleus ($N-1$, $Z-1$) nucleus. The state
$\ket{\psi_{\overline k}^{(q)}}$ is the time-reversed partner of
$\ket{\psi_k^{(q)}}$. By convention, the quantum number $\Omega_k$ is
the positive eigenvalue in $\hbar$ unit of $\widehat J_z$ (projection
on the $z$ axis of the total angular momentum) for the eigenstate
$\ket{\psi_k^{(q)}}$, hence 
\eq{
  \widehat J_z\ket{\psi_k^{(q)}} = \Omega_k \, \ket{\psi_k^{(q)}}
  \quad \mbox{and} \quad \widehat J_z\ket{\psi_{\overline k}^{(q)}} =
  -\Omega_k \, \ket{\psi_{\overline k}^{(q)}}\,.
}
Owing to time-reversal symmetry, the local
spin density $\vect s_q^{(\mathrm{e-e})}(\vect r)$ in the ground state of this
even-even nucleus vanishes
\begin{align}
\vect s_q^{(\mathrm{e-e})}(\vect r) = \sum_{k>0}v_k^2
\Big( & [\psi_k^{(q)}]^{\dagger}(\vect r) \, 
\bm{\sigma}\, [\psi_k^{(q)}](\vect r) \nonumber \\ 
& + [\psi_{\overline k}^{(q)}]^{\dagger}(\vect r)\, \bm{\sigma}\, 
   [\psi_{\overline k}^{(q)}](\vect r)\Big) = 0
\end{align}
where $[\psi_k](\vect r)$ is the spinor at the space point of
coordinate vector $\vect r$ associated with the state
$\ket{\psi_k}$, $v_k^2$ are the BCS occupation probabilities such that
$\sum\limits_{k>0}v_k^2 = N-1$ or $Z-1$ according to the charge index
$q$, and $\bm{\sigma} = (\sigma_x,\sigma_y,\sigma_z)$ is the triplet
of Pauli matrices.

In the odd-odd nucleus ($N$, $Z$) described by perturbative blocking,
the local spin density $\vect s_q^{(\mathrm{o-o})}(\vect r)$ takes the
simple form 
\eq{
  \vect s_q^{(\mathrm{o-o})}(\vect r) = \vect
  s_q^{(\mathrm{e-e})}(\vect r) + \vect s_q^{(i_q)}(\vect r) = \vect
  s_q^{(i_q)}(\vect r) 
}
where the index $i_q$ of the blocked state denotes either an index
$k_q$, hence $\Omega_{i_q} > 0$, or the time-reversed partner
$\overline{k_q}$ so that $\Omega_{i_q} < 0$. The individual local spin
density $\vect s_q^{(i_q)}(\vect r)$ is the contribution from the
unpaired, blocked state $\ket{\psi_{i_q}^{(q)}}$
\eq{
  \vect s_q^{(i_q)}(\vect r) = [\psi_{i_q}^{(q)}]^{\dagger}(\vect r)
\, \bm{\sigma}\, [\psi_{i_q}^{(q)}](\vect r) \,.
}
Because of the time-odd character of the spin density, we have
\eq{
\vect s_q^{(\overline{k_q})}(\vect r) = -\vect s_q^{(k_q)}(\vect r) \,.
}
It is worth recalling that by definition of the local spin density,
the integral over space of its component along the $z$ axis (nucleus'
symmetry axis) is twice the expectation value of the projection
$\widehat S_z$ of the spin angular momentum (in $\hbar$ unit)
\eq{
  \int d^3 \vect r \: s_{z,q}^{(i_q)}(\vect r) =
  2\,\elmx{\psi_{i_q}^{(q)}}{\widehat S_z}{\psi_{i_q}^{(q)}} \,.
}

Let us call $\ket q$ the state among the
Kramers degenerate pair of states $\ket{\psi_{k_q}^{(q)}}$ and
$\ket{\psi_{\overline{k_q}}^{(q)}}$ such that $\elmx{q}{\widehat
  S_z}{q} > 0$. The actual blocked state of charge $q$ is thus either
$\ket q$ or its time-reversed partner $\ket{\overline q}$.
Then we call $\vect s_q$ the local spin density of the state $\ket
q$. Regardless of the $\Omega_n$ and $\Omega_p$ quantum numbers, we
thus have by construction $\elmx{n}{\widehat S_z}{n} \times
\elmx{p}{\widehat S_z}{p} > 0$. According to Ref.~\cite{Bonneau15},
this leads to
$\displaystyle \int d^3\vect r \; \vect s_n \cdot \vect s_p > 0$.
In the Gallagher--Moszkowski doublet built from the 
neutron $\ket{\psi_{i_n}^{(n)}}$ and proton $\ket{\psi_{i_p}^{(p)}}$
blocked states, the spin-aligned configuration has a total local spin
density $\vect s_{\uparrow \uparrow} = \vect s_n + \vect s_p$ or
$\vect s_{\downarrow \downarrow} = -\vect s_{\uparrow \uparrow}$, while 
the spin-anti-aligned configuration has a total local spin density
$\vect s_{\uparrow \downarrow} = \vect s_n - \vect s_p$ or $\vect
s_{\downarrow \uparrow} = -\vect s_{\uparrow \downarrow}$. As will be clear
below, both cases of spin alignment and spin anti-alignment lead to
the same expression of $\Delta E_{\rm GM}$ because of its bilinear
character. We thus choose the combinations $\vect s_{\uparrow
  \uparrow}$ and~$\vect s_{\uparrow \downarrow}$. Similar 
expressions can be derived for the current density, defined for a
single-particle state $\ket{\psi_{i_q}^{(q)}}$ by
\eq{
  \vect j_q^{(i_q)}(\vect r) = 
  \mathrm{Im}\Big(
         [\psi_{i_q}]^{\dagger}(\vect r) {\bm \nabla}
         [\psi_{i_q}](\vect r)\Big) \,.
}
We have $\vect j_{\uparrow \uparrow} = \vect j_n + \vect j_p$ in
the spin-aligned configuration, and $\vect j_{\uparrow \downarrow} =
\vect j_n - \vect j_p$. For the Skyrme EDF in the minimal scheme (see
section II) we thus obtain the following expression of $\Delta E_{G\rm M}$
in such a perturbative-blocking approach
\begin{align}
  \Delta E_{\rm GM} = & \int d^3 \vect r \: \Big[ 
    (B_{10}+B_{12} \, \rho^{\alpha}) \,\big(\vect s_{\uparrow
      \downarrow}^2 - \vect s_{\uparrow \uparrow}^2\big) \nonumber \\
    & \phantom{\int d^3 \vect r \: \Big[ }
      - B_3\, \big(\vect j_{\uparrow \downarrow}^2 -
    \vect j_{\uparrow \uparrow}^2\big)\big] \nonumber \\
  & \phantom{\int d^3 \vect r \: \Big[ }
  + B_9 \, \big(\vect s_{\uparrow \downarrow} \cdot
  \bm{\nabla}\times\vect j_{\uparrow \downarrow}
  -\vect s_{\uparrow \uparrow} \cdot
  \bm{\nabla}\times\vect j_{\uparrow \uparrow}\big) \Big]
\end{align}
After substitution of the spin-aligned and spin-anti-aligned local
spin and current densities the energy splitting $\Delta E_{\rm GM}$
becomes 
\begin{align}
  \Delta E_{\rm GM} = & -4 \, \int d^3 \vect r \;
  \big[(B_{10} + B_{12}\rho^{\alpha}) \,
    \vect s_n \cdot \vect s_p
    - B_{3} \, \vect j_n \cdot \vect j_p \big] \nonumber \\
  & -2\,B_9 \int d^3 \vect r \, \big(\vect s_n \cdot
  \bm{\nabla}\times\vect j_p + \vect s_p \cdot
  \bm{\nabla}\times\vect j_n\big) \,.
\end{align}
The energy-density functional coupling constants $B_3$, $B_9$,
$B_{10}$ and $B_{12}$ are related to the Skyrme SIII parameters by
(see Table~IX of Ref.~\cite{Koh17} or Ref.~\cite{Hellemans12}, with
$x_1=x_2=0$, $x_3=1$)
\eq{
\begin{aligned}
  B_{3} & =  \frac 1 4 \, (t_1+t_2)  & B_{9} & =  -\frac W 2 \\
  B_{10} & =  \frac 1 4 \, t_0x_0  & B_{12} & =  \frac 1 {24} \, t_3
  \,. 
\end{aligned}
}
This leads to
\eq{
  \Delta E_{\rm GM} = \EGM^{(t_0)} + \EGM^{(t_3)} + \EGM^{(t_1+t_2)} +
  \EGM^{(W)}
}
where
\begin{subequations}
\begin{align}
  \EGM^{(t_0)} & =  -t_0\, x_0
  \int d^3 \vect r \; \vect s_n \cdot \vect s_p \\
  \EGM^{(t_3)} & =  -\frac{t_3}6 \int d^3 \vect r \; \rho^{\alpha}
  \, \vect s_n \cdot \vect s_p \\
  \EGM^{(t_1+t_2)} & = (t_1+t_2)
    \int d^3 \vect r \; \vect j_n \cdot \vect j_p \\
  \Delta E_{W} & =  W \int d^3 \vect r \, \big(\vect s_n \cdot
      \bm{\nabla}\times\vect j_p + \vect s_p \cdot
      \bm{\nabla}\times\vect j_n\big) \,.
  \end{align}
\end{subequations}

It is possible to go further in the analysis of the contributions
to $\Delta E_{\rm GM}$ of each term of the Skyrme effective potential
if we approximate the neutron spin-up state $\ket n$ and proton
spin-up state $\ket p$ of the corresponding blocked levels (see their
definition above) by the dominant axially-deformed
harmonic-oscillator basis state
\eq{
\label{psi_HO}
\ket q = \ket{\psi_{i_q}^{(q)}} \approx C_{i_q}^{(q)} \, \ket{N_qn_{z,q}
  \Lambda_q\Sigma_q} \quad \mbox{($q=n$ or $p$)}\,,
}
where $C_{i_q}^{(q)}$ is the expansion coefficient and by definition
$\Sigma_q= 12$. It is worth recalling that either $i_q = k_q$, in
which case $\Omega_{i_q} = \Lambda_q + \Sigma_q > 0$, or $i_q =
\overline{k_q}$, in which case $\Omega_{i_q} = \Lambda_q + \Sigma_q <
0$. The Nilsson quantum numbers $N_q$, $n_{z,q}$, $\Lambda_q$ and
$\Sigma_q$ of the dominant harmonic-oscillator basis state in the
actual neutron and proton blocked states, together with the weights
$|C_{i_q}^{(q)}|^2$, are given in Table~\ref{tab_GM_splitting}.
The wavefunction of the spatial part $\ket{Nn_z\Lambda}$ of the
harmonic-oscillator basis state in cylindrical coordinates
$(\rho,\varphi,z)$ is given by 
\eq{
  \Psi_{Nn_z\Lambda}(\rho,\varphi,z) = \mathcal N\,
  \widetilde H_{n_z}(\xi) \, \widetilde L_{n_r}^{(|\Lambda|)}(\eta) \,
  e^{i\Lambda\varphi} \,, 
}
where $\xi = \beta_zz$, $\eta = (\beta_{\bot}\rho)^2$, $2n_r+|\Lambda| =
N-n_z$, $\widetilde H_n(\xi) = e^{-\xi^2\beta_z^2/2} \, H_n(\xi)$,
$\widetilde L_{n_r}^{(|\Lambda|)}(\eta) =
e^{-\eta/2}\eta^{|\Lambda|/2} L_{n_r}^{(|\Lambda|)}(\eta)$,
and $\mathcal N$ is a real normalization
coefficient. Because $\elmx{\Sigma}{\sigma_z}{\Sigma} = \Sigma$ and
$\elmx{\Sigma}{\sigma_x}{\Sigma} = \elmx{\Sigma}{\sigma_y}{\Sigma} =
0$, we can thus deduce the following approximate expression for
the local spin density associated with the
single-particle state $\ket{q}$ of Eq.~(\ref{psi_HO}) 
\eq{
  \vect s_{q}^{(i_q)}(\vect r) = |C_{i_q}^{(q)}|^2
  \big(\mathcal N_q\widetilde H_{n_{z,q}}(\xi)
  \widetilde L_{n_{r,q}}^{(|\Lambda|)}(\eta)\big)^2
  \, \Sigma_q \, \vect e_z \,,
}
where $\vect e_z$ is the unit vector of the symmetry axis. The
gradient of the harmonic-oscillator wavefunction 
$\Psi_{Nn_z\Lambda}(\vect r)$, expressed in the orthonormal
cylindrical basis $\{\vect e_{\rho},\vect e_{\varphi},\vect e_z\}$,
is given by
\begin{align}
  {\bm \nabla} \Psi_{Nn_z\Lambda}(\vect r) = \,
  \mathcal N \,e^{i\Lambda\varphi} \Big[ &
    \widetilde H_{n_z}(\xi) \,
    \frac{\partial \widetilde L_{n_r}^{(|\Lambda|)}}{\partial
      \rho} \, \vect{e}_{\rho} \nonumber \\
    & + \frac{i\Lambda}{\rho} \, \widetilde H_{n_z}(\xi) \,
    \widetilde L_{n_r}^{(|\Lambda|)}(\eta) \,\vect{e}_{\varphi} \nonumber \\   
    & + \frac{\partial \widetilde H_{n_z}}{\partial z} \,
    \widetilde L_{n_r}^{(|\Lambda|)}(\eta)
     \, \vect{e}_z \Big] \,,
\end{align}
therefore the local current density associated with the
single-particle state $\ket q$ is orthoradial and
reads 
\begin{align}
  \vect j_{q}^{(i_q)}(\vect r) & = |C_{i_q}^{(q)}|^2 \,
  \mathcal N_q \, \Lambda_q \,F_q(\rho,z) \,
  \vect{e}_{\varphi} \,.
  \label{jq}
\end{align}
where
\eq{
  \label{Fq}
  F_q(\rho,z)= \frac{1}{\rho} \,
  \big[\widetilde H_{n_{z,q}}(\xi) \,
    \widetilde L_{n_{r,q}}^{(|\Lambda_q|)}(\eta)\big]^2 \,.
}
As above mentionned, the single-particle $\ket q$ is the one
among the Kramers degenerate pair of states $\ket{\psi_{k_q}^{(q)}}$
and $\ket{\psi_{\overline{k_q}}^{(q)}}$ such that $\elmx{q}{\widehat
  S_z}{q} > 0$, therefore we expect that $\Sigma_q > 0$.
We can deduce that, in the approximation of Eq.~(\ref{psi_HO}),
$\EGM^{(t_0)}$ and $\EGM^{(t_3)}$ are proportional to $\Sigma_n
\Sigma_p$ (expected to be positive)
\begin{align}
  \EGM^{(t_0)} \approx & -t_0x_0 \, |C_{i_n}^{(n)}|^2 \,
  |C_{i_p}^{(p)}|^2 \, \Sigma_n\,\Sigma_p \,
  \frac{\pi\,(\mathcal N_n\mathcal N_p)^2}{\beta_z\beta_{\bot}^2}
  \times \nonumber \\ 
  & \int_{-\infty}^{+\infty} d\xi \,\big[\widetilde H_{n_{z,n}}(\xi) \,
    \widetilde H_{n_{z,p}}(\xi)\big]^2 \, \times \nonumber \\
  & \int_0^{+\infty} d\eta \: \big[\widetilde L_{n_{r,n}}^{(|\Lambda_n|)}(\eta)
    \widetilde L_{n_{r,p}}^{(|\Lambda_p|)}(\eta)\big]^2 
\end{align}
and
\begin{align}
  \EGM^{(t_3)} \approx & -\frac{t_3}6 \, |C_{i_n}^{(n)}|^2 \,
  |C_{i_p}^{(p)}|^2 \, \Sigma_n\,\Sigma_p \,
  \frac{\pi\,(\mathcal N_n\mathcal
    N_p)^2}{\beta_z\beta_{\bot}^2}\times \nonumber \\ 
  & \int_{-\infty}^{+\infty} d\xi \,\int_0^{+\infty} d\eta \:
  \big[\widetilde H_{n_{z,n}}(\xi) \,
    \widetilde H_{n_{z,p}}(\xi)\big]^2 \, \times \nonumber \\
  & \phantom{\int_{-\infty}^{+\infty} d\xi \,\int_0^{+\infty} d\eta \:}
  \big[\widetilde L_{n_{r,n}}^{(|\Lambda_n|)}(\eta)
    \widetilde L_{n_{r,p}}^{(|\Lambda_p|)}(\eta)\big]^2 \,
  \rho^{\alpha}(\vect r) \,.
\end{align}
whereas $\EGM^{(t_1+t_2)}$ is proportional to $\Lambda_n\Lambda_p$
\begin{align}
  \EGM^{(t_1+t_2)} \approx & (t_1+t_2) \, |C_{i_n}^{(n)}|^2 \,
  |C_{i_p}^{(p)}|^2 \, \Lambda_n\,\Lambda_p \,
  \frac{(\mathcal N_n\mathcal N_p)^2}{2\beta_z\beta_{\bot}^2}
  \times \nonumber \\ 
  & \int_{-\infty}^{+\infty} d\xi \,\big[\widetilde H_{n_{z,n}}(\xi) \,
    \widetilde H_{n_{z,p}}(\xi)\big]^2 \, \times \nonumber \\
  & \int_0^{+\infty} \frac{d\eta}{\eta} \:
  \big[\widetilde L_{n_{r,n}}^{(|\Lambda_n|)}(\eta)
    \widetilde L_{n_{r,p}}^{(|\Lambda_p|)}(\eta)\big]^2 \,.
\end{align}

To complete this appendix we also derive the approximate expression
of the spin-orbit contribution $\EGM^{(W)}$. To do so we need to calculate the
curl of the current density (\ref{jq}) in cylindrical coordinates
\begin{align}
{\bm \nabla}\times \vect j_{q}^{(i_q)}(\vect r) = & |C_{i_q}^{(q)}|^2 \,
\mathcal N_q^2 \, \Lambda_q \, \times \nonumber \\
& \Big[\frac 1{\rho} \,
  \frac{\partial}{\partial \rho}\big(\rho \, F_q\big) \, \vect e_z
  - \frac{\partial F_q}{\partial z}\, \vect e_{\rho}\Big] \,,
\end{align}
where $F_q$ was defined by Eq.~(\ref{Fq}). The dot product of
${\bm \nabla}\times j_{i_q}^{(q)}$ with the spin density thus involves
only the $z$ component, and we obtain
\begin{align}
\EGM^{(W)} \approx & W \, |C_{i_n}^{(n)}|^2 \, |C_{i_p}^{(p)}|^2 \,
\frac{2\pi\,(\mathcal N_n\mathcal N_p)^2}{\beta_z} \,
\times \nonumber \\
& \int_{-\infty}^{+\infty} d\xi \,
\big[\widetilde H_{n_{z,n}}(\xi)
  \widetilde H_{n_{z,p}}(\xi) \big]^2 \times \nonumber \\ 
& \big(\Sigma_n\Lambda_p \, I_{np} 
+ \Sigma_p\Lambda_n \, I_{pn}\big)\,,
\end{align}
where
\begin{subequations}
  \begin{align}
    \label{Inp}
I_{np} & = \int_0^{+\infty} d\eta \, \big[
  \widetilde L_{n_{r,n}}^{(|\Lambda_n|)}(\eta) \big]^2\,
  \frac{\partial}{\partial \eta}\big[
    \widetilde L_{n_{r,p}}^{(|\Lambda_p|)}(\eta) \big]^2 \\
    \label{Ipn}
I_{pn} & = \int_0^{+\infty} d\eta \, \big[
    \widetilde L_{n_{r,p}}^{(|\Lambda_p|)}(\eta) \big]^2\,
  \frac{\partial}{\partial \eta}\big[
  \widetilde L_{n_{r,n}}^{(|\Lambda_n|)}(\eta) \big]^2 \,.
\end{align}
\end{subequations}
Using integration by parts we can relate $I_{pn}$ and $I_{np}$ as
follows
\eq{
I_{pn} + I_{np} = -\big[
  \widetilde L_{n_{r,n}}^{(|\Lambda_n|)}(0) \,
  \widetilde L_{n_{r,p}}^{(|\Lambda_p|)}(0) \big]^2 \,
}
and then bring $\EGM^{(W)}$ to the form
\begin{align}
  \EGM^{(W)} \approx & \, W \, |C_{i_n}^{(n)}|^2 \, |C_{i_p}^{(p)}|^2 \,
  \frac{2\pi\,(\mathcal N_n\mathcal N_p)^2}{\beta_z} \,
  \times \nonumber \\
& \int_{-\infty}^{+\infty} d\xi \,
\big[H_{n_{z,n}}(\xi) \, H_{n_{z,p}}(\xi) \big]^2 \times \nonumber \\ 
& \Big\{\big(\Sigma_p\Lambda_n - \Sigma_n\Lambda_p\big)
\, I_{pn}
  \nonumber \\
  & - \Sigma_n\Lambda_p \, \big[
    \widetilde L_{n_{r,n}}^{(|\Lambda_n|)}(0) \,
    \widetilde L_{n_{r,p}}^{(|\Lambda_p|)}(0) \big]^2
  \Big\}\,,
\end{align}
with $\widetilde L_{n_r}^{(|\Lambda|)}(0) = \delta_{\Lambda 0}$ hence
$\Lambda \widetilde L_{n_r}^{(|\Lambda|)}(0) =0$.
Finally, because $\Sigma_q$ is expected to be positive, we have
$\Sigma_n = \Sigma_p =  1 2$ and we can simplify $\EGM^{(W)}$ as
\begin{align}
  \EGM^{(W)} \approx & \, W \, |C_{i_n}^{(n)}|^2 \, |C_{i_p}^{(p)}|^2 \,
  \frac{\pi\,(\mathcal N_n \mathcal N_p)^2}{\beta_z} \,
  \times \nonumber \\
& \big(\Lambda_n - \Lambda_p\big) \, I_{pn}\int_{-\infty}^{+\infty} d\xi \,
  \big[H_{n_{z,n}}(\xi) \, H_{n_{z,p}}(\xi) \big]^2 \,.
  \label{EGM_W_HO}
\end{align}
Given that $W > 0$, the sign of $\EGM^{(W)}$ is the sign of
$\big(\Lambda_n - \Lambda_p\big) \, I_{pn}$ and depends on the quantum
numbers $\Lambda_n$, $\Lambda_p$ as well as $n_{r,n}$ and $n_{r,p}$
through $I_{pn}$. The integral $I_{pn}$ can be analytically
calculated but gives a very complicated function of $\Lambda_n$,
$\Lambda_p$, $n_{r,n}$ and $n_{r,p}$. Therefore the sign of
$\EGM^{(W)}$ is strongly dependent on the nodal and azimuthal
structure of the proton and neutron wavefunctions.

\bigskip\bigskip

\bibliography{mybibfile}

%apsrev4-2.bst 2019-01-14 (MD) hand-edited version of apsrev4-1.bst
%Control: key (0)
%Control: author (8) initials jnrlst
%Control: editor formatted (1) identically to author
%Control: production of article title (0) allowed
%Control: page (0) single
%Control: year (1) truncated
%Control: production of eprint (0) enabled
\begin{thebibliography}{65}%
\makeatletter
\providecommand \@ifxundefined [1]{%
 \@ifx{#1\undefined}
}%
\providecommand \@ifnum [1]{%
 \ifnum #1\expandafter \@firstoftwo
 \else \expandafter \@secondoftwo
 \fi
}%
\providecommand \@ifx [1]{%
 \ifx #1\expandafter \@firstoftwo
 \else \expandafter \@secondoftwo
 \fi
}%
\providecommand \natexlab [1]{#1}%
\providecommand \enquote  [1]{``#1''}%
\providecommand \bibnamefont  [1]{#1}%
\providecommand \bibfnamefont [1]{#1}%
\providecommand \citenamefont [1]{#1}%
\providecommand \href@noop [0]{\@secondoftwo}%
\providecommand \href [0]{\begingroup \@sanitize@url \@href}%
\providecommand \@href[1]{\@@startlink{#1}\@@href}%
\providecommand \@@href[1]{\endgroup#1\@@endlink}%
\providecommand \@sanitize@url [0]{\catcode `\\12\catcode `\$12\catcode
  `\&12\catcode `\#12\catcode `\^12\catcode `\_12\catcode `\%12\relax}%
\providecommand \@@startlink[1]{}%
\providecommand \@@endlink[0]{}%
\providecommand \url  [0]{\begingroup\@sanitize@url \@url }%
\providecommand \@url [1]{\endgroup\@href {#1}{\urlprefix }}%
\providecommand \urlprefix  [0]{URL }%
\providecommand \Eprint [0]{\href }%
\providecommand \doibase [0]{https://doi.org/}%
\providecommand \selectlanguage [0]{\@gobble}%
\providecommand \bibinfo  [0]{\@secondoftwo}%
\providecommand \bibfield  [0]{\@secondoftwo}%
\providecommand \translation [1]{[#1]}%
\providecommand \BibitemOpen [0]{}%
\providecommand \bibitemStop [0]{}%
\providecommand \bibitemNoStop [0]{.\EOS\space}%
\providecommand \EOS [0]{\spacefactor3000\relax}%
\providecommand \BibitemShut  [1]{\csname bibitem#1\endcsname}%
\let\auto@bib@innerbib\@empty
%</preamble>
\bibitem [{\citenamefont {Bohr}\ and\ \citenamefont
  {Mottelson}(1953)}]{Bohr-Mottelson53}%
  \BibitemOpen
  \bibfield  {author} {\bibinfo {author} {\bibfnamefont {{\AA}.}~\bibnamefont
  {Bohr}}\ and\ \bibinfo {author} {\bibfnamefont {B.~R.}\ \bibnamefont
  {Mottelson}},\ }\bibfield  {title} {\bibinfo {title} {{Collective and
  individual-particle aspects of nuclear structure}},\ }\href@noop {}
  {\bibfield  {journal} {\bibinfo  {journal} {Kgl. Danske Videnskab. Selskab,
  Mat.-fys. Medd.}\ }\textbf {\bibinfo {volume} {27}},\ \bibinfo {pages} {No.
  16} (\bibinfo {year} {1953})}\BibitemShut {NoStop}%
\bibitem [{\citenamefont {Peker}(1957)}]{Peker57}%
  \BibitemOpen
  \bibfield  {author} {\bibinfo {author} {\bibfnamefont {L.~K.}\ \bibnamefont
  {Peker}},\ }\href@noop {} {\bibfield  {journal} {\bibinfo  {journal} {Izvest.
  Akad. Nauk. S.S.S.R., Ser. Fiz.}\ }\textbf {\bibinfo {volume} {31}},\
  \bibinfo {pages} {1029} (\bibinfo {year} {1957})}\BibitemShut {NoStop}%
\bibitem [{\citenamefont {Gallagher}\ and\ \citenamefont
  {Moszkowski}(1958)}]{Gallagher-Moszkowski58}%
  \BibitemOpen
  \bibfield  {author} {\bibinfo {author} {\bibfnamefont {C.~J.}\ \bibnamefont
  {Gallagher}}\ and\ \bibinfo {author} {\bibfnamefont {S.~A.}\ \bibnamefont
  {Moszkowski}},\ }\bibfield  {title} {\bibinfo {title} {{Coupling of angular
  momenta in odd-odd nuclei}},\ }\href@noop {} {\bibfield  {journal} {\bibinfo
  {journal} {Phys. Rev.}\ }\textbf {\bibinfo {volume} {111}},\ \bibinfo {pages}
  {1282} (\bibinfo {year} {1958})}\BibitemShut {NoStop}%
\bibitem [{\citenamefont {Boisson}\ \emph {et~al.}(1976)\citenamefont
  {Boisson}, \citenamefont {Piepenbring},\ and\ \citenamefont
  {Ogle}}]{Boisson76}%
  \BibitemOpen
  \bibfield  {author} {\bibinfo {author} {\bibfnamefont {J.~P.}\ \bibnamefont
  {Boisson}}, \bibinfo {author} {\bibfnamefont {R.}~\bibnamefont
  {Piepenbring}},\ and\ \bibinfo {author} {\bibfnamefont {W.}~\bibnamefont
  {Ogle}},\ }\bibfield  {title} {\bibinfo {title} {{The effective
  neutron-proton interaction in rare-earth nuclei}},\ }\href@noop {} {\bibfield
   {journal} {\bibinfo  {journal} {Phys. Rep.}\ }\textbf {\bibinfo {volume}
  {26}},\ \bibinfo {pages} {99} (\bibinfo {year} {1976})}\BibitemShut {NoStop}%
\bibitem [{\citenamefont {Pyatov}(1963)}]{Pyatov63}%
  \BibitemOpen
  \bibfield  {author} {\bibinfo {author} {\bibfnamefont {N.~I.}\ \bibnamefont
  {Pyatov}},\ }\bibfield  {title} {\bibinfo {title} {{Level splitting in
  deformed even-$A$ nuclei}},\ }\href@noop {} {\bibfield  {journal} {\bibinfo
  {journal} {Bull. Acad. Sci. USSR, Phys. Ser.}\ }\textbf {\bibinfo {volume}
  {27}},\ \bibinfo {pages} {1409} (\bibinfo {year} {1963})}\BibitemShut
  {NoStop}%
\bibitem [{\citenamefont {Pinho}\ and\ \citenamefont
  {Picard}(1965{\natexlab{a}})}]{DePinho65_PLB15}%
  \BibitemOpen
  \bibfield  {author} {\bibinfo {author} {\bibfnamefont {A.~G.~D.}\
  \bibnamefont {Pinho}}\ and\ \bibinfo {author} {\bibfnamefont
  {J.}~\bibnamefont {Picard}},\ }\bibfield  {title} {\bibinfo {title}
  {{Pourquoi la règle de Gallagher--Moszkowski est-elle si bien vérifiée
  ?}},\ }\href@noop {} {\bibfield  {journal} {\bibinfo  {journal} {Phys. Lett.
  B}\ }\textbf {\bibinfo {volume} {15}},\ \bibinfo {pages} {250} (\bibinfo
  {year} {1965}{\natexlab{a}})}\BibitemShut {NoStop}%
\bibitem [{\citenamefont {Nunberg}\ and\ \citenamefont
  {Prosperi}(1965)}]{Nunberg65}%
  \BibitemOpen
  \bibfield  {author} {\bibinfo {author} {\bibfnamefont {P.}~\bibnamefont
  {Nunberg}}\ and\ \bibinfo {author} {\bibfnamefont {D.}~\bibnamefont
  {Prosperi}},\ }\bibfield  {title} {\bibinfo {title} {{Neutron-Proton Residual
  Interaction in Odd-Odd Deformed Nuclei}},\ }\href@noop {} {\bibfield
  {journal} {\bibinfo  {journal} {Nuov. Cimento}\ }\textbf {\bibinfo {volume}
  {40}},\ \bibinfo {pages} {318} (\bibinfo {year} {1965})}\BibitemShut
  {NoStop}%
\bibitem [{\citenamefont {Jain}\ \emph {et~al.}(1998)\citenamefont {Jain},
  \citenamefont {Sheline}, \citenamefont {Headly}, \citenamefont {Sood},
  \citenamefont {Burke}, \citenamefont {Hrivnacova}, \citenamefont {Kvasil},
  \citenamefont {Nosek},\ and\ \citenamefont {Hoff}}]{Jain98_RMP}%
  \BibitemOpen
  \bibfield  {author} {\bibinfo {author} {\bibfnamefont {A.~K.}\ \bibnamefont
  {Jain}}, \bibinfo {author} {\bibfnamefont {R.~K.}\ \bibnamefont {Sheline}},
  \bibinfo {author} {\bibfnamefont {D.~M.}\ \bibnamefont {Headly}}, \bibinfo
  {author} {\bibfnamefont {P.~C.}\ \bibnamefont {Sood}}, \bibinfo {author}
  {\bibfnamefont {D.~G.}\ \bibnamefont {Burke}}, \bibinfo {author}
  {\bibfnamefont {I.}~\bibnamefont {Hrivnacova}}, \bibinfo {author}
  {\bibfnamefont {J.}~\bibnamefont {Kvasil}}, \bibinfo {author} {\bibfnamefont
  {D.}~\bibnamefont {Nosek}},\ and\ \bibinfo {author} {\bibfnamefont {R.~W.}\
  \bibnamefont {Hoff}},\ }\bibfield  {title} {\bibinfo {title} {{Nuclear
  structure in odd-odd nuclei, $144 \leqslant$ {\it \uppercase{A}} $\leqslant
  194$}},\ }\href@noop {} {\bibfield  {journal} {\bibinfo  {journal} {Rev. Mod.
  Phys.}\ }\textbf {\bibinfo {volume} {70}},\ \bibinfo {pages} {843} (\bibinfo
  {year} {1998})}\BibitemShut {NoStop}%
\bibitem [{\citenamefont {Covello}\ \emph {et~al.}(1997)\citenamefont
  {Covello}, \citenamefont {Gargano},\ and\ \citenamefont {Itaco}}]{Covello97}%
  \BibitemOpen
  \bibfield  {author} {\bibinfo {author} {\bibfnamefont {A.}~\bibnamefont
  {Covello}}, \bibinfo {author} {\bibfnamefont {A.}~\bibnamefont {Gargano}},\
  and\ \bibinfo {author} {\bibfnamefont {N.}~\bibnamefont {Itaco}},\ }\bibfield
   {title} {\bibinfo {title} {{Tensor force in doubly odd deformed nuclei}},\
  }\href@noop {} {\bibfield  {journal} {\bibinfo  {journal} {Phys. Rev. C}\
  }\textbf {\bibinfo {volume} {56}},\ \bibinfo {pages} {3092} (\bibinfo {year}
  {1997})}\BibitemShut {NoStop}%
\bibitem [{\citenamefont {Robledo}\ \emph {et~al.}(2014)\citenamefont
  {Robledo}, \citenamefont {Bernard},\ and\ \citenamefont
  {Bertsch}}]{Robledo14}%
  \BibitemOpen
  \bibfield  {author} {\bibinfo {author} {\bibfnamefont {L.~M.}\ \bibnamefont
  {Robledo}}, \bibinfo {author} {\bibfnamefont {R.~N.}\ \bibnamefont
  {Bernard}},\ and\ \bibinfo {author} {\bibfnamefont {G.~F.}\ \bibnamefont
  {Bertsch}},\ }\bibfield  {title} {\bibinfo {title} {{Spin constraints on
  nuclear energy density functionals}},\ }\href@noop {} {\bibfield  {journal}
  {\bibinfo  {journal} {Phys. Rev. C}\ }\textbf {\bibinfo {volume} {89}},\
  \bibinfo {pages} {021303(R)} (\bibinfo {year} {2014})}\BibitemShut {NoStop}%
\bibitem [{\citenamefont {Ward}\ \emph {et~al.}(2019)\citenamefont {Ward},
  \citenamefont {Carlsson}, \citenamefont {Möller},\ and\ \citenamefont
  {{\AA}berg}}]{Ward19}%
  \BibitemOpen
  \bibfield  {author} {\bibinfo {author} {\bibfnamefont {D.~E.}\ \bibnamefont
  {Ward}}, \bibinfo {author} {\bibfnamefont {B.~G.}\ \bibnamefont {Carlsson}},
  \bibinfo {author} {\bibfnamefont {P.}~\bibnamefont {Möller}},\ and\ \bibinfo
  {author} {\bibfnamefont {S.}~\bibnamefont {{\AA}berg}},\ }\bibfield  {title}
  {\bibinfo {title} {{Global microscopic calculations of odd-odd nuclei}},\
  }\href@noop {} {\bibfield  {journal} {\bibinfo  {journal} {Phys. Rev. C}\
  }\textbf {\bibinfo {volume} {100}},\ \bibinfo {pages} {034301} (\bibinfo
  {year} {2019})}\BibitemShut {NoStop}%
\bibitem [{\citenamefont {Bennour}\ \emph {et~al.}(1987)\citenamefont
  {Bennour}, \citenamefont {Libert}, \citenamefont {Meyer},\ and\ \citenamefont
  {Quentin}}]{Bennour87}%
  \BibitemOpen
  \bibfield  {author} {\bibinfo {author} {\bibfnamefont {L.}~\bibnamefont
  {Bennour}}, \bibinfo {author} {\bibfnamefont {J.}~\bibnamefont {Libert}},
  \bibinfo {author} {\bibfnamefont {M.}~\bibnamefont {Meyer}},\ and\ \bibinfo
  {author} {\bibfnamefont {P.}~\bibnamefont {Quentin}},\ }\bibfield  {title}
  {\bibinfo {title} {{A self-consistent description of the spectroscopic
  properties of odd-odd nuclei}},\ }\href@noop {} {\bibfield  {journal}
  {\bibinfo  {journal} {Nucl. Phys. A}\ }\textbf {\bibinfo {volume} {465}},\
  \bibinfo {pages} {35} (\bibinfo {year} {1987})}\BibitemShut {NoStop}%
\bibitem [{\citenamefont {Minkov}\ \emph {et~al.}(2022)\citenamefont {Minkov},
  \citenamefont {Bonneau}, \citenamefont {Quentin}, \citenamefont {Bartel},
  \citenamefont {Molique},\ and\ \citenamefont {Ivanova}}]{Minkov22}%
  \BibitemOpen
  \bibfield  {author} {\bibinfo {author} {\bibfnamefont {N.}~\bibnamefont
  {Minkov}}, \bibinfo {author} {\bibfnamefont {L.}~\bibnamefont {Bonneau}},
  \bibinfo {author} {\bibfnamefont {P.}~\bibnamefont {Quentin}}, \bibinfo
  {author} {\bibfnamefont {J.}~\bibnamefont {Bartel}}, \bibinfo {author}
  {\bibfnamefont {H.}~\bibnamefont {Molique}},\ and\ \bibinfo {author}
  {\bibfnamefont {D.}~\bibnamefont {Ivanova}},\ }\bibfield  {title} {\bibinfo
  {title} {{$K$-isomeric states in well-deformed heavy even-even nuclei}},\
  }\href@noop {} {\bibfield  {journal} {\bibinfo  {journal} {Phys. Rev. C}\
  }\textbf {\bibinfo {volume} {105}},\ \bibinfo {pages} {044329} (\bibinfo
  {year} {2022})}\BibitemShut {NoStop}%
\bibitem [{\citenamefont {Maruhn}\ \emph {et~al.}(2014)\citenamefont {Maruhn},
  \citenamefont {Reinhard}, \citenamefont {Stevenson},\ and\ \citenamefont
  {Umar}}]{Maruhn14}%
  \BibitemOpen
  \bibfield  {author} {\bibinfo {author} {\bibfnamefont {J.~A.}\ \bibnamefont
  {Maruhn}}, \bibinfo {author} {\bibfnamefont {P.-G.}\ \bibnamefont
  {Reinhard}}, \bibinfo {author} {\bibfnamefont {P.~D.}\ \bibnamefont
  {Stevenson}},\ and\ \bibinfo {author} {\bibfnamefont {A.~S.}\ \bibnamefont
  {Umar}},\ }\bibfield  {title} {\bibinfo {title} {{The TDHF code Sky3D}},\
  }\href@noop {} {\bibfield  {journal} {\bibinfo  {journal} {Comput. Phys.
  Commun.}\ }\textbf {\bibinfo {volume} {185}},\ \bibinfo {pages} {2195}
  (\bibinfo {year} {2014})}\BibitemShut {NoStop}%
\bibitem [{\citenamefont {Schunck}\ \emph {et~al.}(2017)\citenamefont
  {Schunck}, \citenamefont {Dobaczewski}, \citenamefont {W.~Satu{\l}a},
  \citenamefont {Dudek}, \citenamefont {Gao}, \citenamefont {Konieczka},
  \citenamefont {Sato}, \citenamefont {Shi}, \citenamefont {Wang},\ and\
  \citenamefont {Werner}}]{Schunck17}%
  \BibitemOpen
  \bibfield  {author} {\bibinfo {author} {\bibfnamefont {N.}~\bibnamefont
  {Schunck}}, \bibinfo {author} {\bibfnamefont {J.}~\bibnamefont
  {Dobaczewski}}, \bibinfo {author} {\bibfnamefont {P.~B.}\ \bibnamefont
  {W.~Satu{\l}a}}, \bibinfo {author} {\bibfnamefont {J.}~\bibnamefont {Dudek}},
  \bibinfo {author} {\bibfnamefont {Y.}~\bibnamefont {Gao}}, \bibinfo {author}
  {\bibfnamefont {M.}~\bibnamefont {Konieczka}}, \bibinfo {author}
  {\bibfnamefont {K.}~\bibnamefont {Sato}}, \bibinfo {author} {\bibfnamefont
  {Y.}~\bibnamefont {Shi}}, \bibinfo {author} {\bibfnamefont {X.~B.}\
  \bibnamefont {Wang}},\ and\ \bibinfo {author} {\bibfnamefont {T.~R.}\
  \bibnamefont {Werner}},\ }\bibfield  {title} {\bibinfo {title} {{Solution of
  the Skyrme-Hartree-Fock-Bogolyubov equations in the Cartesian deformed
  harmonic-oscillator basis. (VIII) HFODD (v2.73y): A new version of the
  program}},\ }\href@noop {} {\bibfield  {journal} {\bibinfo  {journal}
  {Comput. Phys. Commun.}\ }\textbf {\bibinfo {volume} {216}},\ \bibinfo
  {pages} {145} (\bibinfo {year} {2017})}\BibitemShut {NoStop}%
\bibitem [{\citenamefont {Ryssens}\ and\ \citenamefont
  {Bender}(2021)}]{Ryssens21}%
  \BibitemOpen
  \bibfield  {author} {\bibinfo {author} {\bibfnamefont {W.}~\bibnamefont
  {Ryssens}}\ and\ \bibinfo {author} {\bibfnamefont {M.}~\bibnamefont
  {Bender}},\ }\bibfield  {title} {\bibinfo {title} {{Skyrme pseudopotentials
  at next-to-next-to-leading order: Construction of local densities and first
  symmetry-breaking calculations}},\ }\href@noop {} {\bibfield  {journal}
  {\bibinfo  {journal} {Phys. Rev. C}\ }\textbf {\bibinfo {volume} {104}},\
  \bibinfo {pages} {044308} (\bibinfo {year} {2021})}\BibitemShut {NoStop}%
\bibitem [{\citenamefont {Bonneau}\ \emph {et~al.}(2015)\citenamefont
  {Bonneau}, \citenamefont {Minkov}, \citenamefont {Duc}, \citenamefont
  {Quentin},\ and\ \citenamefont {Bartel}}]{Bonneau15}%
  \BibitemOpen
  \bibfield  {author} {\bibinfo {author} {\bibfnamefont {L.}~\bibnamefont
  {Bonneau}}, \bibinfo {author} {\bibfnamefont {N.}~\bibnamefont {Minkov}},
  \bibinfo {author} {\bibfnamefont {D.~D.}\ \bibnamefont {Duc}}, \bibinfo
  {author} {\bibfnamefont {P.}~\bibnamefont {Quentin}},\ and\ \bibinfo {author}
  {\bibfnamefont {J.}~\bibnamefont {Bartel}},\ }\bibfield  {title} {\bibinfo
  {title} {{Effect of core polarization on magnetic dipole moments in deformed
  odd-mass nuclei}},\ }\href@noop {} {\bibfield  {journal} {\bibinfo  {journal}
  {Phys. Rev. C}\ }\textbf {\bibinfo {volume} {91}},\ \bibinfo {pages} {054307}
  (\bibinfo {year} {2015})}\BibitemShut {NoStop}%
\bibitem [{\citenamefont {Beiner}\ \emph {et~al.}(1975)\citenamefont {Beiner},
  \citenamefont {Flocard}, \citenamefont {Giai},\ and\ \citenamefont
  {Quentin}}]{Beiner75}%
  \BibitemOpen
  \bibfield  {author} {\bibinfo {author} {\bibfnamefont {M.}~\bibnamefont
  {Beiner}}, \bibinfo {author} {\bibfnamefont {H.}~\bibnamefont {Flocard}},
  \bibinfo {author} {\bibfnamefont {N.~V.}\ \bibnamefont {Giai}},\ and\
  \bibinfo {author} {\bibfnamefont {P.}~\bibnamefont {Quentin}},\ }\bibfield
  {title} {\bibinfo {title} {{Nuclear ground-state calculations properties and
  self-consistent with the Skyrme interaction}},\ }\href@noop {} {\bibfield
  {journal} {\bibinfo  {journal} {Nucl. Phys. A}\ }\textbf {\bibinfo {volume}
  {238}},\ \bibinfo {pages} {29} (\bibinfo {year} {1975})}\BibitemShut
  {NoStop}%
\bibitem [{\citenamefont {Koh}\ \emph {et~al.}(2017)\citenamefont {Koh},
  \citenamefont {Bonneau}, \citenamefont {Quentin}, \citenamefont {Hao},\ and\
  \citenamefont {Wagiran}}]{Koh17}%
  \BibitemOpen
  \bibfield  {author} {\bibinfo {author} {\bibfnamefont {M.-H.}\ \bibnamefont
  {Koh}}, \bibinfo {author} {\bibfnamefont {L.}~\bibnamefont {Bonneau}},
  \bibinfo {author} {\bibfnamefont {P.}~\bibnamefont {Quentin}}, \bibinfo
  {author} {\bibfnamefont {T.~V.~N.}\ \bibnamefont {Hao}},\ and\ \bibinfo
  {author} {\bibfnamefont {H.}~\bibnamefont {Wagiran}},\ }\bibfield  {title}
  {\bibinfo {title} {Fission barriers of two odd-neutron actinide nuclei taking
  into account the time-reversal symmetry breaking at the mean-field level},\
  }\href@noop {} {\bibfield  {journal} {\bibinfo  {journal} {Phys. Rev. C}\
  }\textbf {\bibinfo {volume} {95}},\ \bibinfo {pages} {014315} (\bibinfo
  {year} {2017})}\BibitemShut {NoStop}%
\bibitem [{\citenamefont {Nor}\ \emph {et~al.}(2019)\citenamefont {Nor},
  \citenamefont {Rezle}, \citenamefont {Kelvin-Lee}, \citenamefont {Koh},
  \citenamefont {Bonneau},\ and\ \citenamefont {Quentin}}]{Nor19}%
  \BibitemOpen
  \bibfield  {author} {\bibinfo {author} {\bibfnamefont {N.~M.}\ \bibnamefont
  {Nor}}, \bibinfo {author} {\bibfnamefont {N.-A.}\ \bibnamefont {Rezle}},
  \bibinfo {author} {\bibfnamefont {K.-W.}\ \bibnamefont {Kelvin-Lee}},
  \bibinfo {author} {\bibfnamefont {M.-H.}\ \bibnamefont {Koh}}, \bibinfo
  {author} {\bibfnamefont {L.}~\bibnamefont {Bonneau}},\ and\ \bibinfo {author}
  {\bibfnamefont {P.}~\bibnamefont {Quentin}},\ }\bibfield  {title} {\bibinfo
  {title} {{Consistency of two different approaches to determine the strength
  of a pairing residual interaction in the rare-earth region}},\ }\href@noop {}
  {\bibfield  {journal} {\bibinfo  {journal} {Phys. Rev. C}\ }\textbf {\bibinfo
  {volume} {99}},\ \bibinfo {pages} {064306} (\bibinfo {year}
  {2019})}\BibitemShut {NoStop}%
\bibitem [{\citenamefont {Flocard}\ \emph {et~al.}(1973)\citenamefont
  {Flocard}, \citenamefont {Quentin}, \citenamefont {Kerman},\ and\
  \citenamefont {Vautherin}}]{Flocard73}%
  \BibitemOpen
  \bibfield  {author} {\bibinfo {author} {\bibfnamefont {H.}~\bibnamefont
  {Flocard}}, \bibinfo {author} {\bibfnamefont {P.}~\bibnamefont {Quentin}},
  \bibinfo {author} {\bibfnamefont {A.~K.}\ \bibnamefont {Kerman}},\ and\
  \bibinfo {author} {\bibfnamefont {D.}~\bibnamefont {Vautherin}},\ }\bibfield
  {title} {\bibinfo {title} {{Nuclear deformation energy curves with the
  constrained Hartree-Fock method}},\ }\href@noop {} {\bibfield  {journal}
  {\bibinfo  {journal} {Nucl. Phys. A}\ }\textbf {\bibinfo {volume} {203}},\
  \bibinfo {pages} {433} (\bibinfo {year} {1973})}\BibitemShut {NoStop}%
\bibitem [{\citenamefont {Afanasjev}\ and\ \citenamefont
  {Ragnarsson}(1995)}]{Afanasjev95}%
  \BibitemOpen
  \bibfield  {author} {\bibinfo {author} {\bibfnamefont {A.~V.}\ \bibnamefont
  {Afanasjev}}\ and\ \bibinfo {author} {\bibfnamefont {I.}~\bibnamefont
  {Ragnarsson}},\ }\bibfield  {title} {\bibinfo {title} {{Existence of
  intrinsic reflection-asymmetry at low spin in odd and odd-odd mass nuclei in
  the Pm/Eu region}},\ }\href@noop {} {\bibfield  {journal} {\bibinfo
  {journal} {Phys. Rev. C}\ }\textbf {\bibinfo {volume} {51}},\ \bibinfo
  {pages} {1259} (\bibinfo {year} {1995})}\BibitemShut {NoStop}%
\bibitem [{\citenamefont {Agbemava}\ \emph {et~al.}(2016)\citenamefont
  {Agbemava}, \citenamefont {Afanasjev},\ and\ \citenamefont
  {Ring}}]{Agbemava16}%
  \BibitemOpen
  \bibfield  {author} {\bibinfo {author} {\bibfnamefont {S.~E.}\ \bibnamefont
  {Agbemava}}, \bibinfo {author} {\bibfnamefont {A.~V.}\ \bibnamefont
  {Afanasjev}},\ and\ \bibinfo {author} {\bibfnamefont {P.}~\bibnamefont
  {Ring}},\ }\bibfield  {title} {\bibinfo {title} {{Octupole deformation in the
  ground state of even-even nuclei: a global analysis within the covariant
  density functional theory}},\ }\href@noop {} {\bibfield  {journal} {\bibinfo
  {journal} {Phys. Rev. C}\ }\textbf {\bibinfo {volume} {93}},\ \bibinfo
  {pages} {044304} (\bibinfo {year} {2016})}\BibitemShut {NoStop}%
\bibitem [{\citenamefont {Cao}\ \emph {et~al.}(2020)\citenamefont {Cao},
  \citenamefont {Agbemava}, \citenamefont {Afanasjev}, \citenamefont
  {Nazarewicz},\ and\ \citenamefont {Ring}}]{Cao20}%
  \BibitemOpen
  \bibfield  {author} {\bibinfo {author} {\bibfnamefont {Y.}~\bibnamefont
  {Cao}}, \bibinfo {author} {\bibfnamefont {S.~E.}\ \bibnamefont {Agbemava}},
  \bibinfo {author} {\bibfnamefont {A.~V.}\ \bibnamefont {Afanasjev}}, \bibinfo
  {author} {\bibfnamefont {W.}~\bibnamefont {Nazarewicz}},\ and\ \bibinfo
  {author} {\bibfnamefont {P.}~\bibnamefont {Ring}},\ }\bibfield  {title}
  {\bibinfo {title} {{Landscape of pear-shaped even-even nuclei}},\ }\href@noop
  {} {\bibfield  {journal} {\bibinfo  {journal} {Phys. Rev. C}\ }\textbf
  {\bibinfo {volume} {102}},\ \bibinfo {pages} {024311} (\bibinfo {year}
  {2020})}\BibitemShut {NoStop}%
\bibitem [{\citenamefont {Nica}(2017)}]{Nica17_A=158}%
  \BibitemOpen
  \bibfield  {author} {\bibinfo {author} {\bibfnamefont {N.}~\bibnamefont
  {Nica}},\ }\bibfield  {title} {\bibinfo {title} {{Nuclear Data Sheets for
  $A=158$}},\ }\href@noop {} {\bibfield  {journal} {\bibinfo  {journal} {Nucl.
  Data Sheets}\ }\textbf {\bibinfo {volume} {141}},\ \bibinfo {pages} {1}
  (\bibinfo {year} {2017})}\BibitemShut {NoStop}%
\bibitem [{\citenamefont {Balodis}\ \emph {et~al.}(1972)\citenamefont
  {Balodis}, \citenamefont {Tambergs}, \citenamefont {Alksnis}, \citenamefont
  {Prokofjev}, \citenamefont {Vonach}, \citenamefont {Vonach}, \citenamefont
  {Koch}, \citenamefont {Gruber}, \citenamefont {Maier},\ and\ \citenamefont
  {Schult}}]{Balodis72}%
  \BibitemOpen
  \bibfield  {author} {\bibinfo {author} {\bibfnamefont {M.~K.}\ \bibnamefont
  {Balodis}}, \bibinfo {author} {\bibfnamefont {J.~J.}\ \bibnamefont
  {Tambergs}}, \bibinfo {author} {\bibfnamefont {K.~J.}\ \bibnamefont
  {Alksnis}}, \bibinfo {author} {\bibfnamefont {P.~T.}\ \bibnamefont
  {Prokofjev}}, \bibinfo {author} {\bibfnamefont {W.~G.}\ \bibnamefont
  {Vonach}}, \bibinfo {author} {\bibfnamefont {H.~K.}\ \bibnamefont {Vonach}},
  \bibinfo {author} {\bibfnamefont {H.~R.}\ \bibnamefont {Koch}}, \bibinfo
  {author} {\bibfnamefont {U.}~\bibnamefont {Gruber}}, \bibinfo {author}
  {\bibfnamefont {B.~P.~K.}\ \bibnamefont {Maier}},\ and\ \bibinfo {author}
  {\bibfnamefont {O.~W.~B.}\ \bibnamefont {Schult}},\ }\bibfield  {title}
  {\bibinfo {title} {{The level scheme of $^{176}$Lu investigated by $(n,
  \gamma)$ and $(n, e)$ reactions}},\ }\href@noop {} {\bibfield  {journal}
  {\bibinfo  {journal} {Nucl. Phys. A}\ }\textbf {\bibinfo {volume} {194}},\
  \bibinfo {pages} {305} (\bibinfo {year} {1972})}\BibitemShut {NoStop}%
\bibitem [{\citenamefont {Klay}\ \emph {et~al.}(1991)\citenamefont {Klay},
  \citenamefont {Käppeler}, \citenamefont {Beer}, \citenamefont {Schatz},
  \citenamefont {Borner}, \citenamefont {Hoyler}, \citenamefont {Robinson},
  \citenamefont {Schreckenbach}, \citenamefont {Krusche}, \citenamefont
  {Mayerhofer}, \citenamefont {Hlawatsch}, \citenamefont {Lindner},
  \citenamefont {von Egidy}, \citenamefont {Antdrejtscheff},\ and\
  \citenamefont {Petkov}}]{Klay91}%
  \BibitemOpen
  \bibfield  {author} {\bibinfo {author} {\bibfnamefont {N.}~\bibnamefont
  {Klay}}, \bibinfo {author} {\bibfnamefont {F.}~\bibnamefont {Käppeler}},
  \bibinfo {author} {\bibfnamefont {H.}~\bibnamefont {Beer}}, \bibinfo {author}
  {\bibfnamefont {G.}~\bibnamefont {Schatz}}, \bibinfo {author} {\bibfnamefont
  {H.}~\bibnamefont {Borner}}, \bibinfo {author} {\bibfnamefont
  {F.}~\bibnamefont {Hoyler}}, \bibinfo {author} {\bibfnamefont {S.~J.}\
  \bibnamefont {Robinson}}, \bibinfo {author} {\bibfnamefont {K.}~\bibnamefont
  {Schreckenbach}}, \bibinfo {author} {\bibfnamefont {B.}~\bibnamefont
  {Krusche}}, \bibinfo {author} {\bibfnamefont {U.}~\bibnamefont {Mayerhofer}},
  \bibinfo {author} {\bibfnamefont {G.}~\bibnamefont {Hlawatsch}}, \bibinfo
  {author} {\bibfnamefont {H.}~\bibnamefont {Lindner}}, \bibinfo {author}
  {\bibfnamefont {T.}~\bibnamefont {von Egidy}}, \bibinfo {author}
  {\bibfnamefont {W.}~\bibnamefont {Antdrejtscheff}},\ and\ \bibinfo {author}
  {\bibfnamefont {P.}~\bibnamefont {Petkov}},\ }\bibfield  {title} {\bibinfo
  {title} {{Nuclear structure of $^{176}$Lu and its astrophysical consequences.
  I. Level scheme of $^{176}$Lu}},\ }\href@noop {} {\bibfield  {journal}
  {\bibinfo  {journal} {Phys. Rev. C}\ }\textbf {\bibinfo {volume} {44}},\
  \bibinfo {pages} {2801} (\bibinfo {year} {1991})}\BibitemShut {NoStop}%
\bibitem [{\citenamefont {Achterberg}\ \emph {et~al.}(2009)\citenamefont
  {Achterberg}, \citenamefont {Capurro},\ and\ \citenamefont
  {Marti}}]{A=178_NDS110_2009}%
  \BibitemOpen
  \bibfield  {author} {\bibinfo {author} {\bibfnamefont {E.}~\bibnamefont
  {Achterberg}}, \bibinfo {author} {\bibfnamefont {O.~A.}\ \bibnamefont
  {Capurro}},\ and\ \bibinfo {author} {\bibfnamefont {G.~V.}\ \bibnamefont
  {Marti}},\ }\bibfield  {title} {\bibinfo {title} {{Nuclear Data Sheets for
  $A=178$}},\ }\href@noop {} {\bibfield  {journal} {\bibinfo  {journal} {Nucl.
  Data Sheets}\ }\textbf {\bibinfo {volume} {110}},\ \bibinfo {pages} {1473}
  (\bibinfo {year} {2009})}\BibitemShut {NoStop}%
\bibitem [{\citenamefont {Kondev}\ \emph {et~al.}(2021)\citenamefont {Kondev},
  \citenamefont {Wang}, \citenamefont {Huang}, \citenamefont {Naimi},\ and\
  \citenamefont {Audi}}]{Nubase2020}%
  \BibitemOpen
  \bibfield  {author} {\bibinfo {author} {\bibfnamefont {F.~G.}\ \bibnamefont
  {Kondev}}, \bibinfo {author} {\bibfnamefont {M.}~\bibnamefont {Wang}},
  \bibinfo {author} {\bibfnamefont {W.~J.}\ \bibnamefont {Huang}}, \bibinfo
  {author} {\bibfnamefont {S.}~\bibnamefont {Naimi}},\ and\ \bibinfo {author}
  {\bibfnamefont {G.}~\bibnamefont {Audi}},\ }\bibfield  {title} {\bibinfo
  {title} {{The NUBASE2020 evaluation of nuclear physics properties}},\
  }\href@noop {} {\bibfield  {journal} {\bibinfo  {journal} {Chinese Phys. C}\
  }\textbf {\bibinfo {volume} {45}},\ \bibinfo {pages} {030001} (\bibinfo
  {year} {2021})}\BibitemShut {NoStop}%
\bibitem [{\citenamefont {Cocks}\ \emph {et~al.}(1999)\citenamefont {Cocks},
  \citenamefont {Hawcroft}, \citenamefont {Amzal}, \citenamefont {Butler},
  \citenamefont {Cann}, \citenamefont {Greenlees}, \citenamefont {Jones},
  \citenamefont {Asztalos}, \citenamefont {Clark}, \citenamefont {Deleplanque},
  \citenamefont {Diamond}, \citenamefont {Fallon}, \citenamefont {Lee},
  \citenamefont {Macchiavelli}, \citenamefont {MacLeod}, \citenamefont
  {Stephens}, \citenamefont {Jones}, \citenamefont {Julin}, \citenamefont
  {Broda}, \citenamefont {Fornal}, \citenamefont {Smith}, \citenamefont
  {Lauritsen}, \citenamefont {Bhattacharyya},\ and\ \citenamefont
  {Zhang}}]{Cocks99_NPA645}%
  \BibitemOpen
  \bibfield  {author} {\bibinfo {author} {\bibfnamefont {J.~F.~C.}\
  \bibnamefont {Cocks}}, \bibinfo {author} {\bibfnamefont {D.}~\bibnamefont
  {Hawcroft}}, \bibinfo {author} {\bibfnamefont {N.}~\bibnamefont {Amzal}},
  \bibinfo {author} {\bibfnamefont {P.~A.}\ \bibnamefont {Butler}}, \bibinfo
  {author} {\bibfnamefont {K.~J.}\ \bibnamefont {Cann}}, \bibinfo {author}
  {\bibfnamefont {P.~T.}\ \bibnamefont {Greenlees}}, \bibinfo {author}
  {\bibfnamefont {G.~D.}\ \bibnamefont {Jones}}, \bibinfo {author}
  {\bibfnamefont {S.}~\bibnamefont {Asztalos}}, \bibinfo {author}
  {\bibfnamefont {R.~M.}\ \bibnamefont {Clark}}, \bibinfo {author}
  {\bibfnamefont {M.~A.}\ \bibnamefont {Deleplanque}}, \bibinfo {author}
  {\bibfnamefont {R.~M.}\ \bibnamefont {Diamond}}, \bibinfo {author}
  {\bibfnamefont {P.}~\bibnamefont {Fallon}}, \bibinfo {author} {\bibfnamefont
  {I.~Y.}\ \bibnamefont {Lee}}, \bibinfo {author} {\bibfnamefont {A.~O.}\
  \bibnamefont {Macchiavelli}}, \bibinfo {author} {\bibfnamefont {R.~W.}\
  \bibnamefont {MacLeod}}, \bibinfo {author} {\bibfnamefont {P.~S.}\
  \bibnamefont {Stephens}}, \bibinfo {author} {\bibfnamefont {P.}~\bibnamefont
  {Jones}}, \bibinfo {author} {\bibfnamefont {R.}~\bibnamefont {Julin}},
  \bibinfo {author} {\bibfnamefont {R.}~\bibnamefont {Broda}}, \bibinfo
  {author} {\bibfnamefont {B.}~\bibnamefont {Fornal}}, \bibinfo {author}
  {\bibfnamefont {J.~F.}\ \bibnamefont {Smith}}, \bibinfo {author}
  {\bibfnamefont {T.}~\bibnamefont {Lauritsen}}, \bibinfo {author}
  {\bibfnamefont {P.}~\bibnamefont {Bhattacharyya}},\ and\ \bibinfo {author}
  {\bibfnamefont {C.~T.}\ \bibnamefont {Zhang}},\ }\bibfield  {title} {\bibinfo
  {title} {{Spectroscopy of Rn, Ra and Th isotopes using multi-nucleon transfer
  reactions}},\ }\href@noop {} {\bibfield  {journal} {\bibinfo  {journal}
  {Nucl. Phys. A}\ }\textbf {\bibinfo {volume} {645}},\ \bibinfo {pages} {61}
  (\bibinfo {year} {1999})}\BibitemShut {NoStop}%
\bibitem [{\citenamefont {Kotthaus}\ \emph {et~al.}(2013)\citenamefont
  {Kotthaus}, \citenamefont {Reiter}, \citenamefont {Hess}, \citenamefont
  {Kalkühler}, \citenamefont {Wendt}, \citenamefont {Wiens}, \citenamefont
  {Hertenberger}, \citenamefont {Morgan}, \citenamefont {Thirolf},
  \citenamefont {Wirth},\ and\ \citenamefont {Faestermann}}]{Kotthaus13_PRC87}%
  \BibitemOpen
  \bibfield  {author} {\bibinfo {author} {\bibfnamefont {T.}~\bibnamefont
  {Kotthaus}}, \bibinfo {author} {\bibfnamefont {P.}~\bibnamefont {Reiter}},
  \bibinfo {author} {\bibfnamefont {H.}~\bibnamefont {Hess}}, \bibinfo {author}
  {\bibfnamefont {M.}~\bibnamefont {Kalkühler}}, \bibinfo {author}
  {\bibfnamefont {A.}~\bibnamefont {Wendt}}, \bibinfo {author} {\bibfnamefont
  {A.}~\bibnamefont {Wiens}}, \bibinfo {author} {\bibfnamefont
  {R.}~\bibnamefont {Hertenberger}}, \bibinfo {author} {\bibfnamefont
  {T.}~\bibnamefont {Morgan}}, \bibinfo {author} {\bibfnamefont {P.~G.}\
  \bibnamefont {Thirolf}}, \bibinfo {author} {\bibfnamefont {H.-F.}\
  \bibnamefont {Wirth}},\ and\ \bibinfo {author} {\bibfnamefont
  {T.}~\bibnamefont {Faestermann}},\ }\bibfield  {title} {\bibinfo {title}
  {{Excited states of the odd-odd nucleus $^{230}$Pa}},\ }\href@noop {}
  {\bibfield  {journal} {\bibinfo  {journal} {Phys. Rev. C}\ }\textbf {\bibinfo
  {volume} {87}},\ \bibinfo {pages} {044322} (\bibinfo {year}
  {2013})}\BibitemShut {NoStop}%
\bibitem [{\citenamefont {Reich}(2009)}]{Eu154_NDS110_2009}%
  \BibitemOpen
  \bibfield  {author} {\bibinfo {author} {\bibfnamefont {C.~W.}\ \bibnamefont
  {Reich}},\ }\bibfield  {title} {\bibinfo {title} {{Nuclear Data Sheets for
  $A=154$}},\ }\href@noop {} {\bibfield  {journal} {\bibinfo  {journal} {Nucl.
  Data Sheets}\ }\textbf {\bibinfo {volume} {110}},\ \bibinfo {pages} {2257}
  (\bibinfo {year} {2009})}\BibitemShut {NoStop}%
\bibitem [{\citenamefont {Rotter}\ \emph {et~al.}(1984)\citenamefont {Rotter},
  \citenamefont {Heiser}, \citenamefont {Schilling}, \citenamefont
  {W.~Andrejtscheff},\ and\ \citenamefont {Balodis}}]{Rotter84_NPA417_Eu154}%
  \BibitemOpen
  \bibfield  {author} {\bibinfo {author} {\bibfnamefont {H.}~\bibnamefont
  {Rotter}}, \bibinfo {author} {\bibfnamefont {C.}~\bibnamefont {Heiser}},
  \bibinfo {author} {\bibfnamefont {K.~D.}\ \bibnamefont {Schilling}}, \bibinfo
  {author} {\bibfnamefont {L.~K.~K.}\ \bibnamefont {W.~Andrejtscheff}},\ and\
  \bibinfo {author} {\bibfnamefont {M.~K.}\ \bibnamefont {Balodis}},\
  }\bibfield  {title} {\bibinfo {title} {{Electromagnetic transition
  probabilities in the doubly odd $N=91$ nucleus $^{154}$Eu}},\ }\href@noop {}
  {\bibfield  {journal} {\bibinfo  {journal} {Nucl. Phys. A}\ }\textbf
  {\bibinfo {volume} {417}},\ \bibinfo {pages} {1} (\bibinfo {year}
  {1984})}\BibitemShut {NoStop}%
\bibitem [{\citenamefont {Balodis}\ \emph {et~al.}(1987)\citenamefont
  {Balodis}, \citenamefont {Prokofjev}, \citenamefont {Kramer}, \citenamefont
  {Simonova}, \citenamefont {Schreckenbach}, \citenamefont {Davidson},
  \citenamefont {Pinston}, \citenamefont {Hungerford}, \citenamefont {Schmidt},
  \citenamefont {Scheerer}, \citenamefont {von Egidy}, \citenamefont {van
  Assche}, \citenamefont {Spits}, \citenamefont {Casten}, \citenamefont {Kane},
  \citenamefont {Warner},\ and\ \citenamefont {Kern}}]{Balodis87_NPA472_Eu154}%
  \BibitemOpen
  \bibfield  {author} {\bibinfo {author} {\bibfnamefont {M.~K.}\ \bibnamefont
  {Balodis}}, \bibinfo {author} {\bibfnamefont {P.~T.}\ \bibnamefont
  {Prokofjev}}, \bibinfo {author} {\bibfnamefont {N.~D.}\ \bibnamefont
  {Kramer}}, \bibinfo {author} {\bibfnamefont {L.~I.}\ \bibnamefont
  {Simonova}}, \bibinfo {author} {\bibfnamefont {K.}~\bibnamefont
  {Schreckenbach}}, \bibinfo {author} {\bibfnamefont {W.~F.}\ \bibnamefont
  {Davidson}}, \bibinfo {author} {\bibfnamefont {J.~A.}\ \bibnamefont
  {Pinston}}, \bibinfo {author} {\bibfnamefont {P.}~\bibnamefont {Hungerford}},
  \bibinfo {author} {\bibfnamefont {H.~H.}\ \bibnamefont {Schmidt}}, \bibinfo
  {author} {\bibfnamefont {H.~J.}\ \bibnamefont {Scheerer}}, \bibinfo {author}
  {\bibfnamefont {T.}~\bibnamefont {von Egidy}}, \bibinfo {author}
  {\bibfnamefont {P.~H.~M.}\ \bibnamefont {van Assche}}, \bibinfo {author}
  {\bibfnamefont {A.~M.~J.}\ \bibnamefont {Spits}}, \bibinfo {author}
  {\bibfnamefont {R.~F.}\ \bibnamefont {Casten}}, \bibinfo {author}
  {\bibfnamefont {W.~R.}\ \bibnamefont {Kane}}, \bibinfo {author}
  {\bibfnamefont {D.~D.}\ \bibnamefont {Warner}},\ and\ \bibinfo {author}
  {\bibfnamefont {J.}~\bibnamefont {Kern}},\ }\bibfield  {title} {\bibinfo
  {title} {Levels in $^{154}$eu populated by $(n,\gamma)$ and $(d,p)$
  reactions},\ }\href@noop {} {\bibfield  {journal} {\bibinfo  {journal} {Nucl.
  Phys. A}\ }\textbf {\bibinfo {volume} {472}},\ \bibinfo {pages} {445}
  (\bibinfo {year} {1987})}\BibitemShut {NoStop}%
\bibitem [{\citenamefont {Reich}(2012)}]{Eu156_NDS113_2012}%
  \BibitemOpen
  \bibfield  {author} {\bibinfo {author} {\bibfnamefont {C.~W.}\ \bibnamefont
  {Reich}},\ }\bibfield  {title} {\bibinfo {title} {{Nuclear Data Sheets for
  $A=156$}},\ }\href@noop {} {\bibfield  {journal} {\bibinfo  {journal} {Nucl.
  Data Sheets}\ }\textbf {\bibinfo {volume} {113}},\ \bibinfo {pages} {2537}
  (\bibinfo {year} {2012})}\BibitemShut {NoStop}%
\bibitem [{\citenamefont {Nica}(2021)}]{Nica17_A=160}%
  \BibitemOpen
  \bibfield  {author} {\bibinfo {author} {\bibfnamefont {N.}~\bibnamefont
  {Nica}},\ }\bibfield  {title} {\bibinfo {title} {{Nuclear Data Sheets for
  $A=160$}},\ }\href@noop {} {\bibfield  {journal} {\bibinfo  {journal} {Nucl.
  Data Sheets}\ }\textbf {\bibinfo {volume} {176}},\ \bibinfo {pages} {1}
  (\bibinfo {year} {2021})}\BibitemShut {NoStop}%
\bibitem [{\citenamefont {Singh}(1995)}]{Tm172_NDS75_1995}%
  \BibitemOpen
  \bibfield  {author} {\bibinfo {author} {\bibfnamefont {B.}~\bibnamefont
  {Singh}},\ }\bibfield  {title} {\bibinfo {title} {{Nuclear Data Sheets for
  $A=172$}},\ }\href@noop {} {\bibfield  {journal} {\bibinfo  {journal} {Nucl.
  Data Sheets}\ }\textbf {\bibinfo {volume} {75}},\ \bibinfo {pages} {199}
  (\bibinfo {year} {1995})}\BibitemShut {NoStop}%
\bibitem [{\citenamefont {Hughes}\ \emph {et~al.}(2008)\citenamefont {Hughes},
  \citenamefont {Lane}, \citenamefont {Dracoulis}, \citenamefont {Kibédi},
  \citenamefont {Nieminen},\ and\ \citenamefont {Watanabe}}]{Hughes08_PRC77}%
  \BibitemOpen
  \bibfield  {author} {\bibinfo {author} {\bibfnamefont {R.~O.}\ \bibnamefont
  {Hughes}}, \bibinfo {author} {\bibfnamefont {G.~J.}\ \bibnamefont {Lane}},
  \bibinfo {author} {\bibfnamefont {G.~D.}\ \bibnamefont {Dracoulis}}, \bibinfo
  {author} {\bibfnamefont {T.}~\bibnamefont {Kibédi}}, \bibinfo {author}
  {\bibfnamefont {P.}~\bibnamefont {Nieminen}},\ and\ \bibinfo {author}
  {\bibfnamefont {H.}~\bibnamefont {Watanabe}},\ }\bibfield  {title} {\bibinfo
  {title} {{Two-quasiparticle isomer, $E1$ hindrances and residual interactions
  in $^{172}$Tm}},\ }\href@noop {} {\bibfield  {journal} {\bibinfo  {journal}
  {Phys. Rev. C}\ }\textbf {\bibinfo {volume} {77}},\ \bibinfo {pages} {044309}
  (\bibinfo {year} {2008})}\BibitemShut {NoStop}%
\bibitem [{\citenamefont {Browne}\ and\ \citenamefont
  {Junde}(1995)}]{Lu174_NDS87_1999}%
  \BibitemOpen
  \bibfield  {author} {\bibinfo {author} {\bibfnamefont {E.}~\bibnamefont
  {Browne}}\ and\ \bibinfo {author} {\bibfnamefont {H.}~\bibnamefont {Junde}},\
  }\bibfield  {title} {\bibinfo {title} {{Nuclear Data Sheets for $A=174$}},\
  }\href@noop {} {\bibfield  {journal} {\bibinfo  {journal} {Nucl. Data
  Sheets}\ }\textbf {\bibinfo {volume} {87}},\ \bibinfo {pages} {15} (\bibinfo
  {year} {1995})}\BibitemShut {NoStop}%
\bibitem [{\citenamefont {Basunia}(2006)}]{Lu176_NDS107_2006}%
  \BibitemOpen
  \bibfield  {author} {\bibinfo {author} {\bibfnamefont {M.~S.}\ \bibnamefont
  {Basunia}},\ }\bibfield  {title} {\bibinfo {title} {{Nuclear Data Sheets for
  $A=176$}},\ }\href@noop {} {\bibfield  {journal} {\bibinfo  {journal} {Nucl.
  Data Sheets}\ }\textbf {\bibinfo {volume} {107}},\ \bibinfo {pages} {791}
  (\bibinfo {year} {2006})}\BibitemShut {NoStop}%
\bibitem [{\citenamefont {Burke}\ \emph {et~al.}(1993)\citenamefont {Burke},
  \citenamefont {Sood}, \citenamefont {Garrett}, \citenamefont {Qu},
  \citenamefont {Sheline},\ and\ \citenamefont {Hoff}}]{Burke93_PRC47}%
  \BibitemOpen
  \bibfield  {author} {\bibinfo {author} {\bibfnamefont {D.~G.}\ \bibnamefont
  {Burke}}, \bibinfo {author} {\bibfnamefont {P.~C.}\ \bibnamefont {Sood}},
  \bibinfo {author} {\bibfnamefont {P.~E.}\ \bibnamefont {Garrett}}, \bibinfo
  {author} {\bibfnamefont {T.}~\bibnamefont {Qu}}, \bibinfo {author}
  {\bibfnamefont {R.~K.}\ \bibnamefont {Sheline}},\ and\ \bibinfo {author}
  {\bibfnamefont {R.~W.}\ \bibnamefont {Hoff}},\ }\bibfield  {title} {\bibinfo
  {title} {{Nuclear structure of $^{178}$Lu}},\ }\href@noop {} {\bibfield
  {journal} {\bibinfo  {journal} {Phys. Rev. C}\ }\textbf {\bibinfo {volume}
  {47}},\ \bibinfo {pages} {131} (\bibinfo {year} {1993})}\BibitemShut
  {NoStop}%
\bibitem [{\citenamefont {Kondev}\ \emph {et~al.}(1998)\citenamefont {Kondev},
  \citenamefont {Dracoulis}, \citenamefont {Byrne},\ and\ \citenamefont
  {Kib{\'e}di}}]{Kondev98_NPA632}%
  \BibitemOpen
  \bibfield  {author} {\bibinfo {author} {\bibfnamefont {F.}~\bibnamefont
  {Kondev}}, \bibinfo {author} {\bibfnamefont {G.~D.}\ \bibnamefont
  {Dracoulis}}, \bibinfo {author} {\bibfnamefont {A.~P.}\ \bibnamefont
  {Byrne}},\ and\ \bibinfo {author} {\bibfnamefont {T.}~\bibnamefont
  {Kib{\'e}di}},\ }\bibfield  {title} {\bibinfo {title} {{Intrinsic states and
  rotational bands in $^{176}$Ta and $^{178}$Ta}},\ }\href@noop {} {\bibfield
  {journal} {\bibinfo  {journal} {Nucl. Phys. A}\ }\textbf {\bibinfo {volume}
  {632}},\ \bibinfo {pages} {473} (\bibinfo {year} {1998})}\BibitemShut
  {NoStop}%
\bibitem [{\citenamefont {McCutchan}(2015)}]{Ta180_NDS126_2015}%
  \BibitemOpen
  \bibfield  {author} {\bibinfo {author} {\bibfnamefont {E.~A.}\ \bibnamefont
  {McCutchan}},\ }\bibfield  {title} {\bibinfo {title} {{Nuclear Data Sheets
  for $A=180$}},\ }\href@noop {} {\bibfield  {journal} {\bibinfo  {journal}
  {Nucl. Data Sheets}\ }\textbf {\bibinfo {volume} {126}},\ \bibinfo {pages}
  {151} (\bibinfo {year} {2015})}\BibitemShut {NoStop}%
\bibitem [{\citenamefont {Browne}\ and\ \citenamefont
  {Tuli}(2012)}]{A=230_NDS113_2012}%
  \BibitemOpen
  \bibfield  {author} {\bibinfo {author} {\bibfnamefont {E.}~\bibnamefont
  {Browne}}\ and\ \bibinfo {author} {\bibfnamefont {J.~K.}\ \bibnamefont
  {Tuli}},\ }\bibfield  {title} {\bibinfo {title} {{Nuclear Data Sheets for
  $A=230$}},\ }\href@noop {} {\bibfield  {journal} {\bibinfo  {journal} {Nucl.
  Data Sheets}\ }\textbf {\bibinfo {volume} {113}},\ \bibinfo {pages} {2113}
  (\bibinfo {year} {2012})}\BibitemShut {NoStop}%
\bibitem [{\citenamefont {Browne}\ and\ \citenamefont
  {Tuli}(2007)}]{A=234_NDS108_2007}%
  \BibitemOpen
  \bibfield  {author} {\bibinfo {author} {\bibfnamefont {E.}~\bibnamefont
  {Browne}}\ and\ \bibinfo {author} {\bibfnamefont {J.~K.}\ \bibnamefont
  {Tuli}},\ }\bibfield  {title} {\bibinfo {title} {{Nuclear Data Sheets for
  $A=234$}},\ }\href@noop {} {\bibfield  {journal} {\bibinfo  {journal} {Nucl.
  Data Sheets}\ }\textbf {\bibinfo {volume} {108}},\ \bibinfo {pages} {681}
  (\bibinfo {year} {2007})}\BibitemShut {NoStop}%
\bibitem [{\citenamefont {Godart}\ and\ \citenamefont
  {Gizon}(1973)}]{Godart73_NPA217}%
  \BibitemOpen
  \bibfield  {author} {\bibinfo {author} {\bibfnamefont {J.}~\bibnamefont
  {Godart}}\ and\ \bibinfo {author} {\bibfnamefont {A.}~\bibnamefont {Gizon}},\
  }\bibfield  {title} {\bibinfo {title} {{Niveaux de $^{234}$Pa atteints par la
  désintégration de $^{234}$Th}},\ }\href@noop {} {\bibfield  {journal}
  {\bibinfo  {journal} {Nucl. Phys. A}\ }\textbf {\bibinfo {volume} {217}},\
  \bibinfo {pages} {159} (\bibinfo {year} {1973})}\BibitemShut {NoStop}%
\bibitem [{\citenamefont {Pinho}\ and\ \citenamefont
  {Picard}(1965{\natexlab{b}})}]{DePinho65_NP65}%
  \BibitemOpen
  \bibfield  {author} {\bibinfo {author} {\bibfnamefont {A.~G.~D.}\
  \bibnamefont {Pinho}}\ and\ \bibinfo {author} {\bibfnamefont
  {J.}~\bibnamefont {Picard}},\ }\bibfield  {title} {\bibinfo {title} {{Les
  bandes de rotation $K=0^-$ et $K=1^-$ dans le $^{234}$Pa}},\ }\href@noop {}
  {\bibfield  {journal} {\bibinfo  {journal} {Nucl. Phys.}\ }\textbf {\bibinfo
  {volume} {65}},\ \bibinfo {pages} {426} (\bibinfo {year}
  {1965}{\natexlab{b}})}\BibitemShut {NoStop}%
\bibitem [{\citenamefont {Browne}\ and\ \citenamefont
  {Tuli}(2015)}]{A=238_NDS127_2015}%
  \BibitemOpen
  \bibfield  {author} {\bibinfo {author} {\bibfnamefont {E.}~\bibnamefont
  {Browne}}\ and\ \bibinfo {author} {\bibfnamefont {J.~K.}\ \bibnamefont
  {Tuli}},\ }\bibfield  {title} {\bibinfo {title} {{Nuclear Data Sheets for
  $A=238$}},\ }\href@noop {} {\bibfield  {journal} {\bibinfo  {journal} {Nucl.
  Data Sheets}\ }\textbf {\bibinfo {volume} {127}},\ \bibinfo {pages} {191}
  (\bibinfo {year} {2015})}\BibitemShut {NoStop}%
\bibitem [{\citenamefont {Browne}\ and\ \citenamefont
  {Tuli}(2008)}]{A=240_NDS109_2008}%
  \BibitemOpen
  \bibfield  {author} {\bibinfo {author} {\bibfnamefont {E.}~\bibnamefont
  {Browne}}\ and\ \bibinfo {author} {\bibfnamefont {J.~K.}\ \bibnamefont
  {Tuli}},\ }\bibfield  {title} {\bibinfo {title} {{Nuclear Data Sheets for
  $A=240$}},\ }\href@noop {} {\bibfield  {journal} {\bibinfo  {journal} {Nucl.
  Data Sheets}\ }\textbf {\bibinfo {volume} {109}},\ \bibinfo {pages} {2439}
  (\bibinfo {year} {2008})}\BibitemShut {NoStop}%
\bibitem [{\citenamefont {Martin}\ and\ \citenamefont
  {Nesaraja}(2022)}]{A=242_NDS186_2022}%
  \BibitemOpen
  \bibfield  {author} {\bibinfo {author} {\bibfnamefont {M.~J.}\ \bibnamefont
  {Martin}}\ and\ \bibinfo {author} {\bibfnamefont {C.~D.}\ \bibnamefont
  {Nesaraja}},\ }\bibfield  {title} {\bibinfo {title} {{Nuclear Data Sheets for
  $A=242$}},\ }\href@noop {} {\bibfield  {journal} {\bibinfo  {journal} {Nucl.
  Data Sheets}\ }\textbf {\bibinfo {volume} {186}},\ \bibinfo {pages} {261}
  (\bibinfo {year} {2022})}\BibitemShut {NoStop}%
\bibitem [{\citenamefont {Nesaraja}(2017)}]{A=244_NDS146_2017}%
  \BibitemOpen
  \bibfield  {author} {\bibinfo {author} {\bibfnamefont {C.~D.}\ \bibnamefont
  {Nesaraja}},\ }\bibfield  {title} {\bibinfo {title} {{Nuclear Data Sheets for
  $A=244$}},\ }\href@noop {} {\bibfield  {journal} {\bibinfo  {journal} {Nucl.
  Data Sheets}\ }\textbf {\bibinfo {volume} {146}},\ \bibinfo {pages} {387}
  (\bibinfo {year} {2017})}\BibitemShut {NoStop}%
\bibitem [{\citenamefont {von Egidy}\ \emph {et~al.}()\citenamefont {von
  Egidy}, \citenamefont {Hoff}, \citenamefont {Logheed},\ and\ \citenamefont
  {D.~H.~White}}]{vonEgidy84_PRC29}%
  \BibitemOpen
  \bibfield  {author} {\bibinfo {author} {\bibfnamefont {T.}~\bibnamefont {von
  Egidy}}, \bibinfo {author} {\bibfnamefont {R.~W.}\ \bibnamefont {Hoff}},
  \bibinfo {author} {\bibfnamefont {R.~W.}\ \bibnamefont {Logheed}},\ and\
  \bibinfo {author} {\bibfnamefont {H.~G.~B.}\ \bibnamefont {D.~H.~White}},\
  }\href@noop {} {\ }\BibitemShut {NoStop}%
\bibitem [{\citenamefont {Akovali}(2001)}]{A=250_NDS94_2001}%
  \BibitemOpen
  \bibfield  {author} {\bibinfo {author} {\bibfnamefont {Y.~A.}\ \bibnamefont
  {Akovali}},\ }\bibfield  {title} {\bibinfo {title} {{Nuclear Data Sheets for
  $A=250$}},\ }\href@noop {} {\bibfield  {journal} {\bibinfo  {journal} {Nucl.
  Data Sheets}\ }\textbf {\bibinfo {volume} {94}},\ \bibinfo {pages} {131}
  (\bibinfo {year} {2001})}\BibitemShut {NoStop}%
\bibitem [{\citenamefont {Hellemans}\ \emph {et~al.}(2012)\citenamefont
  {Hellemans}, \citenamefont {Heenen},\ and\ \citenamefont
  {Bender}}]{Hellemans12}%
  \BibitemOpen
  \bibfield  {author} {\bibinfo {author} {\bibfnamefont {V.}~\bibnamefont
  {Hellemans}}, \bibinfo {author} {\bibfnamefont {P.-H.}\ \bibnamefont
  {Heenen}},\ and\ \bibinfo {author} {\bibfnamefont {M.}~\bibnamefont
  {Bender}},\ }\bibfield  {title} {\bibinfo {title} {{Tensor part of the Skyrme
  energy density functional. III. Time-odd terms at high spin}},\ }\href@noop
  {} {\bibfield  {journal} {\bibinfo  {journal} {Phys. Rev. C}\ }\textbf
  {\bibinfo {volume} {85}},\ \bibinfo {pages} {014326} (\bibinfo {year}
  {2012})}\BibitemShut {NoStop}%
\bibitem [{\citenamefont {Dietrich}\ \emph {et~al.}(1964)\citenamefont
  {Dietrich}, \citenamefont {Mang},\ and\ \citenamefont {Pradal}}]{Dietrich64}%
  \BibitemOpen
  \bibfield  {author} {\bibinfo {author} {\bibfnamefont {K.}~\bibnamefont
  {Dietrich}}, \bibinfo {author} {\bibfnamefont {H.~J.}\ \bibnamefont {Mang}},\
  and\ \bibinfo {author} {\bibfnamefont {J.~H.}\ \bibnamefont {Pradal}},\
  }\bibfield  {title} {\bibinfo {title} {{Conservation of particle number in
  the nuclear pairing model}},\ }\href@noop {} {\bibfield  {journal} {\bibinfo
  {journal} {Phys. Rev.}\ }\textbf {\bibinfo {volume} {135}},\ \bibinfo {pages}
  {B22} (\bibinfo {year} {1964})}\BibitemShut {NoStop}%
\bibitem [{\citenamefont {Egido}\ and\ \citenamefont {Ring}(1982)}]{Egido82}%
  \BibitemOpen
  \bibfield  {author} {\bibinfo {author} {\bibfnamefont {J.~L.}\ \bibnamefont
  {Egido}}\ and\ \bibinfo {author} {\bibfnamefont {P.}~\bibnamefont {Ring}},\
  }\bibfield  {title} {\bibinfo {title} {{Symmetry conserving
  Hartree--Fock--Bogoliubov theory (I). On the solution of variationnal
  equations.}},\ }\href@noop {} {\bibfield  {journal} {\bibinfo  {journal}
  {Nucl. Phys. A}\ }\textbf {\bibinfo {volume} {383}},\ \bibinfo {pages} {189}
  (\bibinfo {year} {1982})}\BibitemShut {NoStop}%
\bibitem [{\citenamefont {Heenen}\ \emph {et~al.}(1993)\citenamefont {Heenen},
  \citenamefont {Bonche}, \citenamefont {Dobaczewski},\ and\ \citenamefont
  {Flocard}}]{Heenen93}%
  \BibitemOpen
  \bibfield  {author} {\bibinfo {author} {\bibfnamefont {P.-H.}\ \bibnamefont
  {Heenen}}, \bibinfo {author} {\bibfnamefont {P.}~\bibnamefont {Bonche}},
  \bibinfo {author} {\bibfnamefont {J.}~\bibnamefont {Dobaczewski}},\ and\
  \bibinfo {author} {\bibfnamefont {H.}~\bibnamefont {Flocard}},\ }\bibfield
  {title} {\bibinfo {title} {{Generator-coordinate method for triaxial
  quadrupole dynamics in Sr isotopes (II). Results for particle-number
  projected states.}},\ }\href@noop {} {\bibfield  {journal} {\bibinfo
  {journal} {Nucl. Phys. A}\ }\textbf {\bibinfo {volume} {561}},\ \bibinfo
  {pages} {367} (\bibinfo {year} {1993})}\BibitemShut {NoStop}%
\bibitem [{\citenamefont {Bender}\ and\ \citenamefont
  {Heenen}(2008)}]{Bender08}%
  \BibitemOpen
  \bibfield  {author} {\bibinfo {author} {\bibfnamefont {M.}~\bibnamefont
  {Bender}}\ and\ \bibinfo {author} {\bibfnamefont {P.-H.}\ \bibnamefont
  {Heenen}},\ }\bibfield  {title} {\bibinfo {title} {{Configuration mixing of
  angular-momentum and particle-number projected triaxial
  Hartree--Fock--Bogoliubov states using the Skyrme energy-density
  functional}},\ }\href@noop {} {\bibfield  {journal} {\bibinfo  {journal}
  {Phys. Rev. C}\ }\textbf {\bibinfo {volume} {78}},\ \bibinfo {pages} {024308}
  (\bibinfo {year} {2008})}\BibitemShut {NoStop}%
\bibitem [{\citenamefont {Bender}\ \emph {et~al.}(2009)\citenamefont {Bender},
  \citenamefont {Duguet},\ and\ \citenamefont {Lacroix}}]{Bender09}%
  \BibitemOpen
  \bibfield  {author} {\bibinfo {author} {\bibfnamefont {M.}~\bibnamefont
  {Bender}}, \bibinfo {author} {\bibfnamefont {T.}~\bibnamefont {Duguet}},\
  and\ \bibinfo {author} {\bibfnamefont {D.}~\bibnamefont {Lacroix}},\
  }\bibfield  {title} {\bibinfo {title} {{Particle-number restoration within
  the energey density functional formalism}},\ }\href@noop {} {\bibfield
  {journal} {\bibinfo  {journal} {Phys. Rev. C}\ }\textbf {\bibinfo {volume}
  {79}},\ \bibinfo {pages} {044319} (\bibinfo {year} {2009})}\BibitemShut
  {NoStop}%
\bibitem [{\citenamefont {Simenel}(2010)}]{Simenel10}%
  \BibitemOpen
  \bibfield  {author} {\bibinfo {author} {\bibfnamefont {C.}~\bibnamefont
  {Simenel}},\ }\bibfield  {title} {\bibinfo {title} {{Particle transfer
  reactions with the time-dependent Hartree--Fock theory using a
  particle-number projection technique}},\ }\href@noop {} {\bibfield  {journal}
  {\bibinfo  {journal} {Phys. Rev. Lett.}\ }\textbf {\bibinfo {volume} {105}},\
  \bibinfo {pages} {192701} (\bibinfo {year} {2010})}\BibitemShut {NoStop}%
\bibitem [{\citenamefont {Rodriguez}\ and\ \citenamefont
  {Egido}(2011)}]{Rodriguez11}%
  \BibitemOpen
  \bibfield  {author} {\bibinfo {author} {\bibfnamefont {T.~R.}\ \bibnamefont
  {Rodriguez}}\ and\ \bibinfo {author} {\bibfnamefont {J.~L.}\ \bibnamefont
  {Egido}},\ }\bibfield  {title} {\bibinfo {title} {{Configuration mixing
  description of the nucleus $^{44}$S}},\ }\href@noop {} {\bibfield  {journal}
  {\bibinfo  {journal} {Phys. Rev. C}\ }\textbf {\bibinfo {volume} {84}},\
  \bibinfo {pages} {051307(R)} (\bibinfo {year} {2011})}\BibitemShut {NoStop}%
\bibitem [{\citenamefont {Jia}(2013)}]{Jia13}%
  \BibitemOpen
  \bibfield  {author} {\bibinfo {author} {\bibfnamefont {L.~Y.}\ \bibnamefont
  {Jia}},\ }\bibfield  {title} {\bibinfo {title} {{Particle-number-conserving
  theory for nuclear pairing}},\ }\href@noop {} {\bibfield  {journal} {\bibinfo
   {journal} {Phys. Rev. C}\ }\textbf {\bibinfo {volume} {88}},\ \bibinfo
  {pages} {044303} (\bibinfo {year} {2013})}\BibitemShut {NoStop}%
\bibitem [{\citenamefont {Zeng}\ and\ \citenamefont {Cheng}(1983)}]{Zeng83}%
  \BibitemOpen
  \bibfield  {author} {\bibinfo {author} {\bibfnamefont {J.~Y.}\ \bibnamefont
  {Zeng}}\ and\ \bibinfo {author} {\bibfnamefont {T.~S.}\ \bibnamefont
  {Cheng}},\ }\bibfield  {title} {\bibinfo {title} {{Particle-number-conserving
  method for treating the nuclear pairing correlation}},\ }\href@noop {}
  {\bibfield  {journal} {\bibinfo  {journal} {Nucl. Phys. A}\ }\textbf
  {\bibinfo {volume} {405}},\ \bibinfo {pages} {1} (\bibinfo {year}
  {1983})}\BibitemShut {NoStop}%
\bibitem [{\citenamefont {Pillet}\ \emph {et~al.}(2002)\citenamefont {Pillet},
  \citenamefont {Quentin},\ and\ \citenamefont {Libert}}]{Pillet02}%
  \BibitemOpen
  \bibfield  {author} {\bibinfo {author} {\bibfnamefont {N.}~\bibnamefont
  {Pillet}}, \bibinfo {author} {\bibfnamefont {P.}~\bibnamefont {Quentin}},\
  and\ \bibinfo {author} {\bibfnamefont {J.}~\bibnamefont {Libert}},\
  }\bibfield  {title} {\bibinfo {title} {{Pairing correlations in an explicitly
  particle-number conserving approach}},\ }\href@noop {} {\bibfield  {journal}
  {\bibinfo  {journal} {Nucl. Phys. A}\ }\textbf {\bibinfo {volume} {697}},\
  \bibinfo {pages} {141} (\bibinfo {year} {2002})}\BibitemShut {NoStop}%
\bibitem [{\citenamefont {N.Pillet}\ \emph {et~al.}(2008)\citenamefont
  {N.Pillet}, \citenamefont {Berger},\ and\ \citenamefont
  {Caurier}}]{Pillet08}%
  \BibitemOpen
  \bibfield  {author} {\bibinfo {author} {\bibnamefont {N.Pillet}}, \bibinfo
  {author} {\bibfnamefont {J.-F.}\ \bibnamefont {Berger}},\ and\ \bibinfo
  {author} {\bibfnamefont {E.}~\bibnamefont {Caurier}},\ }\bibfield  {title}
  {\bibinfo {title} {{Variational multiparticle-multihole configuration mixing
  method applied to pairing correlations in nuclei}},\ }\href@noop {}
  {\bibfield  {journal} {\bibinfo  {journal} {Phys. Rev. C}\ }\textbf {\bibinfo
  {volume} {78}},\ \bibinfo {pages} {024305} (\bibinfo {year}
  {2008})}\BibitemShut {NoStop}%
\end{thebibliography}%

\end{document}